# Driven by Excess? Climatic Implications of New Global Mapping of Near-Surface Water-Equivalent Hydrogen on Mars

Asmin V. Pathare[1], William C. Feldman[1],

Thomas H. Prettyman[1], Sylvestre Maurice[2]

[1]Planetary Science Institute, Tucson, AZ 91106; [2]Institut de Recherche en Astrophysique et Planétologie, Université Paul Sabatier, Toulouse, France.

**Editorial correspondence to:**

Asmin Pathare

Planetary Science Institute

1700 E. Fort Lowell, Suite 106

Tucson, AZ 85719

Email: pathare@psi.edu

**Abstract**

We present improved Mars Odyssey Neutron Spectrometer (MONS) maps of near-surface Water-Equivalent Hydrogen (WEH) on Mars that have intriguing implications for the global distribution of "excess" ice, which occurs when the mass fraction of water ice exceeds the threshold amount needed to saturate the pore volume in normal soils. We have refined the crossover technique of *Feldman et al.* (2011) by using spatial deconvolution and Gaussian weighting to create the first globally self-consistent map of WEH. At low latitudes, our new maps indicate that WEH exceeds 15% in several near-equatorial regions, such as Arabia Terra, which has important implications for the types of hydrated minerals present at low latitudes. At high latitudes, we demonstrate that the disparate MONS and Phoenix Robotic Arm (RA) observations of near surface WEH can be reconciled by a three-layer model incorporating dry soil over fully saturated pore ice over pure excess ice: such a three-layer model can also potentially explain the strong anticorrelation of subsurface ice content and ice table depth observed at high latitudes. At moderate latitudes, we show that the distribution of recently formed impact craters is also consistent with our latest MONS results, as both the shallowest ice-exposing crater and deepest non-ice-exposing crater at each impact site are in good agreement with our predictions of near-surface WEH. Overall, we find that our new mapping is consistent with the widespread presence at mid-to-high Martian latitudes of recently deposited shallow excess ice reservoirs that are not yet in equilibrium with the atmosphere.

**1. Introduction**

There is abundant geomorphic evidence of a large ancient inventory of surface water on Mars, but how much remains extant and where it currently resides is largely unknown: presently, water ice is only stable on the Martian surface near the poles (e.g. *Carr*, 1996). One of the primary objectives of the Mars Odyssey Neutron Spectrometer (MONS), which measured neutron leakage fluxes from the upper meter of the Martian surface, is to generate global maps of near-surface hydrogen abundance. Although MONS provides an unambiguous indicator of the presence of Water-Equivalent Hydrogen (WEH), quantitative details of its magnitude and burial depth depend on the model of the host regolith utilized to interpret the data. Typically, a two-layer near-surface regolith model is assumed that expresses WEH concentration in terms of an upper layer of weight fraction $W_{up}$ having thickness $D$ overlying a semi-infinite lower layer of weight fraction $W_{dn}$ (e.g. *Prettyman et al.*, 2004). By assuming a constant value of $W_{up}$, the relative proportions of WEH and regolith in the lower layer can be derived, thereby allowing for the creation of global Martian maps of near-surface WEH (*Maurice et al.*, 2011).

Here, we present improved global MONS-derived maps of near-surface WEH, based on the "crossover" technique developed by *Feldman et al.* (2011), which self-consistently determines – entirely from MONS neutron flux observations – the WEH content of the upper layer ($W_{up}$). We have refined the *Feldman et al.* (2011) crossover technique by utilizing spatial deconvolution of MONS counting rate data (e.g. *Prettyman et al.*, 2009) and employing a Gaussian-weighted least squares fit parameterization of the deconvolved MONS counting rates to improve determinations of $W_{up}$. These enhancements have allowed us to create ***the most definitive global map to date of the vertical distribution of WEH in the upper meter of the Martian surface*** – which we will utilize to inform fundamental questions related to the global distribution and long-term stability of near-surface water ice and hydrates on Mars.

## 2. Previous MONS Mapping

The first Mars Odyssey Neutron Spectrometer results were reported by *Feldman et al.* (2002), who from just one month of MONS data found evidence for buried water ice at high latitudes (poleward of ±60°) tens of cm below the surface, as well as lower-latitude deposits (in regions such as Arabia Terra) that are consistent with chemically or physically bound subsurface $H_2O$ and/or OH. *Feldman et al.* (2003a) mapped the global distribution of near-surface WEH, assuming a single uniform layer, and found a total inventory of near-surface hydrogen in the upper meter of the Martian surface equivalent to a 13-cm thick global layer of $H_2O$. *Feldman et al.* (2003b) calculated MONS-derived $CO_2$ frost cap thickness variations during northern winter and spring, obtaining values that were significantly lower than those inferred from Mars Orbiter Laser Altimeter (MOLA) measurements and more consistent with Global Circulation Model (GCM) estimates. This detailed analysis of $CO_2$ frost thickness provided an essential calibration (see Fig. 4 in *Feldman et al.* [2003b]) for the Monte Carlo N-Particle eXtended (MCNPX) simulation code more fully described in *Prettyman et al.* (2004) that is used to convert top-of-the-atmosphere neutron current MONS observations to estimates of (sub)surface hydrogen composition. *Prettyman et al.* (2004), which focused on MONS results for high southern latitudes, also presented the first two-layer model (described by three independent attributes: $W_{up}$, $W_{dn}$, and $D$) of near-surface hydrogen content.

*Feldman et al.* (2004a) refined the single layer (i.e., $D = 0$) global MONS mapping of *Feldman et al.* (2003a) by incorporating a two-layer surface model (assuming $W_{up} = 2\%$) equatorward of ±45°, where predicted WEH contents ranged from 2% to 10% (by weight, which is how all WEH / $W_{up}$ / $W_{dn}$ values are expressed in this paper). *Feldman et al.* (2007; 2008a) applied the two-layer surface model (for $W_{up} = 1\%$) to high northern and southern Martian latitudes, finding a strong anti-correlation between lower layer WEH content $W_{dn}$ and apparent depth $D$, as well as north-to-south asymmetries in the high latitude WEH distribution. In order to study the hydrogen content of sand dunes within Olympia Undae, *Feldman et al.* (2008b) applied deconvolution techniques to MONS neutron flux data; similarly, *Prettyman et al.* (2009) utilized Jansson's algorithm to deconvolve MONS epithermal counting rates to better characterize the evolution of Martian seasonal caps. *Diez et al.* (2008) explored the effects of compositional variations upon MONS-derived WEH maps by assessing the range of macroscopic absorption cross sections measured in situ by the Mars Exploration Rovers Spirit and Opportunity, and found that an average value ("Composition 21") generated from soils at five Mars mission landing sites (Viking 1, Viking 2, Pathfinder, Spirit, and Opportunity) minimized MONS mapping variance associated with compositional uncertainties.

*Maurice et al.* (2011) developed new data reduction and analysis techniques for the raw MONS data collected by the four Prism sensors, thereby generating thermal, epithermal, and fast neutron counting rates – which represent the three primary MONS-derived data sets – with significantly reduced counting-rate variances (*Maurice et al.*, 2011). In theory, since there are three MONS-derived measurables (thermal, epithermal, and fast neutron counting rates), and three MCNPX model unknowns ($W_{up}$, $D$, and $W_{dn}$), it should be possible to derive a unique solution for the vertical distribution of WEH. However, the actual response of the fast neutrons to surficial hydrogen is extremely similar to that of the epithermals, making it difficult to disentangle the two data sets (*Feldman et al.*, 2002; *Prettyman et al.*, 2004). Therefore, most previous workers assumed a constant value for $W_{up}$, and then used thermal and epithermal counting rates to produce global maps of Martian $D$ and $W_{dn}$ (e.g. *Feldman et al.*, 2004a; 2007; 2008a; *Maurice et al.*, 2011).

*Feldman et al.* (2011) devised a crossover technique that self-consistently determines $W_{up}$ from the MONS fast and epithermal neutron counting rates. *Feldman et al.* (2011) generated the first-ever global map of $W_{up}$ by applying their crossover technique to MONS fast and epithermal counting rates from large sliding 1800-km diameter Regions of Interest (ROIs). However, their initial crossover-derived $W_{up}$ mapping (see Fig. 7 of *Feldman et al.*, 2011) contained obvious inconsistencies: for example, several regions at lower latitudes were mapped as having *negative* values of $W_{up}$, which is of course unphysical.

In this work, we improve upon previous MONS-derived global mapping of $W_{dn}$, $D$, and $W_{up}$ (*Maurice et al.*, 2011; *Feldman et al.*, 2011) in two ways. First, we apply an iterated spherical harmonic expansion of the latitude-longitude count-rate matrix to the *Maurice et al.* (2011) global mapping of epithermal, thermal, and fast neutron count rates using a newly-developed deconvolution algorithm (described in the **Appendix**). Secondly, we improve upon the *Feldman et al.* (2011) crossover technique to calculate $W_{up}$, which utilized unweighted sliding 1800-km ROIs, by incorporating a Gaussian-weighted least squares fit parameterization. Since we self-consistently derive $W_{up}$ directly from MONS data (instead of assuming a constant value like most previous workers), our resulting $W_{up}$-dependent maps of $W_{dn}$ and $D$ represent the most definitive global mapping to date of the vertical distribution of WEH in the upper meter of the Martian surface.

## 3. Methodology
### 3.1 Neutron Count Rates

The starting point for our global mapping of near-surface WEH on Mars is the apportionment by *Maurice et al.* (2011) of epithermal, thermal, and fast neutron count rates to equal-area grids equivalent to 1° of latitude x 1° of longitude at the equator. The comprehensive analysis of *Maurice et al.* (2011) included MONS observations of frost-free Martian surfaces, beginning in February of 2002 with the attainment of a near-polar (~93° inclination) 400 km altitude circular orbit and extending for approximately 3.5 Martian years through July of 2009. The accumulation time for individual data samples is 1 second, and the total number of samples in these data sets is sufficiently large that uncertainties in each of the count-rate entries in the count-rate arrays are limited by systematic effects, not statistics (*Maurice et al.*, 2011). Undeconvolved global maps of epithermal, thermal, and fast neutron count rates are shown in **Fig. 1a**, **Fig. 2a**, and **Fig. 3a**, respectively (note that these global maps have been generated from the same epithermal/thermal/fast neutron count rate data sets utilized to produce the global maps shown in Fig. 17 / Fig. 25 / Fig. 30 of *Maurice et al.*, 2011).

In order to improve upon the *Maurice et al.* (2011) global mapping of epithermal, thermal, and fast neutron count rates, we spatially deconvolved all three of these MONS data sets. Whereas previously we have utilized the Pixon method (*Puetter*, 1995; *Feldman et al.*, 2008a) and an iterated Jansson algorithm (*Jansson*, 1997; *Prettyman et al.*, 2009), in this work we implement a new deconvolution methodology (detailed in the **Appendix**) involving a tesseral spherical harmonics (TSH) expansion (a.k.a. "real" spherical harmonics) of the latitude-longitude count rate matrices. As discussed below, we find that a deconvolution solution utilizing a TSH order of {N=16} provides the most physically credible representation of the MONS data (where "order" is defined in **Appendix Eq. A2**). Our resulting deconvolved {N=16} global maps of epithermal, thermal, and fast neutron count rates are shown in **Fig. 1b**, **Fig. 2b**, and **Fig. 3b**, respectively.

*Epithermal neutron counting rates* $C_E$ can be calculated directly from the MONS downward-facing Prism-1 sensor (*Feldman et al.*, 2003b). The global maps of undeconvolved (**Fig. 1a**) and {N=16} deconvolved (**Fig. 1b**) epithermal neutron count rates are fairly similar at lower latitudes: the primary difference is that the deconvolved map exhibits slightly higher local maxima – as evident by the increased intensity within the deconvolved $C_E > 8$ counts/s contours (**Fig. 1b**). For example, the global maximum undeconvolved neutron count rate value of $C_E = 10.7$ counts/s (located at 25°S, 95°W in Solis Planum) increases by 6% to $C_E = 11.4$ counts/s in the deconvolved map (**Figs. 1a, 1b**: note all deconvolved {N=16} and undeconvolved mapping data are provided in supplemental **Table S1** and **Table S2**, respectively). At higher latitudes, the undeconvolved and deconvolved {N=16} epithermal maps vary more significantly from one another (**Fig. 1**). For instance, the global minimum value of $C_E$ that occurs near 87°N, 0°E in the nearly pure water ice North Polar Layered Deposits (NPLD) decreases by 29% from $C_E = 1.22$ counts/s in the undeconvolved solution (**Fig. 1a**) to $C_E = 0.87$ counts/s in the deconvolved {N=16} solution (**Fig. 1b**).

*Thermal neutron counting rates* $C_T$ can be calculated from the difference between the MONS forward-facing Prism-2 and backward-facing Prism-4 sensors (*Feldman et al.*, 2003b). As noted by *Maurice et al.* (2011), the undeconvolved global thermal neutron flux map (**Fig. 2a**) is similar overall to the undeconvolved global epithermal neutron flux map (**Fig. 1a**) – aside from the sensitivity of thermal neutrons to carbon dioxide, resulting in elevated values of $C_T$ centered on the permanent $CO_2$ deposit in the South Polar Residual Cap (SPRC). The global maps of undeconvolved (**Fig. 2a**) and {N=16} deconvolved (**Fig. 2b**) thermal neutron count rates exhibit far more relative variation to one another than do their epithermal counterparts. Perhaps the most noticeable difference is how much sharper features appear in the smoother deconvolved

{N=16} global thermal neutron flux map (**Fig. 2b**) relative to the undeconvolved solution (**Fig. 2a**). Another notable deconvolved/undeconvolved distinction occurs at high latitudes in the deconvolved {N=16} map, where the local minima within the $C_T < 3.5$ counts/s contours extend across most of the 60-70°N and 60-70°S latitude bands (**Fig. 2b**). Lastly, the deconvolved {N=16} map exhibits slightly higher local maxima at low-to-mid latitudes, as indicated by the increased intensity within $C_E > 6.5$ counts/s contours (**Fig. 2b**): for example, the deconvolved global maximum of $C_T = 8.28$ counts/s in eastern Elysium Planitia at 7°N, 165°E represents an increase of 9% from the same location on the undeconvolved map (**Fig. 2a**).

*Fast neutron counting rates* $C_F$ can be calculated directly from the MONS downward-facing Prism-1 sensor category-2 data (*Maurice et al.*, 2011). As with the thermal neutron counting rates (**Fig. 2**), features in the smoother deconvolved {N=16} fast neutron flux map (**Fig. 3b**) are significantly sharper than corresponding undeconvolved features (**Fig. 3a**). The deconvolved {N=16} global minimum value of $C_F = 0.463$ counts/s at 87°N, 1°E is 11% lower than at the same place in the North PLD in the undeconvolved map, whereas the deconvolved global maximum value in Solis Planum of $C_F = 1.62$ counts/s at 23°S, 91°W is only 1.4% greater compared to the undeconvolved map (**Fig. 3**). That the global maximum and minimum values of $C_F$ occur in nearly the same locations as those for $C_E$ (**Fig. 1**) is no coincidence, since as *Maurice et al.* (2011) noted, the epithermal and fast neutron counting rates are remarkably similar (compare **Fig. 1b** to **Fig. 3b**). A global linear fit to both deconvolved {N=16} data sets yields the following equation: $C_F = 0.51+0.11*C_E$, with a correlation coefficient of 98% (for comparison, the epithermal and thermal neutron count rates, $C_E$ and $C_T$, only exhibit an 80% correlation). The very high global correlation of the epithermal and fast neutron count rate data sets led *Maurice et al.* (2011) to conclude that "both maps yield nearly the same information about the single-layer distribution of hydrogen on Mars." However, we note that the ratio of $R_{E/F} = C_E / C_F$ (the global mean of which is 4.98) appears to be sensitive to the near-surface WEH content, ranging from $R_{E/F} = 1.86$ in high-latitude regions of low $C_E$ to $R_{E/F} = 7.10$ in low-latitude areas of high $C_E$ – this sensitivity is the foundation of the crossover technique that self-consistently calculates $W_{up}$ directly from MONS data (**Sec. 3.3**).

**3.2 Model and Calibration**

In order to derive actual Martian near-surface WEH content from raw neutron count rates, the instrument response of MONS to simulated Martian surfaces must be modeled. Such simulations were first conducted by *Prettyman et al.* (2004), who utilized the Monte Carlo N-Particle eXtended (MCNPX) Version 2.5 radiation transport code of *Waters* (1999) to calculate the energy distribution and angular dependence of the epithermal, thermal, and fast neutron current at the top of the atmosphere (TOA) for a variety of Martian atmospheric masses and spatially heterogeneous surface compositions (see Appendix B of *Prettyman et al.*, 2004). Additionally, *Prettyman et al.* (2004) performed parallel Monte Carlo simulations to convert TOA neutron leakage fluxes to MONS-measured count rates, accounting for factors such as neutron ballistic trajectories and mean lifetimes, spacecraft orbit and velocity, and the laboratory-measured MONS response function for epithermal, thermal, and fast neutrons (see Appendix C of *Prettyman et al.*, 2004).

The dependence of neutron counting rate on leakage flux for thermal, epithermal, and fast neutrons can be expressed as (*Feldman et al.*, 2003b; *Prettyman et al.*, 2004):

$$C_T = \{P_2 - P_4\} = \delta_T \cdot (0.081 + 16.8 \, J_T) \tag{1}$$

$$C_E = \{P_1\} = \gamma_E \cdot (0.224 + 9.17\, J_E - 1.07\, J_E^2) \qquad (2)$$

$$C_F = \{P_{1b}\} = \tau_F \cdot (1.08\, J_F + 0.628\, J_F^2) \qquad (3)$$

where $J_T$, $J_E$, $J_F$ are TOA neutron leakage currents, $P_x$ refers to a neutron count rate measured at a MONS Prism detector, and $\delta_T$, $\gamma_E$, $\tau_F$ are conversion constants for thermal, epithermal, and fast neutrons, respectively. (Note that Eq. 2 in the similar formulation of *Maurice et al.* [2011] mistakenly reprints their Eq. 1 instead of recapitulating Eq. 2 from *Feldman et al.* [2003b] – an equation which also contains a typo, as it mistakenly omitted the exponent in the final term.) The conversion constants can be derived by calibrating the MCNPX-based modeling to high-latitude wintertime circumpolar observations when the entire MONS field of view is covered in $CO_2$ (*Feldman et al.*, 2003b; *Prettyman et al.*, 2004).

In this work, we have revised the *Prettyman et al.* (2004) radiative transport modeling by utilizing an updated atmospheric composition (*Williams*, 2016) and employing Version 2.7 of MCNPX (*Pelowitz et al.*, 2011). These refinements yield values for the conversion constants of $\delta_T = 2.05$, $\gamma_E = 2.51$, and $\tau_F = 2.54$. Once the exact relationship between MONS-detected neutron count rates and TOA neutron fluxes is known, then the near-surface WEH content can be derived from MCNPX-generated grids, as discussed below in **Sec. 3.4** for a two-layer surface model.

***One-Layer Model.*** For a uniform single-layer model, the relationship between the WEH mass fraction $M_{H2O}$ of the soil and the observed epithermal, fast, and thermal count rates can be expressed as a polynomial series of the form:

$$M_{H2O} = \exp\left(\sum_{i=0}^{7} A_{ix}\, C_x^i\right) \qquad (4)$$

For $0.01 < M_{H2O} < 1.0$ and $0 \leq i \leq 7$, the most robust result was obtained by fitting the natural log of the WEH mass fraction (hence the exponential). **Table 1** lists all the $A_{ix}$ parameters in **Eq. 4** for $x = E$ (epithermal), $x = F$ (fast), and $x = T$ (thermal) neutron count rates. **Fig. 4** and **Fig. S4** show that the polynomial fits (red dashed lines) to the modeled mass fractions (black lines) are excellent, as the epithermal (**Fig. 4a**) and fast (**Fig. 4b**) neutron flux fits exhibit less than 2% and 4% error, respectively, over the entire WEH range of mass fractions, and the thermal neutron flux fit (which is only applicable to the monotonically increasing portion of the MCNPX model curve corresponding to WEH > 10%) exhibits less than 1% error (**Fig. S4**).

Global maps of WEH abundance can be constructed from **Eq. 4** and **Table 1** assuming a single uniform layer – or, equivalently, an upper layer with depth $D = 0$ (henceforth referred to as "WEHD0"). As noted by *Feldman et al.* (2004a), such one-dimensional WEHD0 maps correspond to lower-limit abundances of the $H_2O$ content that exists at greater depths in a two-layer model (in the upper layer closer to the surface, the $H_2O$ content will likely be lower than the WEHD0 values due to dessication by insolation-induced thermal variations: *Mellon and Jakosky*, 1993). **Fig. 5** shows the WEHD0 mapping for both undeconvolved and deconvolved {N=16} data derived from epithermal and fast neutron count rates: the significant differences between WEHD0 (epi) and WEHD0 (fast) evident in these figures (compare **Fig. 5a/b** to **Fig. 5c/d**) are the basis of the *Feldman et al.* (2011) crossover technique for calculating $W_{up}$ (**Sec. 3.3**).

Interestingly, a boot-shaped local maximum centered in Arabia at 0°N, 21°E in the deconvolved epithermal WEHD0 map (**Fig. 5b**) occurs at the same location as a similarly-shaped maximum in the deconvolved fast neutron count rate map (**Fig. 3b**) – neither of which are evident in the corresponding undeconvolved maps (**Figs. 3a, 5a**). The fact that two entirely separate deconvolutions of two independent MONS-measured data sets (i.e., epithermal and fast neutrons count rates) both reveal a common hitherto unseen feature supports the accuracy of our deconvolution methodology for TSH order {N=16}.

*Higher order deconvolution.* We have also produced global neutron counting rate maps for TSH deconvolution solutions at higher orders corresponding to {N=20} and {N=25}, as shown in the supplemental figures for epithermal (**Fig. S1**), thermal (**Fig. S2**), and fast (**Fig. S3**) neutron fluxes. The improvement in spatial resolution with increasing TSH deconvolution order $N$ is particularly apparent when comparing low-to-mid latitude features in the {N=16} global maps (**Figs. 1b, 2b, 3b**) to the {N=25} global maps (**Figs. S1b, S2b, S3b**). However, a potential pitfall of deconvolution is the creation of spurious artifacts at higher orders.

Consider **Fig. 6**, which plots global minimum epithermal, thermal, and fast neutron count rates (**Fig. 6a**), along with corresponding global maximum WEHD0 values (**Fig. 6b**), as a function of deconvolution harmonic order {N}. Global minimum epithermal fluxes are less than $C_E = 0.61$ counts/s for all {N} < 12 and {N} > 16 (**Fig. 6a**), resulting in unphysically large epithermal WEHD0 values that exceed 150% for these {N} values (**Fig. 6b**). The spatial extent of such overfitting is shown in supplemental **Fig. S5**, as WEHD0 > 100% regions (indicated by maroon contours) span much of the North PLD for both the {N=20} and {N=25} deconvolution solutions. Hence, **Fig. 6** implies that only TSH deconvolution orders of $12 \leq$ {N} $\leq 16$ will yield credible global solutions of the epithermal WEHD0 data (e.g. **Fig. 5b**). This same range of $12 \leq$ {N} $\leq 16$ also yields reasonable maximum thermal WEHD0 values, unlike higher {N} $\geq 19$ orders that exhibit thermal WEHD0 maxima exceeding 150% (**Fig. 6b**). Therefore, the global deconvolved {N=16} mapping of WEHD0 is fundamentally more physically plausible than the global {N=20} and {N=25} deconvolution solutions.

### 3.3 Crossover Calculations of $W_{up}$

*Feldman et al.* (2011) devised a crossover technique that self-consistently determines $W_{up}$ from the MONS fast and epithermal neutron counting rates. This approach relies on the fact that conversion of MONS-derived fast and epithermal counting rates to the near-surface WEH content of a single-layer deposit assumed to be homogeneous (WEHD0) follows a linear trend having slope and ordinate-intercept parameters that can be used to estimate $W_{up}$ (*Feldman et al.*, 2011). Effectively, if a spatial region of interest (ROI) contains a sufficient range of $W_{dn}$ and $D$ values within its perimeter, then the MONS data can be used to determine $W_{up}$ (*Feldman et al.*, 2011). For example, **Fig. 7a** shows the application of the crossover technique for MCNPX simulations of a typical $W_{up}$ = 2% Martian surface (for two-layer subsurface parameters ranging from $W_{dn}$ = 2-100% and $D$ = 5-100 g/cm$^2$): the predicted "crossover" – i.e., the location where the best-fit linear solution (solid line) to WEHD0 (epi) vs. WEHD0 (fast) intersects with the $y = x$ identity function (dashed line) – occurs at a value of $X_o$ = 1.93%, which is very close to the value of $W_{up}$ = 2% used to create this simulated data. As shown in **Fig. 7b**, such crossover calculations can also approximately reproduce $W_{up}$ values over an MCNPX simulation range of $W_{up}$ = 1-10%, giving us great confidence in the accuracy of the general crossover approach.

*Feldman et al.* (2011) produced the first-ever global map of $W_{up}$ by applying the crossover technique to MONS fast and epithermal neutron counting rate data from large sliding 1800-km diameter ROIs. However, their initial crossover-derived $W_{up}$ mapping (see Fig. 7 of *Feldman et al.*, 2011) was clearly erroneous, as several regions equatorward of ±45° were mapped as having unphysical *negative* values of $W_{up}$. *Feldman et al.* (2011) attributed these errors to the equal weighting of every fast and epithermal neutron data point within the large ROIs needed to sufficiently constrain the crossover calculations, which degraded the accuracy of their final crossover-derived $W_{up}$ maps.

We have improved upon the *Feldman et al.* (2011) crossover technique to calculate $W_{up}$ by weighting each MONS-derived WEHD0 measurement based on its distance $R$ (measured along the surface of a Mars-sized sphere) from the center of the ROI. We utilize a Gaussian least squares fit parameterization that employs the standard chi-square goodness-of-fit test modified to account for weighted data points. The Gaussian weight, $w_g$, for every point within the ROI is given by:

$$w_g = 1/ \sqrt{\exp((R/R_0)^2)/2} \qquad (5)$$

where $R_0$ = a normalization factor set by trial and error to $R_0$ = 1300 km, and the radius of the ROI is given by $R_{max} = 4*R_0$. Although the size of the resulting ROIs, $R_{max}$ = 5200 km, is much larger than the $R_{max}$ = 1800 km ROIs used by *Feldman et al.* (2011), the inverse distance-squared Gaussian weighting results in a more physically coherent solution.

**Fig. 8** shows the global maps of $W_{up}$ derived from our modification of the *Feldman et al.* (2011) crossover technique. For the undeconvolved case (**Fig. 8a**), we obtain a minimum of $W_{up}$ = -0.06% just north of Valles Marineris and a maximum of $W_{up}$ = 3.1% along the western flanks of Olympus Mons. For the deconvolved {N=16} case (**Fig. 8b**), the minimum, which is still near Valles Marineris, is $W_{up}$ = 0.34%, while the maximum of $W_{up}$ = 3.8% occurs in Promethei Terra just off the South PLD. Generally speaking, the lower the value of the distance normalization factor $R_0$, the greater the effective spatial resolution of the global $W_{up}$ maps: these calculations assume $R_0$ = 1300 km, which is the lowest value that yields non-negative $W_{up}$ across ALL of Mars in the deconvolved {N=16} solution. (Note that even though the undeconvolved global minimum has a slightly negative value of $W_{up}$ = -0.06%, this is well within our expected error for Gaussian-weighted crossover calculations of $W_{up}$, so for simplicity we also employ $R_0 = 1300$ km for the undeconvolved case to facilitate more direct comparisons with the deconvolved solution.)

The undeconvolved and {N=16} deconvolved $W_{up}$ maps are fairly similar at low latitudes (i.e., equatorward of 30°), but begin to manifest significant differences at "moderate" latitudes (i.e., from 30°-60°), and then deviate considerably from one another at high latitudes (i.e., poleward of 60°), where the deconvolved {N=16} map exhibits a stronger latitudinal dependence, especially in the northern hemisphere (**Fig. 8b**). These {N=16} deconvolved $W_{up}$ results are broadly consistent with the global map of Martian $H_2O$ derived from infrared OMEGA (Observatoire pour la Minéralogie, l'Eau, les Glaces et l'Activité) observations (see Fig. 4 of *Audouard et al.*, 2014), where a strong dependence on latitude is evident (note the much larger OMEGA latitudinal signal is expected, given the shallower depth of the OMEGA map [upper few mm] relative to the $W_{up}$ map [upper tens of cm]). In stark contrast, higher order {N=20} and {N=25} deconvolution maps (**Fig. S8**) exhibit $W_{up}$ values at high northern latitudes that are actually lower than the undeconvolved case (**Fig. 8a**), with $W_{up}$ even going unphysically negative in the {N=25} solution (**Fig. S8b**, pink contour). Therefore, the more physically reasonable latitudinal dependence of $W_{up}$ at mid-to-high latitudes shown in **Fig. 8b** suggests that the global deconvolved {N=16} map not only improves upon the corresponding undeconvolved case (**Fig. 8a**), but is also more credible than higher order {N=20} and {N=25} deconvolution solutions (**Fig. S8**).

### 3.4 Gridded Interpolation of $W_{dn}$ and *Depth*

Once $W_{up}$ has been calculated, $W_{dn}$ and $D$ can be determined from observed epithermal and thermal neutron count rates via interpolation of simulated $W_{up}$-dependent grids. Following the general approach of *Prettyman et al.* (2004), we utilized version 2.7 of the MCNPX Monte Carlo code (*Pelowitz et al.*, 2011) to convert thermal and epithermal neutron count rates to $W_{dn}$ and $D$ contours, assuming standard atmospheric abundances (*Prettyman et al.*, 2004) and a soil chemistry corresponding to the average regolith composition measured at the MER, Pathfinder, and Viking landing sites (i.e., "Soil 21" from *Diez et al.*, 2008).

We generated such MCNPX grids for $W_{up}$ = 1, 1.5, 2, 2.5, 3, 3.5, 4, 4.5, 5, 6, 7, 8.5, 10 %: our model results are shown in **Fig. 9** for $W_{up}$ = 1% (**Fig. 9a**) and $W_{up}$ = 2% (**Fig. 9b**). Within each $W_{up}$-dependent grid, constant depth contours for $D$ = 0, 5, 10, 20, 30, 60, 100 g/cm² (red labels) start along the lower boundary – ranging from $D$ = 0 g/cm² at lower left up to $D$ = 100 g/cm² along the bottom right – and rise upward to the $W_{dn}$ = $W_{up}$ termination point of 1% in **Fig. 9a** and 2% in **Fig. 9b** (since in our two-layer model $W_{up}$ cannot exceed $W_{dn}$). Similarly, contours of constant $W_{dn}$ = 1, 1.5, 2, 2.5, 3, 3.5, 4, 4.5, 5, 6, 7, 8.5, 10, 12.5, 15, 20, 30, 60, 100 % (black labels) start along the upper zero-depth boundary (ranging from $W_{dn}$ = $W_{up}$ at upper right to $W_{dn}$ = 100% at lower left) and curve downward before rising back up to the $D$ = 100 g/cm² boundary on the right-hand side. (Incidentally, the modeled values of $W_{dn}$ and $C_E$ along the upper $D$ = 0 g/cm² boundary were used to calculate WEHD0 [epi] in **Fig. 4a**.)

We can calculate $W_{dn}$ and $D$ from MONS-derived neutron counting rates by converting the irregularly-spaced MCNPX grid simulations (**Fig. 9**) via interpolation to finer grids spaced at regular 0.01/s counting rate intervals. This allows for more precise determinations of $W_{dn}$ and $D$, both of which are sensitive to the value of $W_{up}$. For example, a MONS-derived data point of $C_E$ = 5.00 epithermal neutron counts/s and $C_T$ = 4.00 thermal neutron counts/s (denoted by asterisks in **Fig. 9**) corresponds to $W_{dn}$ = 21.5% and $D$ = 29.6 g/cm² for $W_{up}$ = 1% (**Fig. 9a**), but $W_{dn}$ = 32.5% and $D$ = 35.7 g/cm² for $W_{up}$ = 2% (**Fig. 9b**).

Such increases in the lower-layer WEH content, $W_{dn}$, are particularly significant because only the larger value for $W_{dn}$ exceeds the threshold for "excess ice" – i.e., the mass fraction of water ice that is greater than that needed to saturate the pore volume in normal soils (assuming mineral grains contain no bound water):

$$W_{sat} = \rho_{ice}P_o / (\rho_g(1-P_o) + \rho_{ice}P_o) \tag{6}$$

where $\rho_{ice}$ is the density of ice = 0.92 g/cm³ and $\rho_g$ is the intrinsic soil density of the regolith grains = 2.75 g/cm³ (*Mellon et al.*, 2004). Viking and Phoenix lander measurements indicate that the Martian soils have a maximum porosity $P_o$ of roughly 50% by volume (*Moore et al.*, 1979; *Zent et al.*, 2010). For porosity $P_o$ = 0.50, **Eq. 6** yields pore ice saturation at $W_{sat}$ = 25% (by weight). Therefore, $W_{dn}$ values that exceed 25% cannot have been emplaced in the subsurface via simple atmospheric diffusion.

For every pair of MONS-measured epithermal and thermal neutron count rates (mapped in **Figs. 1,2**), we calculate $W_{dn}$ and $D$ for each of our MCNPX-simulated $W_{up}$-dependent grids, and then conduct a subsequent interpolation utilizing our crossover-derived value of $W_{up}$ (**Fig. 8**). For example, at 57°S and 171°W, our deconvolved {N=16} solution gives epithermal and thermal neutron counts of $C_E$ = 5.00 counts/s and $C_T$ = 4.03 counts/s, respectively, and a crossover-derived $W_{up}$ = 1.72% (see supplemental **Table S1**). This then yields interpolated values of $W_{dn}$ = 28.7% and $D$ = 33.5 g/cm$^2$ that are intermediate to the $W_{dn}$ and $D$ values at $W_{up}$ = 1% and $W_{up}$ = 2% listed above for $C_E$ = 5.00 counts/s and $C_T$ = 4.00 counts/s.

Lastly, the fidelity of the MONS-observed neutron count rate data points to the MCNPX-simulated grids can be used to assess whether spatial deconvolution has resulted in overfitting. **Fig. 10a** displays MONS-mapped neutron count rates corresponding to 1.5% < $W_{up}$ < 2% for the deconvolved {N=16} solution, mapped onto the MCNPX-simulated $W_{up}$ = 1.5% grid: similar maps are shown for 2% < $W_{up}$ < 2.5% data points on the $W_{up}$ = 2% model grid (**Fig. 10b**), 2.5% < $W_{up}$ < 3% data points on the $W_{up}$ = 2.5% model grid (**Fig. 10c**), and 3% < $W_{up}$ < 3.5% data points on the $W_{up}$ = 3% model grid (**Fig. 10d**). These plots demonstrate that almost every single data point derived from MONS using the deconvolved {N=16} solution and the crossover technique falls within the predicted boundaries of the $W_{up}$-dependent model grids at all latitudes north of 75ºS. Note that the southern circumpolar regions have been excluded because the presence of a $CO_2$ cap may be inconsistent with the assumption of a standard two-layer near-surface WEH model (*Maurice et al.,* 2011), which makes the utility of any flux or WEH results poleward of 75°S highly uncertain.

Our calculations indicate that this nearly uniform data-to-grid concordance (**Fig. 10**) decreases for lower order deconvolutions, due to generally elevated $W_{up}$ values shrinking the average grid size, AND for higher order deconvolutions, because overfitted thermal and epithermal neutron count rates produce more extreme flux values that fall outside of the grids (as shown for {N=20} and {N=25} in supplemental **Fig. S10**). Therefore, we conclude that a TSH order of {N=16} is the most accurate deconvolution solution (in terms of yielding physically plausible results) for our global MONS mapping of near-surface water-equivalent hydrogen.

## 4. Results and Discussion

### 4.1 Global Mapping

The results of our global mapping of $W_{dn}$ are shown in **Fig. 11** for both the undeconvolved and deconvolved {N=16} solutions. Compared to the single-layer WEHD0 maps (**Fig. 5**), the two-layer model $W_{dn}$ maps (**Fig. 11**) exhibit significantly larger average WEH values over much of the planet.

For the undeconvolved solution, the maximum value of $W_{dn}$ = 91.2% occurs in the NPLD (**Fig. 11a**). By contrast, not only does the deconvolved {N=16} solution yield a maximum $W_{dn}$ of 100% near the very center of the NPLD (see **Table S1**), but $W_{dn}$ values also exceed 90% over much of the latitude band between 60°-70°S (**Fig. 11b**). Additionally, there are prominent local $W_{dn}$ maxima – corresponding to deconvolved $W_{dn}$ > 80% (**Fig. 11b**) and undeconvolved $W_{dn}$ > 70% (**Fig. 11a**) – both in the Vastitas Borealis Formation encompassing the Phoenix landing site (68.2°N, 125.7°W) and in a nearly antipodal area in Promethei Terra (65°S, 100°E) just off the South Polar Layered Deposits (SPLD). The most significant high northern latitude minimum occurs in Acidalia Planitia (60-70°N, 30-60°W), where $W_{dn}$ < 50% (**Fig. 11**).

In order to highlight low-latitude variations, **Fig. 11c** and **Fig. 11d** re-plot the global undeconvolved and {N=16} deconvolved $W_{dn}$ maps (**Fig. 11a,b**) on a scale of 0-20%. In both solutions (**Fig. 11c,d**), $W_{dn}$ exceeds 10% across a broad rise in Arabia Terra (centered at 0°N, 20°E) and throughout much of a large U-shaped region (first identified by *Maurice et al.*, 2011) bounded by Utopia Planitia (site of the Viking 2 lander: 48.0°N, 134.3°E) that extends down through Elysium Mons (25°N, 147°E) and western Elysium Planitia to Gale Crater (landing site of the Curiosity Rover: 4.6°S, 137.5°E), traverses laterally across Gusev Crater (landing site of the Spirit Rover: 14.6°S, 175.5°E) and Medusae Fossae (3°S, 160°W), and then rises back up the western flanks of the Tharsis highlands and Olympus Mons to Arcadia Planitia (45°N, 150°W) and Alba Patera (45°N, 110°W).

Equatorial maxima are significantly greater in the deconvolved {N=16} solution (**Fig. 11d**), with $W_{dn}$ values greater than 15% in Arabia Terra and nearly approaching $W_{dn}$ = 20% in Medusae Fossae. In both maps (**Fig. 11c,d**), low-latitude WEH minima corresponding to regions of $W_{dn}$ < 6% occur in eastern Elysium Planitia (5°N, 170°E), Nilosyrtis Mensae (45°N, 75°E), Isidis Planitia (10°N, 100°E), northeastern Hellas Planitia (35°S, 75°E), northern Argyre Planitia (40°S, 40°W), eastern Tempe Terra (45°N, 50°W), north of Valles Marineris (15°N, 80°W), and within the ring of highlands around Solis Planum (15-45°S, 30-100°W) south of Valles Marineris. Note that almost all of the local maxima and minima evident in the **Fig. 11** $W_{dn}$ maps exhibit corresponding minimum or maximum values in the epithermal, thermal, and/or fast neutron count rate maps (**Figs. 1-3**).

**Fig. 12** shows our global mapping of the depth $D$ of the upper layer for both the undeconvolved and deconvolved {N=16} solutions. The deconvolved map (**Fig. 12b**) incorporates a slight smoothing factor: since $D$ becomes effectively meaningless when $W_{dn}$ ~ $W_{up}$, we set the depth to $D$ = 30 g/cm$^2$ (i.e., the midpoint of the range of calculated $D$) in regions where ($W_{dn}$ − $W_{up}$) < 1% to minimize potential errors arising from the relative coarseness of our MCNPX grids at higher $D$ (**Fig. 9**). This mostly occurs in low $W_{dn}$ regions within Elysium Planitia and along the rim of Hellas (**Fig. 11b,d**), corresponding to less than 0.5% of the total area of the global deconvolved $D$ map (**Fig. 12b**).

For the undeconvolved solution, the maximum depth of $D$ = 50.6 g/cm$^2$ occurs at southern mid-latitudes (53°S, 21°E: **Fig. 12a**). This is also approximately where the maximum deconvolved {N=16} depth of $D$ = 59.1 g/cm$^2$ occurs (**Fig. 12b**). Similarly, the relative maxima in depth at northern mid-latitudes (30°N-60°N) are much more prominent in the deconvolved {N=16} solution (**Fig. 12b**), with depths exceeding $D = 50$ g/cm$^2$ in western Acidalia Planitia (50°N, 50°W).

At higher latitudes, both solutions yield a depth of zero (dark purple contour) over most of the SPLD (**Fig. 12**). This is consistent with the low (relative to the NPLD) values of $W_{dn} = 40$-50% measured over much of the SPLD (**Fig. 11a,b**), since MONS is only measuring the water content within the ice-rich dust cover instead of the purer underlying ice. Interestingly, only the deconvolved {N=16} solution yields a similar $D = 0$ result (dark purple contour) at the center of the NPLD (**Fig. 12b**), which of course is completely consistent with the $W_{dn} = 100\%$ results that it produces there (**Fig. 11b**). Instead, the undeconvolved solution produces its other $D = 0$ result (dark purple) close to the equator (**Fig. 12a**). We suspect that this anomalous undeconvolved zero depth result is an artifact of the steep topography associated with Valles Marineris, ***and thus conclude that the deconvolved {N=16} solution produces a more accurate global map of WEH.***

Supplemental **Figs. S11** and **S12** show our higher order {N=20} and {N=25} deconvolved global maps of $W_{dn}$ and $D$, respectively. The various deconvolution solutions are quite robust at low-to-mid latitudes ranging from 45°N to 45°S, where $W_{dn}$ values are fairly similar to one another (compare **Fig. 11d** {N=16} to **Fig. S11c** {N=20} and **Fig. S11d** {N=25}), as are depths (compare **Fig. 12b** {N=16} to **Fig. S12a** {N=20} and **Fig. S12b** {N=25}).

However, at higher latitudes, the {N=20} and {N=25} deconvolved maps of $W_{dn}$ and $D$ diverge significantly from the deconvolved {N=16} solution in unexpected ways. From north to south: (a) both the {N=20} and {N=25} maps unphysically exceed $W_{dn} = 100\%$ on the NPLD (**Fig. S11a,b**); (b) just off the NPLD (~75°N–85°N), both the {N=20} and {N=25} solutions in the eastern hemisphere exhibit minima of $W_{dn} < 30\%$ (**Fig. S11a,b**), even though the undeconvolved map stays above $W_{dn}$ = 50% throughout this region (**Fig 11a**), which is attributable to an anomalous increase (relative to the undeconvolved solution) in deconvolved {N=20} and {N=25} neutron flux (**Figs. S1-S3**) due to overfitting in adjacent areas; (c) at 45°N-60°N, both the {N=20} and {N=25} $D$ maps (**Fig. S12**) exhibit values that are lower than the corresponding undeconvolved depths (**Fig. 12a**), probably because of the implausibly low {N=20} and {N=25} $W_{up}$ values that extend down to these depth maxima (**Fig. S8**); and (d) at about 60°S, the {N=20} map unphysically exceeds $W_{dn}$ = 100% in Promethei Terra (**Fig. S11a**).

In contrast, the deconvolved {N=16} solution exhibits no such high-latitude anomalies. ***Therefore, we conclude that the deconvolved {N=16} global flux and WEH maps are more physically accurate than the deconvolved {N=20} and {N=25} solutions.*** Hence, our subsequent regional analyses will only refer to results from our preferred deconvolved {N=16} solution.

## 4.2 Regional Maps

### 4.2.1 Arabia Terra & Aeolis Planum

The global MONS map shown in **Fig. 11d** indicates that $W_{dn}$ abundances within several near-equatorial local maxima are considerably enhanced relative to prior mapping (e.g. *Feldman et al.*, 2004a,b; *Maurice et al.*, 2011). For example, deconvolved $W_{dn}$ in Arabia Terra (**Fig. 13a**) not only attains a maximum value of 16.5% within Henry Crater (10°N, 23°E), but also exhibits three other local maxima surpassing 14% – all of which are 10-20% higher than the previous $W_{dn}$ maximum of ~13% calculated by *Maurice et al.* (2011). Such elevated WEH abundances greatly constrain the types of hydrated minerals present at low latitudes, as stability calculations indicate that magnesium sulfates such as epsomite ($MgSO_4 \cdot 7H_2O$) are much more likely than either zeolites or clays to be abundant enough to explain very high water contents (e.g. *Feldman et al.*,

2004c; *Fialips et al.*, 2005); alternatively, meridianite ($MgSO_4 \cdot 11H_2O$) may be stable within Henry Crater beneath a thin, low permeability cover layer (*Wang et al.*, 2013). Regardless of the specific magnesium sulfate hydrate that is present, our new $W_{dn}$ results in Arabia Terra appear to be consistent with a mixture of $MgSO_4$ hydrate mass fractions as high as 10 +/- 5% (*Feldman et al.*, 2004c; *Fialips et al.*, 2005; *Wang et al.*, 2013). Our deconvolved {N=16} global mapping (**Fig. 11d**) also exhibits two other notable low latitude maxima (**Fig. 13b**), as $W_{dn}$ exceeds 17% in Aeolis Planum just southeast of Gale Crater (landing site of Curiosity), and $W_{dn}$ approaches 20% in western Medusae Fossae near Gusev Crater (landing site of Spirit).

Recently, *Wilson et al.* (2018) mapped low-latitude equatorial locations of water on Mars by applying a new pixon-based deconvolution technique to the MONS epithermal neutron flux. Overall, their single-layer results are similar to our deconvolved {N=16} one-dimensional WEHD0 estimates (**Fig. 5b**), with one glaring exception: they predict a WEH abundance exceeding 40% in Aeolis Planum (*Wilson et al.*, 2018), which is well above the $W_{sat}$ = 25% pore ice threshold. The pixon-based deconvolution of *Wilson et al.* (2018) yields an epithermal neutron flux in Aeolis Planum of $C_E$ < 2 counts/s, corresponding to WEH > 40% (**Fig. 4a**).

However, our tesseral spherical harmonics deconvolution does not predict epithermal neutron fluxes lower than $C_E$ = 4 counts/s in Aeolis Planum (located along 0-10°S in **Fig. S6**) – even at orders as high as {N=25} – resulting in maximum WEHD0 abundances ranging from 10-13% (**Fig. S7**). Furthermore, epithermal neutron fluxes of $C_E$ < 2 counts/s can only stay within our two-layer MCNPX-modeled grids (**Figs. 9, 10**) if thermal neutron fluxes are lower than $C_T$ = 5 counts/s, which does not occur at low-to-mid latitudes equatorward of 45° in any of our deconvolution solutions (**Figs. 2, S2**). This implies that the *Wilson et al.* (2018) WEH > 40% maximum in Aeolis Planum must be a single layer extending to the actual surface, which is problematic because the absence of a protective lag is difficult to reconcile with the present-day instability of equatorial water ice (e.g. *Mellon and Jakosky*, 1993; *Schorghofer*, 2007).

Returning to our deconvolved {N=16} two-layer MONS model, which predicts low-latitude maxima ranging from 15-20% in Arabia Terra, Aeolis Planum, and Medusae Fossae (**Fig. 13**), it is important to note that these results were calculated for a global average soil composition ("Soil 21" of *Diez et al.*, 2008) that may not be reflective of local conditions. *Diez et al.* (2008) systematically studied the sensitivity of MONS-derived WEH estimates to surficial compositional variations, and found that the key parameter is the macroscopic absorption cross section ($\Sigma$), which based on Mars Exploration Rover (MER) and Gamma Ray Spectrometer (GRS) observations can vary by a factor of 2 from $0.007 < \Sigma < 0.014$. *Diez et al.* (2008) further showed that the macroscopic absorption cross section is most sensitive to the elemental concentrations of Fe and Cl – both of which are elevated in the vicinity of the local $W_{dn}$ maxima associated with Aeolis Planum and Medusae Fossae, according to the global GRS-derived maps of *Boynton et al.* (2007).

Since the implied increase in macroscopic absorption cross section could significantly reduce MONS-derived WEH abundance estimates (*Diez et al.,* 2008), the elevated $W_{dn}$ results shown in **Fig. 13b** for Aeolis Planum and Medusae Fossae need to be verified by conducting MCNPX simulations for a more regionally-appropriate near-surface soil composition (an exercise that is beyond the scope of this paper). In contrast, regional concentrations of Fe and Cl within Arabia Terra are not notably enhanced (*Boynton et al.*, 2007). Therefore, we can confidently conclude that our new deconvolved {N=16} Arabia Terra $W_{dn}$ results (**Fig. 13a**) correspond to significantly larger hydrated water contents than previously predicted by two-layer MONS models of near-surface WEH at low latitudes (e.g. *Feldman et al.*, 2004a; *Maurice et al.*, 2011).

### 4.2.2 Vastitas Borealis

At the Phoenix landing site (68.22°N, 125.75°W: denoted by triangle in **Fig. 14**), our deconvolved two-layer model solution yields $W_{up}$ = 3.24% (**Fig. 8b**), $W_{dn}$ = 72.2% (**Fig. 14a**), and $D$ = 10.1 g/cm² (**Fig. 14b**). Assuming $\rho_{ice}$ = 0.92 g/cm³, $\rho_g$ = 2.75 g/cm³, and $P_o$ = 50%, the effective density of a $W_{up}$ = 3.24% upper layer is $\rho_e$ = 1.40 g/cm³, so the thickness of the upper layer is $T_{up}$ = $D$ / $\rho_g$ = (10.1 g/cm²) / (1.40 g/cm³) = 7.2 cm. For comparison, Phoenix Robotic Arm (RA) in situ trench observations measured an average depth to ice of 4.6 cm (*Mellon et al.*, 2009). However, the RA only detected ice approximately two-thirds of the time: if the ice table is assumed to lie 5 cm below the 28 of 91 RA non-ice detections (the depths of which are listed in Table 3 of *Mellon et al.*, 2009), then the average depth to the ice table would be 7.3 cm, which is nearly equal to the MONS-derived upper layer thickness of $T_{up}$ = 7.2 cm.

However, MONS measurements of $W_{dn}$ are incompatible with Phoenix's RA trench observations (*Mellon et al.*, 2009), given that the lander primarily detected darker pore ice (90% of detections) instead of light toned excess ice (10% of detections). According to the Stereo Surface Imager (SSI) spectral modeling of *Cull et al.* (2010), the darker pore ice at the Phoenix landing site is comprised of 30 ± 20% ice, in contrast to the almost pure 99+% light toned ice. Accounting for the 30% of non-ice detections (*Mellon et al.*, 2009), the Phoenix-detected ice content of the lower layer averages out to (0.9*0.3 + 0.1*0.99)*(28/91) = 25.5%, which is well below the value of $W_{dn}$ = 72.2% derived from MONS.

How can these disparate MONS and Phoenix observations be reconciled? The simplest answer is scale: i.e., the Phoenix landing site (the RA work space spanned roughly 1 m x 1.5 m:

*Mellon et al.*, 2009) may be unrepresentative of MONS' large ~550 km footprint (*Maurice et al.*, 2011). But this answer is unsatisfying, given that (a) the upper layer thicknesses derived from MONS and the Phoenix RA are potentially consistent with one another; and (b) the Phoenix landing site in Green Valley was chosen precisely because it seemed to have more subsurface water ice than the surrounding terrains (*Mellon et al.*, 2008), so for it to have a nearly 50% local deficit in $W_{dn}$ seems highly unlikely.

Therefore, we posit that instead of the standard MONS two-layer near-surface model, MONS data at high latitudes may be better fit by a three-layer near-surface model (e.g. **Fig. 15**). The simplified three-layer near-surface model shown in **Fig. 15** is comprised of:

(a) an ice-poor upper layer ($W_1 = W_{up}$, $D_1 = D$),

(b) a pore-saturated middle layer ($W_2 = W_{sat} = 25\%$,

$$D_2 = 60 \text{ g/cm}^2 * ((W_3 - W_{dn})/(W_3 - W_2)), \tag{7}$$

(c) and a semi-infinite pure ice lower layer ($W_3 = 100\%$).

Since a three-layer model has five parameters and there are still only three independent MONS-derived data sets (epithermal, thermal, and fast neutron fluxes), we made two simplifying assumptions: the middle layer contains fully saturated $W_2 = 25\%$ pore ice, and the lower layer consists of pure $W_3 = 100\%$ ice (which is consistent with Phoenix SSI observations of light toned ice: *Cull et al.*, 2010). As discussed in **Sec. 4.4**, this three-layer near-surface model is conceptually similar to the models of saturated pore ice overlying almost pure ice sheets simulated by *Schorhgofer* (2007) and *Schorghofer and Forget* (2012).

According to **Eq. 7**, the depth of the middle layer ranges from $D_2 = 0$ (for $W_{dn} = 100\%$) to a maximum of $D_2 = 60$ g/cm² (for $W_{dn} = 25\%$), an upper limit that we imposed based on two-layer MCNPX simulations indicating an effective maximum measurable depth of $D = 60$ g/cm² (**Fig. 12b**). The effective density of a $W_2 = 25\%$ layer is $\rho_e = 1.60$ g/cm³, so the thickness of the middle layer ranges from $T_2 = 0$ to a maximum of $T_2 = (60 \text{ g/cm}^2) / (1.60 \text{ g/cm}^3) = 37.4$ cm.

This three-layer model can be used to compute a minimum depth to the excess ice table, $T_E = T_1 + T_2$, which for the Phoenix landing site is equal to $T_E = 7.2$ cm + 13.9 cm = 21.1 cm (**Fig. 14c**). This result is completely consistent with the measurements of the Phoenix RA, which only dug down to a maximum trench depth of 18.3 cm (*Mellon et al.*, 2009). Hence, a simple three-layer model (e.g. **Fig. 15**) can potentially **resolve the long-standing discrepancy between MONS and Phoenix RA observations**. Though the actual structure of near-surface ice at high northern latitudes may be even more complicated by the presence of shallow ice lenses (see Fig. 19 of *Sizemore et al.*, 2015) or unsaturated pore ice, thereby requiring more free parameters and/or layers than shown in **Fig. 15**.

### 4.2.3 Arcadia Planitia

Our simplified three-layer model results are also more consistent (relative to the two-layer model) with the excess ice table depths implied by recent ice-exposing craters at mid-latitudes (*Byrne et al.*, 2009; *Dundas et al.*, 2014). For example, **Fig. 16** shows a recent ice-exposing crater cluster at 46°N, 177°E (corresponding to "Site 1" from *Byrne et al.*, 2009) that impacted Mars within the past 15 years, based on before and after CTX (Context Camera) pictures. *Dundas and Byrne* (2010) modeled the sublimation of ice from mid-latitude recent impacts, and concluded that the persistence of such exposed ice indicates it is most likely relatively pure instead of pore-filling. The comprehensive mapping of *Dundas et al.* (2014) identified 20 recent ice-exposing impacts at latitudes ranging from 39-64°N and 71-74°S, suggesting the presence of widespread subsurface excess ice at mid-to-high Martian latitudes. For example, **Fig. 14c** indicates that two such ice-exposing impacts (white circles) have occurred in Vastitas Borealis, thereby supporting our earlier MONS-derived conclusion of buried excess ice at the nearby Phoenix landing site (grey triangle).

As shown in **Fig. 16**, many of the recent ice-exposing impacts identified by *Byrne et al.* (2009) and *Dundas et al.* (2014) are part of crater clusters, thus further constraining the distribution of near-surface ice, since generally only the deeper craters in a given cluster expose excess ice. Hence, the shallowest ice-exposing crater and the deepest *non*-ice-exposing crater in a cluster can be jointly used to place upper and lower bounds on the minimum depth ($T_E$) to the excess ice table; similarly, lower bounds on $T_E$ can also be imposed by recent impact sites that do not expose any ice at all (*Dundas et al.*, 2014).

**Fig. 17a** plots a region of Arcadia Planitia (spanning 36-56°N, 132-176°W) that contains twelve recent impact sites: six of which did not expose ice (brown circles), three of which did (white circles), and three of which are clusters in which the deeper craters exposed ice and the shallower craters did not (brown-in-white circles). **Fig. 17b** also displays depths for each of these impacts, which can either be determined directly via shadow measurements (as done by *Daubar et al.*, 2014) or estimated using the standard *Pike* (1974; 1977) parameterization of crater depth $d_c = 0.20 * D_c$ (where $D_c$ is the crater diameter), which is consistent with the average crater depth of $d_c = 0.23 * D_c$ measured for recent Martian impacts by *Daubar et al.* (2014). Where such *Daubar et al.* (2014) shadow-derived depths were not available, we utilized *Dundas et al.* (2014)

measurements of crater diameter (though note that our crater depth estimates differ from the shallower crater excavation depths $d_e = 0.084 * D_c$ assumed in Fig. 4 of *Dundas et al.*, 2014).

Using this crater depth data, we can now apply three different tests for our MONS-constrained models of near-surface ice distribution:

*(1) Are ice-exposing impacts in regions of MONS-predicted excess ice ($W_{dn} > 25\%$) deeper than the predicted minimum excess ice table?* As shown in **Fig. 17a**, four ice-exposing impacts (white circles) occur poleward of the dark blue $W_{dn} = 25\%$ contour. Since these craters are all at least 80 cm deep, which is below the deepest minimum excess ice table depth in this region (**Fig. 17b**), the presence of these ice-exposing impacts is consistent with both our two-layer and three-layer near-surface MONS models.

*(2) Do ice-exposing impacts in regions where MONS does not predict excess ice ($W_{dn} < 25\%$) penetrate deeper than the predicted 1-meter MONS sensitivity depth?* As seen in **Fig. 17a**, two ice-exposing impacts (white circles) occur equatorward of the $W_{dn} = 25\%$ contour: since both are much deeper than the estimated ~1 m sensitivity of MONS, neither is inconsistent with our near-surface MONS models. Indeed, the presence of deep excess ice at such low latitudes is concordant with the results of *Bramson et al.* (2015), who concluded based on SHARAD (Shallow Radar) soundings and terraced crater observations that there is a widespread, decameters-thick layer of excess ice throughout much of Arcadia Planitia. Interestingly, only the deeper craters in the cluster at the southernmost impact expose ice (**Fig. 17b**), suggesting that the subsurface excess ice layer predicted by *Bramson et al.* (2015) does not extend to the surface – which is of course consistent with our near-surface MONS models.

*(3) Are non-ice-exposing impacts in regions of MONS-predicted excess ice ($W_{dn} > 25\%$) shallower than the predicted minimum excess ice table?* Including clusters, there are five impacts poleward of the $W_{dn} = 25\%$ contour that do not expose ice (**Fig. 17a**). In order of increasing $W_{dn}$:

(a) $W_{dn} = 25.5\%$, $d_c = 1.1$ m. Given how close this crater is to the $W_{dn} = 25\%$ excess ice threshold, the lack of exposed ice at this impact site can easily be attributed to local variations in $W_{dn}$, porosity $P_o$, and/or soil density $\rho_g$ (note both $P_o$ and $\rho_g$ affect the excess ice threshold via **Eq. 6**).

(b) $W_{dn} = 28.0\%$, $d_c = 35$ cm. The MONS-derived depth (**Fig. 12b**) is $D = 38$ g/cm$^2$, corresponding to a thickness of 27 cm, so the lack of exposed ice at this impact site is NOT consistent with the standard two-layer MONS near-surface model, with an "error" of $E_2 = 35$ cm $- 27$ cm $= 8$ cm. However, the predicted excess ice table depth is $T_E = 63$ cm (**Fig. 17b**), which means the lack of exposed ice here is completely consistent with our three-layer MONS near-surface model (**Fig. 15**).

(c) $W_{dn} = 30.3\%$, $d_c = 40$ cm. The MONS-derived depth (**Fig. 12b**) is $D = 26$ g/cm$^2$, corresponding to a thickness of 19 cm, so the lack of exposed ice at this impact site is inconsistent with the two-layer MONS model, with an "error" of $E_2 = 40$ cm $- 19$ cm $= 21$ cm. However, the predicted excess ice table depth is $T_E = 53$ cm (**Fig. 17b**), which means the lack of exposed ice here is completely consistent with our three-layer MONS model.

(d) $W_{dn} = 42.6\%$, $d_c = 70$ cm. The MONS-derived depth (**Fig. 12b**) is $D = 26$ g/cm$^2$, corresponding to a thickness of 18 cm, so the lack of exposed ice at this impact site is not consistent with the two-layer MONS model, with an "error" of $E_2 = 70$ cm $- 18$ cm $= 52$ cm. The predicted excess ice table depth is $T_E = 47$ cm (**Fig. 17b**), which means the lack of exposed ice here is also inconsistent with our three-layer MONS model; however, the "error" of $E_3 = 70$ cm – 47 cm = 23 cm is significantly less than the two-layer "error" of $E_2 = 52$ cm.

(e) $W_{dn} = 58.1\%$, $d_c = 55$ cm. The MONS-derived depth (**Fig. 12b**) is $D = 36$ g/cm$^2$, corresponding to a thickness of 26 cm, so the lack of exposed ice at this impact site is inconsistent with the two-layer MONS model, with an "error" of $E_2 = 55$ cm – 26 cm = 29 cm. The predicted excess ice table depth is $T_E = 46$ cm (**Fig. 17b**), which means the lack of exposed ice here is also not consistent with our three-layer MONS model; however, the "error" of $E_3 = 55$ cm – 46 cm = 9 cm is significantly less than the two-layer "error" of $E_2 = 29$ cm.

Therefore, we conclude that our three layer MONS model (**Fig. 15**) is a better representation of the near-surface at Martian mid-latitudes than the standard two-layer model, based primarily on its ability to better predict where impacts that do NOT expose ice should occur. But once again, we add the caveat that the actual structure of near-surface ice at Martian mid-latitudes may be complicated by the presence of unsaturated pore ice or shallow ice lenses (*Sizemore et al.*, 2015), thereby requiring more free parameters and/or layers than shown in **Fig. 15**.

### 4.3 High Latitude Anticorrelation of $W_{dn}$ and Depth

*Feldman et al.* (2007) noted the anticorrelation of $W_{dn}$ and $D$ at high northern latitudes between 60°N-70°N, which *Feldman et al.* (2008a) showed also applied to the 60°S-70°S latitude band. According to our new deconvolved {N = 16} solution, this inverse correlation extends to nearly ALL mid-to-high latitudes (**Fig. 18**), as the $W_{dn}$ and $D$ of all points between 50°N-75°N (purple dots) and 50°S-75°S (green dots) are strongly anticorrelated ($R = -0.80$), with a linear best fit of $W_{dn} = 92.3 - 1.65*D$. Although the southern hemispheric $W_{dn}$ vs. $D$ data is significantly steeper ($R = -0.82$, best fit: $W_{dn} = 97.5 - 1.93*D$), the northern hemispheric $W_{dn}$ - $D$ anticorrelation is almost as strong ($R = -0.79$, best fit: $W_{dn} = 87.3 - 1.42*D$). In the context of a two-layer model, this puzzling mid-to-high latitude anticorrelation is difficult to explain, since there is no obvious reason why varying $W_{dn}$ in the lower, buried layer should affect the integrated depth $D$ of the mass of material in the upper layer, given that the precipitation event that deposited the excess ice in the lower layer and its subsequent burial by a drier, upper layer were presumably governed by separate processes (*Feldman et al.*, 2008a).

However, the $W_{dn}$ - $D$ anticorrelation is much easier to understand in terms of the three-layer model shown in **Fig. 15**. That's because the depth $D_2$ of the pore-saturated $W_2 = 25\%$ middle layer is inversely dependent on $W_{dn}$ (**Eq. 7**), varying from $D_2 = 0$ for $W_{dn} = 100\%$ to $D_2 = 60$ g/cm$^2$ for $W_{dn} = 25\%$. So the three-layer model version of **Fig. 18** would simply show that the depth of the upper layer ($D_1$) is proportional to the depth of the middle layer ($D_2$). This makes sense if the WEH content of the upper layer ($W_1 = W_{up}$) is ultimately derived not from the atmosphere by downward diffusion (a process that should result in $D_1$ depths that are independent of $D_2$ depths, the values of which in this three-layer scenario are governed by interactions with the $W_3$ lower layer), but rather from the pure ice ($W_3 = 100\%$) lower layer via upward diffusion through the intervening pore-saturated middle layer (**Fig. 15**).

**4.4 Discussion**

*Schorghofer and Forget* (2012) modeled the history and anatomy of subsurface ice on Mars. They identified two distinct modes of subsurface pore ice growth (depicted in Fig. 4 of *Schorghofer and Forget*, 2012): "volumetric" deposition, involving downward diffusion of atmospheric water vapor driven by thermal cycles (*Mellon and Jakosky*, 1993) that results in partial pore ice deposition both above and below the ice table; and "vertical" deposition upon an impermeable ice layer, in which pore ice is "pasted on" layer by layer, thereby depositing fully saturated pore ice above the ice table. Such pasted-on vertical pore ice growth may have occurred not only at the Phoenix landing site – resulting in a three-layer configuration (much like **Fig. 15**) of relatively dry soil over fully saturated pore ice over nearly pure excess ice (as postulated by *Schorghofer*, 2007) – but also at lower latitudes, where the intermediate pore ice layer can be much thicker (*Schorghofer and Forget*, 2012).

Therefore, we propose that global subsurface pore ice deposition on Mars is driven by the ubiquitous presence of an impermeable excess ice layer at high latitudes. In our scenario, pure ice was deposited via precipitation at higher obliquities wherever MONS now detects evidence for excess ice – basically, everywhere poleward of the dark blue $W_{dn} = 25\%$ contours in **Fig. 11b**. We further suggest that the subsequent evolution of the ice table at a given locale was primarily governed by the obliquity-dependent insolation when ice was last stable on the surface: the greater the insolation, the thicker the resultant sublimation lag, and thus the greater the burial depth of the excess ice layer. So for example the deconvolved MONS-derived maximum depth (**Fig. 12b**) of $D = 59.1$ g/cm$^2$ in Noachis Terra (53°S, 21°E) indicates surface ice has not been present there since an episode of intense high obliquity sublimation, whereas the local depth minimum ($D < 10$ g/cm$^2$) in nearby Promethei Terra (65°S, 100°E) suggests that surface ice has been stable in this region at much more moderate obliquities. Thus, the high latitude anticorrelation of $W_{dn}$ and $D$ plotted in **Fig. 18** is, within the context of our conceptual three-layer model (**Fig. 15**), a natural consequence of sublimation lag formation burying an impermeable excess ice layer that then drives subsequent pore ice deposition.

## 5. Conclusions and Future Work

(a) We have produced the most definitive global map to date of the vertical distribution of WEH in the upper meter of the Martian surface. We improved upon the crossover technique of *Feldman et al.* (2011) by using spatial deconvolution and Gaussian weighting to create the first globally self-consistent (i.e., no unphysical negative values) map of $W_{up}$. This in turn allowed us to improve upon MCNPX-simulated estimates of $W_{dn}$ across Mars, as well as create the first truly global map of depth $D$. All of the global {N=16} deconvolved and undeconvolved MONS mapping results presented in this paper are enumerated in supplemental **Table S1** and **Table S2**, respectively. Although these maps represent the best global solution of the near-surface that can be simultaneously applied to all of Mars, future work simulating regional variations (depending on factors such as albedo and thermal inertia) may produce more accurate local results.

(b) At low latitudes, our new maps indicate that $W_{dn}$ exceeds 15% in several near-equatorial regions, including Arabia Terra. These elevated $W_{dn}$ values – which are higher than previous MONS-derived two-layer results at low latitudes – have important implications for the types of hydrated minerals (e.g. epsomite, meridianite, zeolite, clay, etc.) and their hydration states present at low latitudes. However, the dependence of macroscopic absorption cross section upon Fe and Cl must be explicitly modeled by composition-dependent MCNPX simulations in order to confirm the high predicted values of $W_{dn}$ in Aeolis Planum and Medusae Fossae.

(c) At high latitudes, we demonstrated that the disparate MONS and Phoenix RA observations of near surface ice content can be reconciled by a three-layer model incorporating dry soil over fully saturated pore ice over pure excess ice. Such a three-layer model can also potentially explain the strong anticorrelation of $W_{dn}$ and $D$ observed at high latitudes. One possible way to confirm these intriguing results might be extensive inverse modeling of MONS neutron flux data via generation of lookup tables involving at least three independent layers and five free parameters.

(d) At moderate latitudes, we showed that the distribution of recently formed craters is also consistent with our latest MONS results, as both the shallowest ice-exposing crater and deepest non-ice-exposing crater at each impact site are in good agreement with our predictions of near-surface WEH (with our three-layer model providing a slightly better fit than the standard two-layer model). As noted by *Schorghoger and Forget* (2012), the presence of recent ice-exposing craters at mid-latitudes implies that the martian subsurface is not yet in equilibrium with the atmosphere. Therefore, it is important to conduct more detailed MONS-constrained three-layer modeling of recent crater impact sites, along with large-scale simulations of atmospheric exchange with the regolith, in order to constrain both the timing of subsurface excess ice deposition and the extent to which such excess ice drives martian climate over the past few million years.

**Table 1: Parameters for Equation 4 Polynomial Fits**

| *Parameter* | *Epithermal Neutrons* | *Fast Neutrons* | *Thermal Neutrons (WEHD0 > 10%)* |
|:---:|:---:|:---:|:---:|
| $A_o$ | 6.75007 | 12.0840 | 214.032 |
| $A_1$ | -3.45924 | -52.7100 | -252.865 |
| $A_2$ | 1.66615 | 141.829 | 123.700 |
| $A_3$ | -0.530839 | -210.635 | -31.4565 |
| $A_4$ | 0.103738 | 170.436 | 4.39026 |
| $A_5$ | -0.0123588 | -70.7658 | -0.319609 |
| $A_6$ | 8.72551e-4 | 11.7499 | 9.50064e-3 |
| $A_7$ | -3.34624e-5 | 0.0 | 0.0 |
| $A_8$ | 5.35954e-7 | 0.0 | 0.0 |

**Appendix: Spatial Deconvolution**

The spatial resolution of the spectrometer is roughly 550 km full-width-at-half-maximum (FWHM) arc length on the surface (*Prettyman et al.*, 2004; *Maurice et al.*, 2011). Consequently, neutrons emitted from broad surface regions contribute to the counts measured at each map location. The response of the neutron spectrometer is nearly omnidirectional (*Boynton et al.*, 2004; *Prettyman et al.*; 2009; *Maurice et al.,* 2011). As shown by *Prettyman et al.* (2009), the Prism detector count rate P1 (from which both epithermal and fast neutron fluxes are derived) is roughly symmetrical about the nadir, whereas Prism detector count rates P2 and P4 are asymmetrical (see also *Boynton et al.*, 2004). However, offsets along the ground track were removed by the data reduction process (*Maurice et al.*, 2011); in addition, subtracting P4 from P2 results in a roughly symmetrical response. Figs. 3 and 4 from *Prettyman et al.* (2009) show that the P1, P2, and P4 response functions have similar widths along the ground track. In this work, the same P1-equivalent response function has been assumed for all three neutron bands, which is valid for epithermal and fast neutron fluxes, and approximately correct for thermal neutron fluxes (*Prettyman et al.*, 2009).

Given a known spatial response function and an understanding of the statistical structure of the measurements, de-blurring of map data can be attempted using numerical methods. Spatial deconvolution was successfully applied to neutron and gamma ray flux data acquired by Lunar

Prospector (*Lawrence et al.*, 2007; *Eke et al.*, 2009; *Teodoro et al.*, 2010) and neutron counting rate data acquired by Mars Odyssey (*Prettyman et al.*, 2009; *Wilson et al.*, 2018). In this study, we use a spherical harmonics expansion to deconvolve global frost-free thermal, epithermal, and fast neutron flux maps (*Maurice et al.*, 2011).

Following *Prettyman et al.* (2009), the data can be modeled as

$$\mathbf{d} = \mathbf{i} \otimes \mathbf{r} + \boldsymbol{\varepsilon} \tag{A1}$$

where $\mathbf{d}$ is a global map of neutron count rates determined by binning measurements at subsatellite points on pixels using the methods described by *Maurice et al.* (2011), $\mathbf{i}$ is the map of counts that would have been obtained if the spatial response function was a delta function, $\mathbf{r}$ is the unit response function of the spectrometer, and $\boldsymbol{\varepsilon}$ represents the uncertainty in the map data (also estimated by *Maurice et al.*, 2011). The symbol $\otimes$ denotes convolution. The objective of spatial deconvolution is to find values of $\mathbf{i}$ that are consistent with the data; however, the inversion problem is ill-posed and regularization is required to prevent amplification of noise. Here, the deconvolved image is approximated using spherical harmonics and the lowest order expansion that fits the measured map data is found. This approach limits the introduction of noise in the deconvolved map.

Real spherical harmonics have been applied to smooth mapped neutron counting rate data (*Prettyman et al.*, 2012; 2017) and are used here. The deconvolved image is approximated by

$$\mathbf{i} = \sum_{l=0}^{N} \sum_{m=-l}^{l} a_{lm} \mathbf{Y}_{lm} \tag{A2}$$

where the $a_{lm}$ are the scalar coefficients, the $\mathbf{Y}_{lm}$ are the spherical harmonics represented as maps, and $N$ is the order of the expansion. Substituting **Eq. A1** into **Eq. A2** and rearranging some of the terms gives the following expression for the errors:

$$\boldsymbol{\varepsilon} = \mathbf{d} - \sum_{l=0}^{N} \sum_{m=-l}^{l} a_{lm} \mathbf{S}_{lm} \tag{A3}$$

where $\mathbf{S}_{lm} = \mathbf{Y}_{lm} \otimes \mathbf{r}$ are precomputed convolutions of the instrument unit response function with the spherical harmonics.

The reconstruction algorithm is formulated as a weighted least squares problem. Given a selected order of expansion $N$, the coefficients $a_{lm}$ that minimize the weighted sum of the differences between the data and the model (**Eq. A3**) are found. For each pixel, the difference is weighted by the square of the uncertainty in the map data estimated by *Maurice et al.* (2011). The deconvolved image is computed using **Eq. A2** given the fitted coefficients $a_{lm}$.

The longitudinal spatial resolution $R_s$ is given (in degrees) by the half wavelength of the basis function, i.e., $R_s = 360/(2*N)$, where $N$ is the order of the expansion. Assuming an average volumetric mean radius of 3390 km for Mars, this works out to $R_s = 1065$ km for $\{N=10\}$, $R_s = 666$ km for $\{N=16\}$, $R_s = 532$ km for $\{N=20\}$, and $R_s = 426$ km for $\{N=25\}$. Hence, all $R_s$ corresponding to deconvolved $\{N \leq 18\}$ solutions are greater than the intrinsic spatial resolution of the MONS spectrometer (~550 km: *Maurice et al.*, 2011) – i.e., the resolution that would be obtained in the absence of statistical fluctuations in the counts.

**Fig. A1** plots the reduced chi-squared statistic ($\chi^2$), which is used to evaluate the goodness of fit, for deconvolved epithermal, thermal, and fast neutron count rates as a function of $N$ (ideally,

*N* is increased until $\chi^2$ is about 1). For epithermal neutrons (blue circles), $\chi^2$ reaches 1 for {N=24} – though note that according to **Fig. 6a**, minimum epithermal neutron fluxes fall below zero for {N=22} to {N=24}, which is clearly unphysical. For thermal neutrons (red squares), $\chi^2$ reaches 1 for {N=14}, corresponding to a spatial resolution of $R_s$ = 760 km that is larger than the intrinsic ~550 km resolution of MONS; for fast neutrons (purple asterisks), $\chi^2$ = 0.53 for the minimum {N=10} and decreases for higher N. These results imply that the deconvolution algorithm cannot improve the spatial resolution of the fast and thermal neutron maps and is instead effectively smoothing the data. However, such deconvolution-derived smoothing, when combined with the Gaussian inverse-squared weighting algorithm, significantly improves upon the *Feldman et al.* (2011) "crossover" technique for calculating $W_{up}$ (compare our **Fig. 8b** to Fig. 7 in *Feldman et al.* [2011]) – which in turns results in more accurate mapping of $W_{dn}$ and *D* (since the values of both parameters are sensitive to $W_{up}$).

**Fig. A1** clearly shows that no single value of *N* can simultaneously satisfy the $\chi^2$ ~ 1 criterion for all three neutron fluxes. The global WEH maps resulting from the multiple-*N* deconvolution solution that comes closest to meeting the $\chi^2$ ~ 1 criterion – comprised of {N=24} for the epithermal neutron flux, {N=14} for the thermal neutron flux, and {N=10} for the fast neutron flux – are shown for $W_{up}$, $W_{dn}$, and *D* in **Figs. S9**, **S13**, and **S14**, respectively. Although this solution's elevated $W_{up}$ values at high northern latitudes are intriguing (**Fig. S9**), the corresponding $W_{dn}$ results unphysically exceed 100% on the North PLD and are very noisy at low-to-mid latitudes (**Fig. S13**): similarly, the corresponding *D* results are also noisy and unphysically fall to zero in Vastitas Borealis (**Fig. S14**). Therefore, we conclude that this multiple-*N* deconvolution solution, despite being the best statistical fit to the neutron flux data (as measured by $\chi^2$), is not a physically credible solution.

A possible explanation for this incongruity may be the integrated nature of our two-layer modeling: since our MCNPX grid-derived estimates of $W_{dn}$ and $D$ are dependent upon crossover calculations of $W_{up}$, our "combined" WEH solution is effectively dependent on the relative values of the epithermal, thermal, and fast neutron fluxes to one another. Therefore, it may be more appropriate to consider a "combined" $\chi^2$ derived from simultaneously solving for epithermal, thermal, and fast maps from all of the data: such a combined $\chi^2$ would likely reach 1 at a significantly lower order than the {N=24} value at which epithermal neutron $\chi^2$ reaches 1 (**Fig. A1**), thereby possibly justifying the choice of a lower $N$ in our preferred {N=16} deconvolution solution. Alternatively, since the crossover technique requires weighted averaging over large length scales, it may simply work best when the intrinsic spatial resolutions of the epithermal, thermal, and fast neutron counting rate maps are similar – i.e., either for undeconvolved neutron flux maps (**Figs. 1a, 2a, 3a**), or for deconvolved neutron flux maps with identical $N$ (e.g. **Figs. 1b, 2b, 3b**).

**Acknowledgements**

This work was funded by the NASA Mars Data Analysis Program (grant number NNX13AK61G S03), and conducted under the auspices of the Planetary Science Institute. We thank both of the anonymous reviewers for insightful comments that greatly improved the paper. One of us (WCF) also wishes to thank the Los Alamos National Laboratory for providing office space and access to their library and the Internet while spending summers in Los Alamos.

**References**

Audouard, J., Poulet, F., Vincendon, M., Milliken, R.E., Jouglet, D., Bibring, J.P., Gondet, B. and Langevin, Y., 2014. Water in the Martian regolith from OMEGA/Mars Express. Journal of Geophysical Research: Planets, 119(8), pp.1969-1989.

Boynton, W.V., Feldman, W.C., Mitrofanov, I.G., Evans, L.G., Reedy, R.C., Squyres, S.W., Starr, R., Trombka, J.I., d'Uston, C., Arnold, J.R. and Englert, P.A.J., 2004. The Mars Odyssey gamma-ray spectrometer instrument suite. In 2001 Mars Odyssey (pp. 37-83). Springer Netherlands.

Boynton, W.V., Taylor, G.J., Evans, L.G., Reedy, R.C., Starr, R., Janes, D.M., Kerry, K.E., Drake, D.M., Kim, K.J., Williams, R.M.S. and Crombie, M.K., 2007. Concentration of H, Si, Cl, K, Fe, and Th in the low-and mid-latitude regions of Mars. Journal of Geophysical Research: Planets, 112(E12).

Bramson, A.M., Byrne, S., Putzig, N.E., Sutton, S., Plaut, J.J., Brothers, T.C. and Holt, J.W., 2015. Widespread excess ice in Arcadia Planitia, Mars. Geophysical Research Letters, 42(16), pp.6566-6574.

Byrne, S., Dundas, C.M., Kennedy, M.R., Mellon, M.T., McEwen, A.S., Cull, S.C., Daubar, I.J., Shean, D.E., Seelos, K.D., Murchie, S.L. and Cantor, B.A., 2009. Distribution of mid-latitude ground ice on Mars from new impact craters. Science, 325(5948), pp.1674-1676.

Carr, M.H., 1996. Water on Mars, 229 pp., Oxford Univ. Press, New York.

Cull, S., Arvidson, R.E., Mellon, M.T., Skemer, P., Shaw, A. and Morris, R.V., 2010. Compositions of subsurface ices at the Mars Phoenix landing site. Geophysical Research Letters, 37(24).

Daubar, I.J., Atwood-Stone, C., Byrne, S., McEwen, A.S. and Russell, P.S., 2014. The morphology of small fresh craters on Mars and the Moon. Journal of Geophysical Research: Planets, 119(12), pp.2620-2639.

Diez, B., Feldman, W.C., Maurice, S., Gasnault, O., Prettyman, T.H., Mellon, M.T., Aharonson, O. and Schorghofer, N., 2008. H layering in the top meter of Mars. Icarus, 196(2), pp.409-421.

Dundas, C.M. and Byrne, S., 2010. Modeling sublimation of ice exposed by new impacts in the martian mid-latitudes. Icarus, 206(2), pp.716-728.

Dundas, C.M., Byrne, S., McEwen, A.S., Mellon, M.T., Kennedy, M.R., Daubar, I.J. and Saper, L., 2014. HiRISE observations of new impact craters exposing Martian ground ice. Journal of Geophysical Research: Planets, 119(1), pp.109-127.

Eke, V.R., Teodoro, L.F.A. and Elphic, R.C., 2009. The spatial distribution of polar hydrogen deposits on the Moon. Icarus, 200(1), pp.12-18.

Feldman, W.C., Boynton, W.V., Tokar, R.L., Prettyman, T.H., Gasnault, O., Squyres, S.W., Elphic, R.C., Lawrence, D.J., Lawson, S.L., Maurice, S. and McKinney, G.W., 2002. Global distribution of neutrons from Mars: Results from Mars Odyssey. Science, 297(5578), pp.75-78.

Feldman, W.C., 2003a, July. The global distribution of near-surface hydrogen on Mars, paper presented at. In Sixth International Conference on Mars, Lunar and Planet. Inst., Pasadena, Calif (pp. 20-25).

Feldman, W.C., Prettyman, T.H., Boynton, W.V., Murphy, J.R., Squyres, S., Karunatillake, S., Maurice, S., Tokar, R.L., McKinney, G.W., Hamara, D.K. and Kelly, N., 2003b. $CO_2$ frost cap thickness on Mars during northern winter and spring. Journal of Geophysical Research: Planets, 108(E9).

Feldman, W.C., Prettyman, T.H., Maurice, S., Plaut, J.J., Bish, D.L., Vaniman, D.T., Mellon, M.T., Metzger, A.E., Squyres, S.W., Karunatillake, S. and Boynton, W.V., 2004a. Global distribution of near-surface hydrogen on Mars. Journal of Geophysical Research: Planets, 109(E9).

Feldman, W.C., Head, J.W., Maurice, S., Prettyman, T.H., Elphic, R.C., Funsten, H.O., Lawrence, D.J., Tokar, R.L. and Vaniman, D.T., 2004b. Recharge mechanism of near-equatorial hydrogen on Mars: Atmospheric redistribution or sub-surface aquifer. Geophysical research letters, 31(18).

Feldman, W.C., Mellon, M.T., Maurice, S., Prettyman, T.H., Carey, J.W., Vaniman, D.T., Bish, D.L., Fialips, C.I., Chipera, S.J., Kargel, J.S. and Elphic, R.C., 2004c. Hydrated states of $MgSO_4$ at equatorial latitudes on Mars. Geophysical Research Letters, 31(16).

Feldman, W.C., Mellon, M.T., Gasnault, O., Diez, B., Elphic, R.C., Hagerty, J.J., Lawrence, D.J., Maurice, S. and Prettyman, T.H., 2007. Vertical distribution of hydrogen at high northern latitudes on Mars: The Mars Odyssey Neutron Spectrometer. Geophysical research letters, 34(5).

Feldman, W.C., Bandfield, J.L., Diez, B., Elphic, R.C., Maurice, S. and Nelli, S.M., 2008a. North to south asymmetries in the water-equivalent hydrogen distribution at high latitudes on Mars. Journal of Geophysical Research: Planets, 113(E8).

Feldman, W.C., Bourke, M.C., Elphic, R.C., Maurice, S., Bandfield, J., Prettyman, T.H., Diez, B. and Lawrence, D.J., 2008b. Hydrogen content of sand dunes within Olympia Undae. Icarus, 196(2), pp.422-432.

Feldman, W.C., Pathare, A., Maurice, S., Prettyman, T.H., Lawrence, D.J., Milliken, R.E. and Travis, B.J., 2011. Mars Odyssey neutron data: 2. Search for buried excess water ice deposits at nonpolar latitudes on Mars. Journal of Geophysical Research: Planets, 116(E11).

Fialips, C.I., Carey, J.W., Vaniman, D.T., Bish, D.L., Feldman, W.C. and Mellon, M.T., 2005. Hydration state of zeolites, clays, and hydrated salts under present-day martian surface conditions: Can hydrous minerals account for Mars Odyssey observations of near-equatorial water-equivalent hydrogen?. Icarus, 178(1), pp.74-83.

Jansson, P.A., 1996, October. Modern constrained nonlinear methods. In Deconvolution of images and spectra (2nd ed.), pp. 107-181. Academic Press, Inc.

Lawrence, D.J., Puetter, R.C., Elphic, R.C., Feldman, W.C., Hagerty, J.J., Prettyman, T.H. and Spudis, P.D., 2007. Global spatial deconvolution of Lunar Prospector Th abundances. Geophysical Research Letters, 34(3).

Maurice, S., Feldman, W., Diez, B., Gasnault, O., Lawrence, D.J., Pathare, A. and Prettyman, T., 2011. Mars Odyssey neutron data: 1. Data processing and models of water-equivalent-hydrogen distribution. Journal of Geophysical Research: Planets, 116(E11).

Mellon, M.T. and Jakosky, B.M., 1993. Geographic variations in the thermal and diffusive stability of ground ice on Mars. Journal of Geophysical Research: Planets, 98(E2), pp.3345-3364.

Mellon, M.T., Feldman, W.C. and Prettyman, T.H., 2004. The presence and stability of ground ice in the southern hemisphere of Mars. Icarus, 169(2), pp.324-340.

Mellon, M.T., Boynton, W.V., Feldman, W.C., Arvidson, R.E., Titus, T.N., Bandfield, J.L., Putzig, N.E. and Sizemore, H.G., 2008. A prelanding assessment of the ice table depth and ground ice characteristics in Martian permafrost at the Phoenix landing site. Journal of Geophysical Research: Planets, 113(E3).

Mellon, M.T., Arvidson, R.E., Sizemore, H.G., Searls, M.L., Blaney, D.L., Cull, S., Hecht, M.H., Heet, T.L., Keller, H.U., Lemmon, M.T. and Markiewicz, W.J., 2009. Ground ice at the Phoenix landing site: Stability state and origin. Journal of Geophysical Research: Planets, 114(E1).

Moore, H.J., Spitzer, C.R., Bradford, K.Z., Cates, P.M., Hutton, R.E., and Shorthill, R.W., 1979. Sample fields of the Viking landers, physical properties, and aeolian processes. Journal of Geophysical Research 87, 10,043-10,050.

Pelowitz, D.B., Durkee, J.W., Elson, J.S., Fensin, M.L., Hendricks, J.S., James, M.R., Johns, R.C., Mc Kinney, F.W., Mashnik, S.G., Waters, L.S. and Wilcox, T.A., 2011. MCNPX 2.7 E extensions (No. LA-UR-11-01502; LA-UR-11-1502). Los Alamos National Laboratory (LANL).

Pike, R.J., 1974. Depth/diameter relations of fresh lunar craters: Revision from spacecraft data. Geophysical Research Letters, 1(7), pp.291-294.

Pike, R.J., 1977. Apparent depth/apparent diameter relation for lunar craters. In Lunar and Planetary Science Conference Proceedings (Vol. 8, pp. 3427-3436).

Prettyman, T.H., Feldman, W.C., Mellon, M.T., McKinney, G.W., Boynton, W.V., Karunatillake, S., Lawrence, D.J., Maurice, S., Metzger, A.E., Murphy, J.R. and Squyres, S.W., 2004. Composition and structure of the Martian surface at high southern latitudes from neutron spectroscopy. Journal of Geophysical Research: Planets, 109(E5).

Prettyman, T.H., Feldman, W.C. and Titus, T.N., 2009. Characterization of Mars' seasonal caps using neutron spectroscopy. Journal of Geophysical Research: Planets, 114(E8).

Prettyman, T.H., Mittlefehldt, D.W., Yamashita, N., Lawrence, D.J., Beck, A.W., Feldman, W.C., McCoy, T.J., McSween, H.Y., Toplis, M.J., Titus, T.N., Tricarico, P., Reedy, R.C., Hendricks, J.S., Forni, O., Le Corre, L., Li, J.-L., Mizzon, H., Reddy, V., Raymond, C.A. and Russell, C.T., 2012. Elemental mapping by Dawn reveals exogenic H in Vesta's regolith. Science, 338(6104), pp.242-246.

Prettyman, T.H., Yamashita, N., Toplis, M.J., McSween, H.Y., Schorghofer, N., Marchi, S., Feldman, W.C., Castillo-Rogez, J., Forni, O., Lawrence, D.J., Ammannito, E., Ehlmann, B.L., Sizemore, H.G., Joy, S.P., Polanskey, C.A., Rayman, M.D., Raymond, C.A. and Russell, C.T., 2016. Extensive water ice within Ceres' aqueously altered regolith: Evidence from nuclear spectroscopy. Science, p.aah6765.

Puetter, R.C., 1995. Pixon-based multiresolution image reconstruction and the quantification of picture information content. International Journal of Imaging Systems and Technology, 6(4), pp.314-331.

Schorghofer, N., 2007. Dynamics of ice ages on Mars. Nature, 449(7159), pp.192-194.

Schorghofer, N. and Forget, F., 2012. History and anatomy of subsurface ice on Mars. Icarus, 220(2), pp.1112-1120.

Sizemore, H.G., Zent, A.P. and Rempel, A.W., 2015. Initiation and growth of martian ice lenses. Icarus, 251, pp.191-210.

Teodoro, L.F.A., Eke, V.R. and Elphic, R.C., 2010. Spatial distribution of lunar polar hydrogen deposits after KAGUYA (SELENE). Geophysical Research Letters, 37(12).

Wang, A., Feldman, W.C., Mellon, M.T. and Zheng, M., 2013. The preservation of subsurface sulfates with mid-to-high degree of hydration in equatorial regions on Mars. Icarus, 226(1), pp.980-991.

Waters, L.S. (Ed.), 1999. MCNPX User's Guide. Doc. LA-UR-99-6058. Los Alamos Natl. Lab., Los Alamos, N. M.

Williams, D., 2016. Mars Fact Sheet, https://nssdc.gsfc.nasa.gov/planetary/factsheet/marsfact.html.

Wilson, J.T., Eke, V.R., Massey, R.J., Elphic, R.C., Feldman, W.C., Maurice, S. and Teodoro, L.F., 2017. Equatorial locations of water on Mars: Improved resolution maps based on Mars Odyssey Neutron Spectrometer data. Icarus, 299, pp.148-160.

Zent, A.P., Hecht, M.H., Cobos, D.R., Wood, S.E., Hudson, T.L., Milkovich, S.M., DeFlores, L.P. and Mellon, M.T., 2010. Initial results from the thermal and electrical conductivity probe (TECP) on Phoenix. Journal of Geophysical Research: Planets, 115(E3).

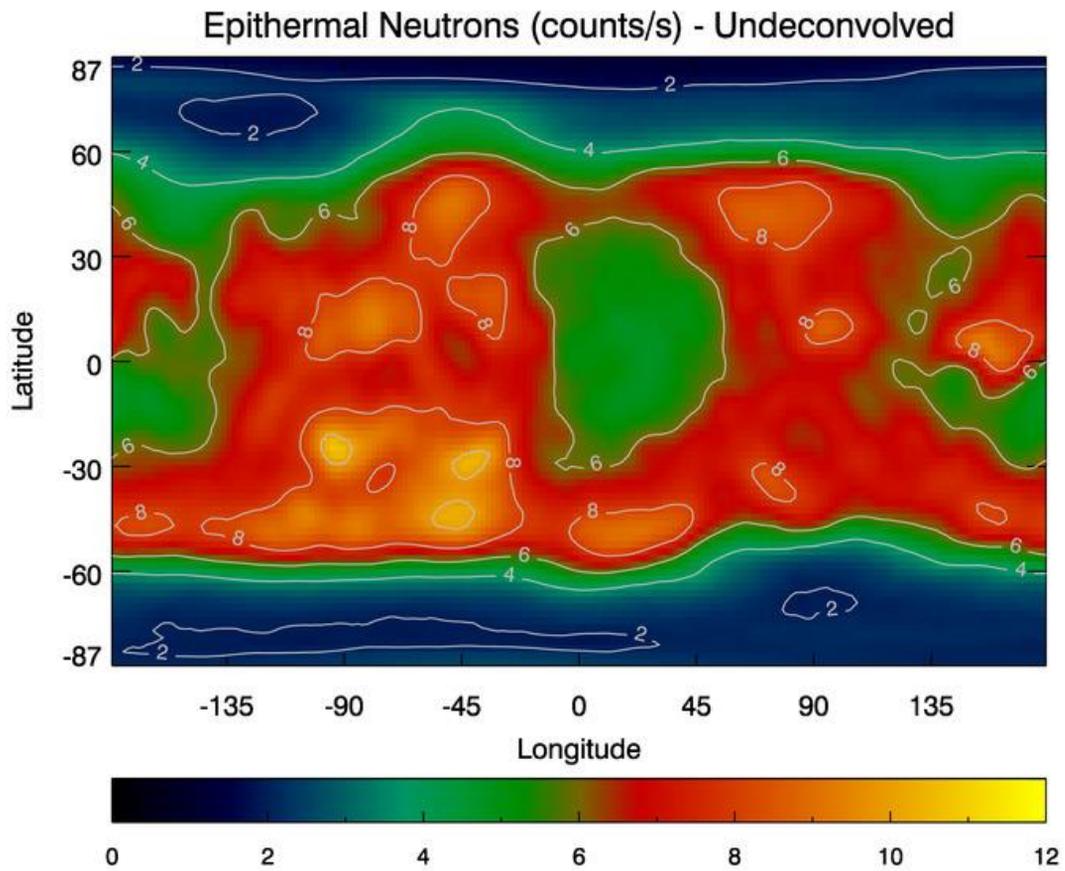

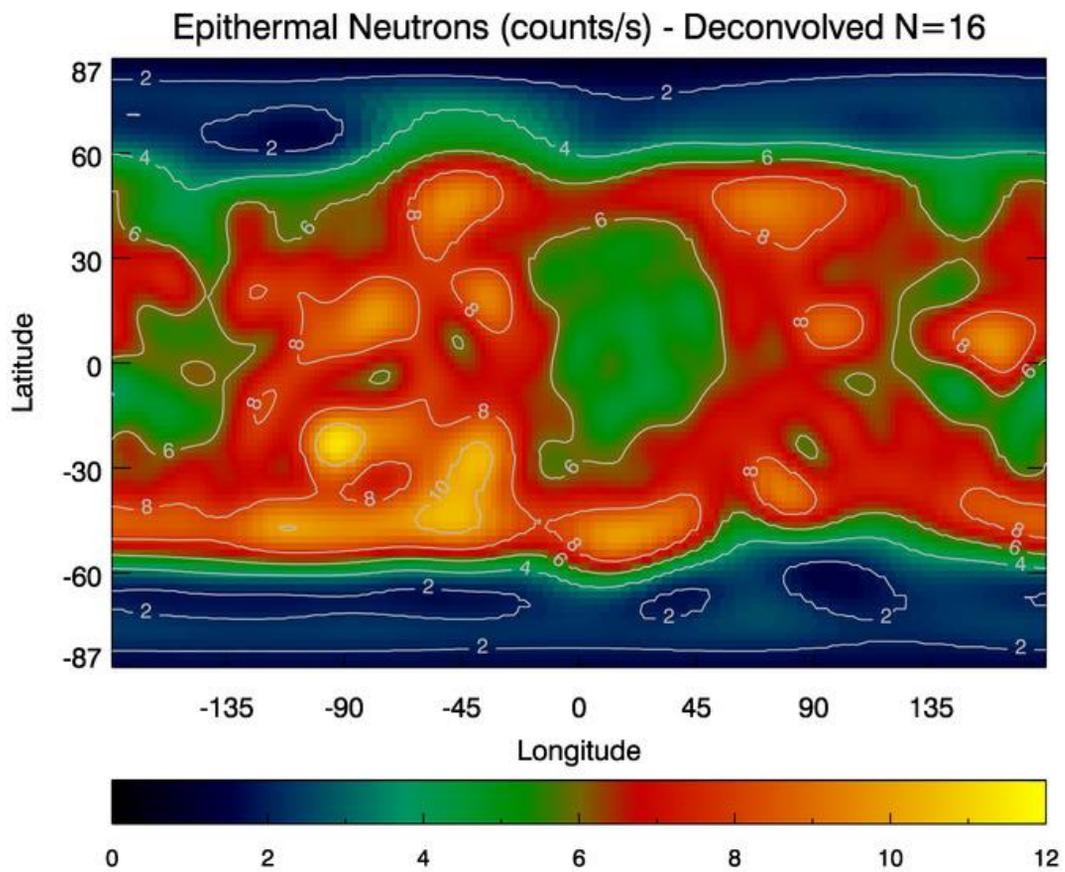

**Figure 1:** MONS frost-free map of epithermal neutron fluxes (counts/s) for **(A)** undeconvolved and **(B)** deconvolved {N = 16} solutions.

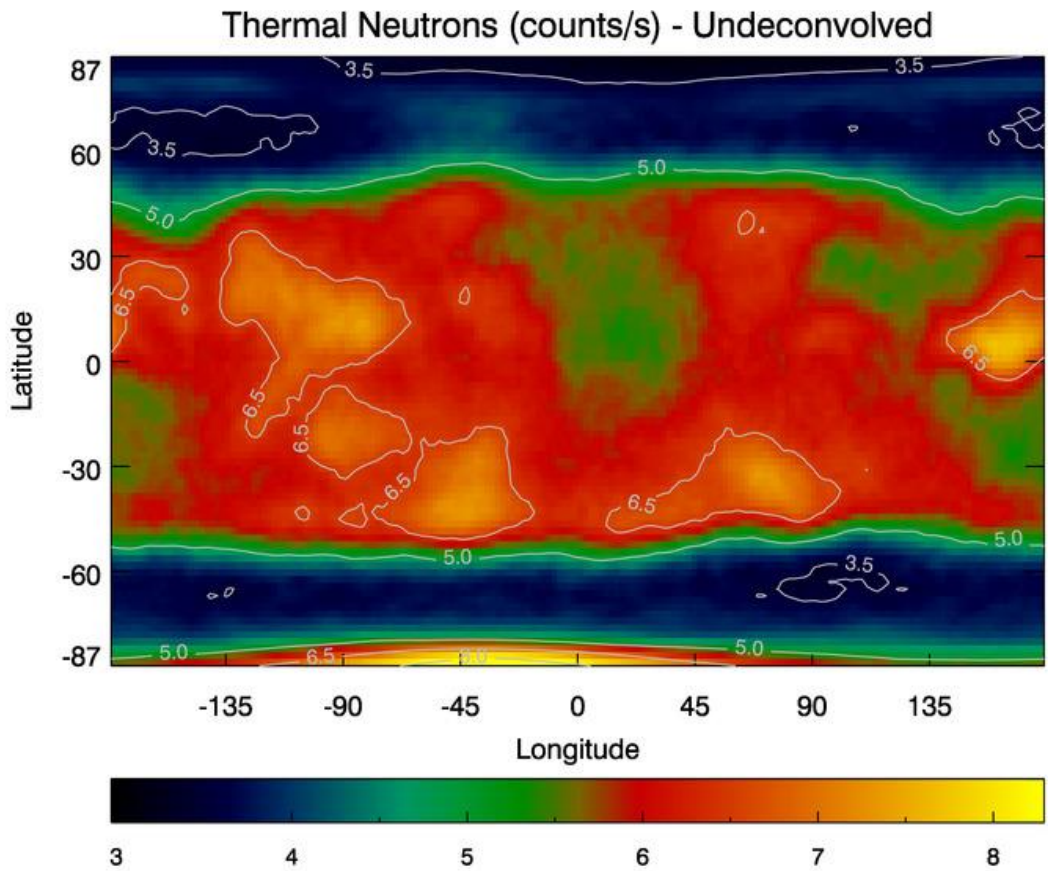
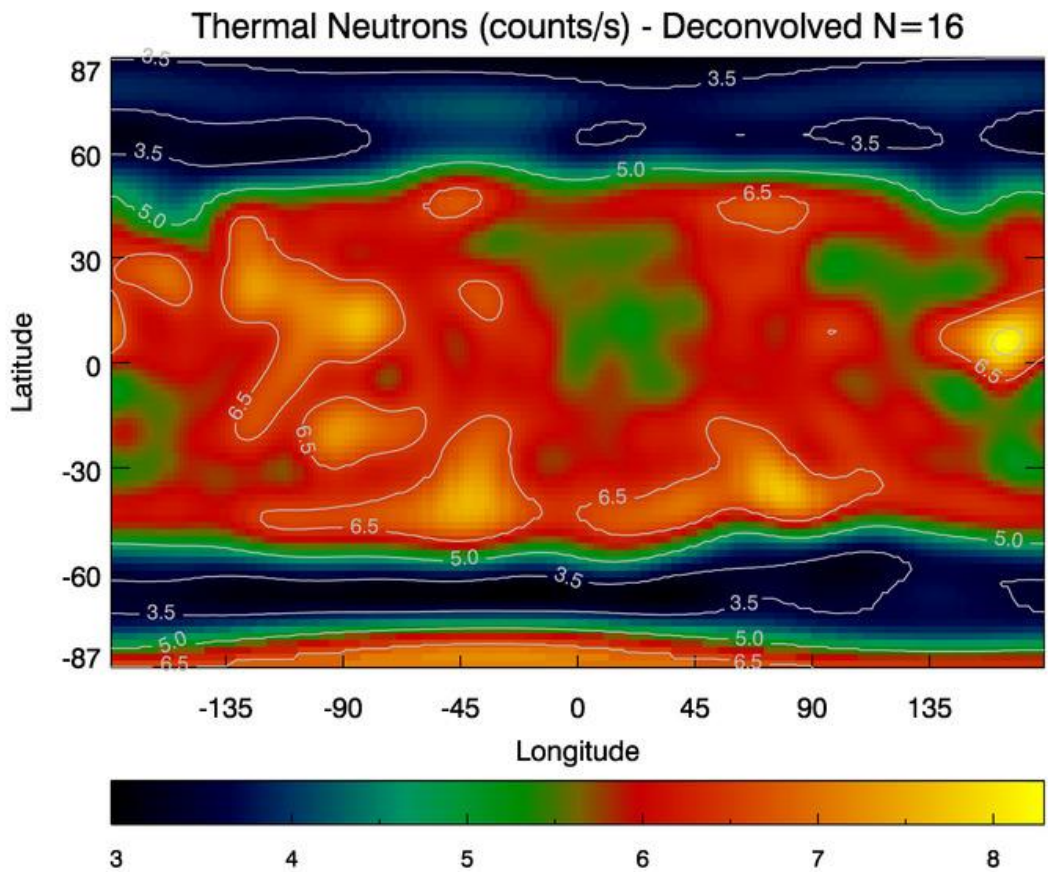

**Figure 2:** MONS frost-free map of thermal neutron fluxes (counts/s) for **(A)** undeconvolved and **(B)** deconvolved {N = 16} solutions.

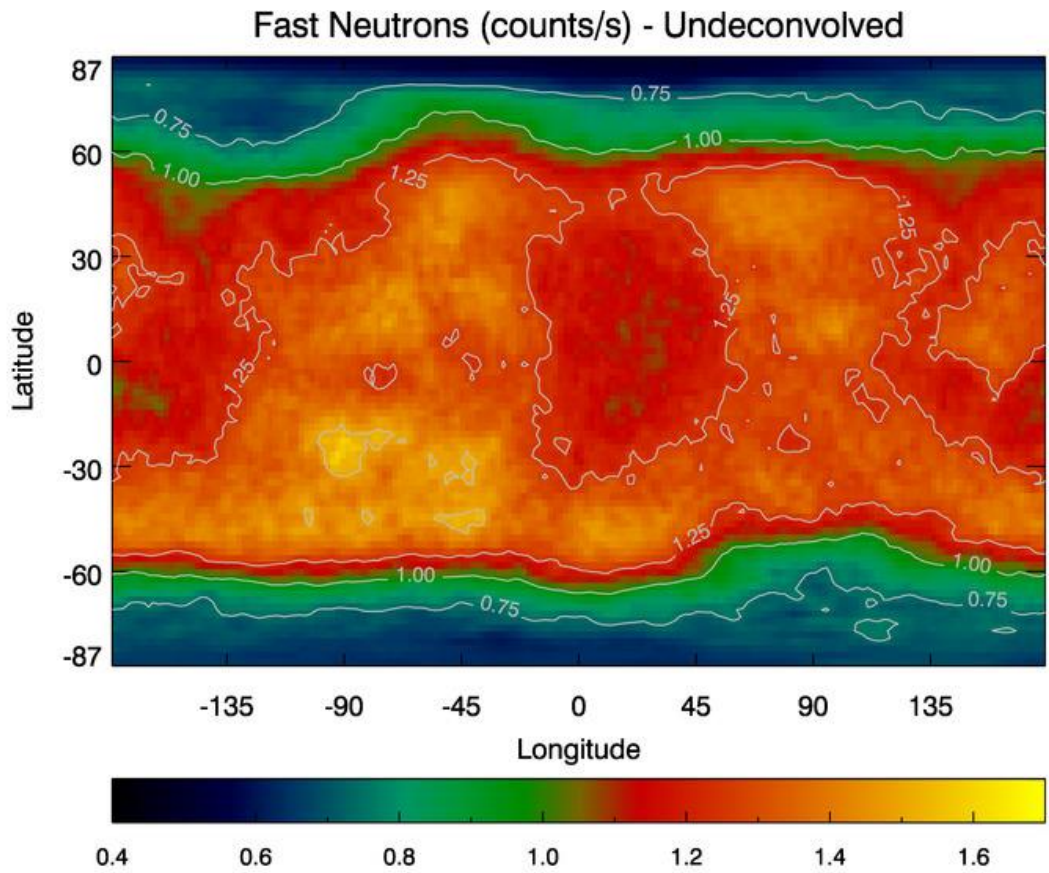
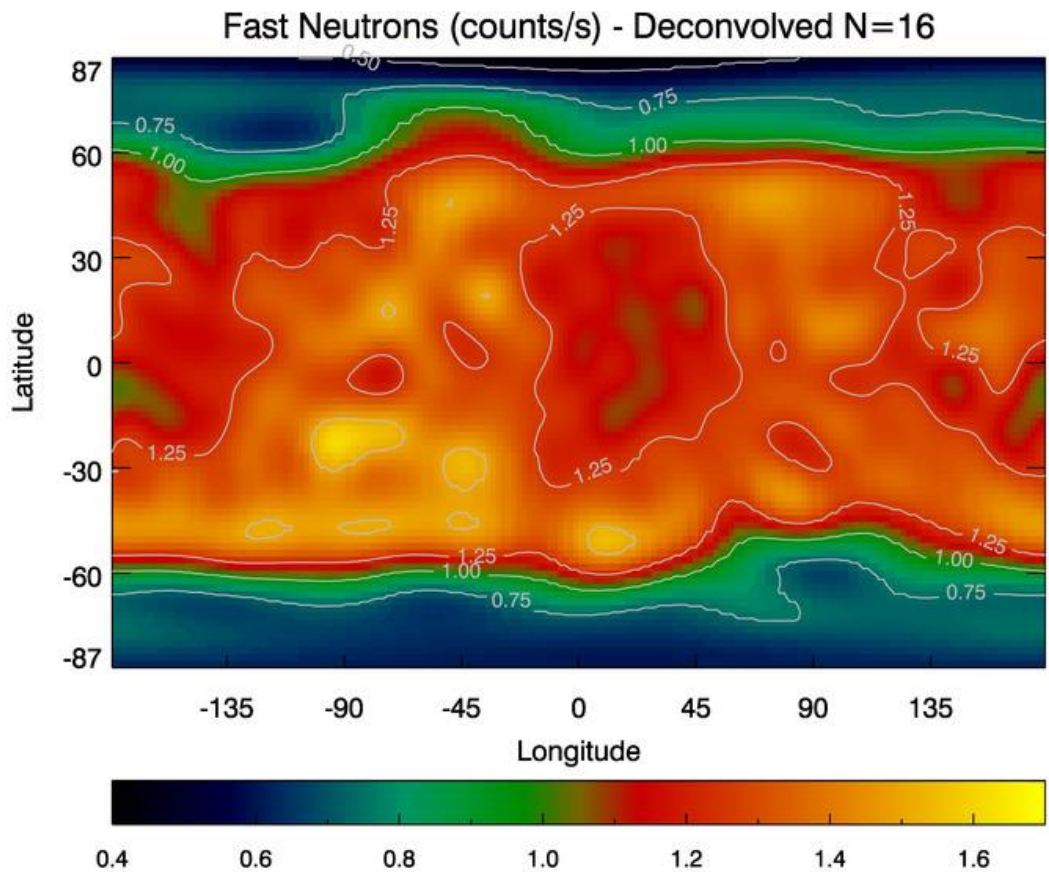

**Figure 3:** MONS frost-free map of fast neutron fluxes (counts/s) for **(A)** undeconvolved and **(B)** deconvolved {N = 16} solutions.

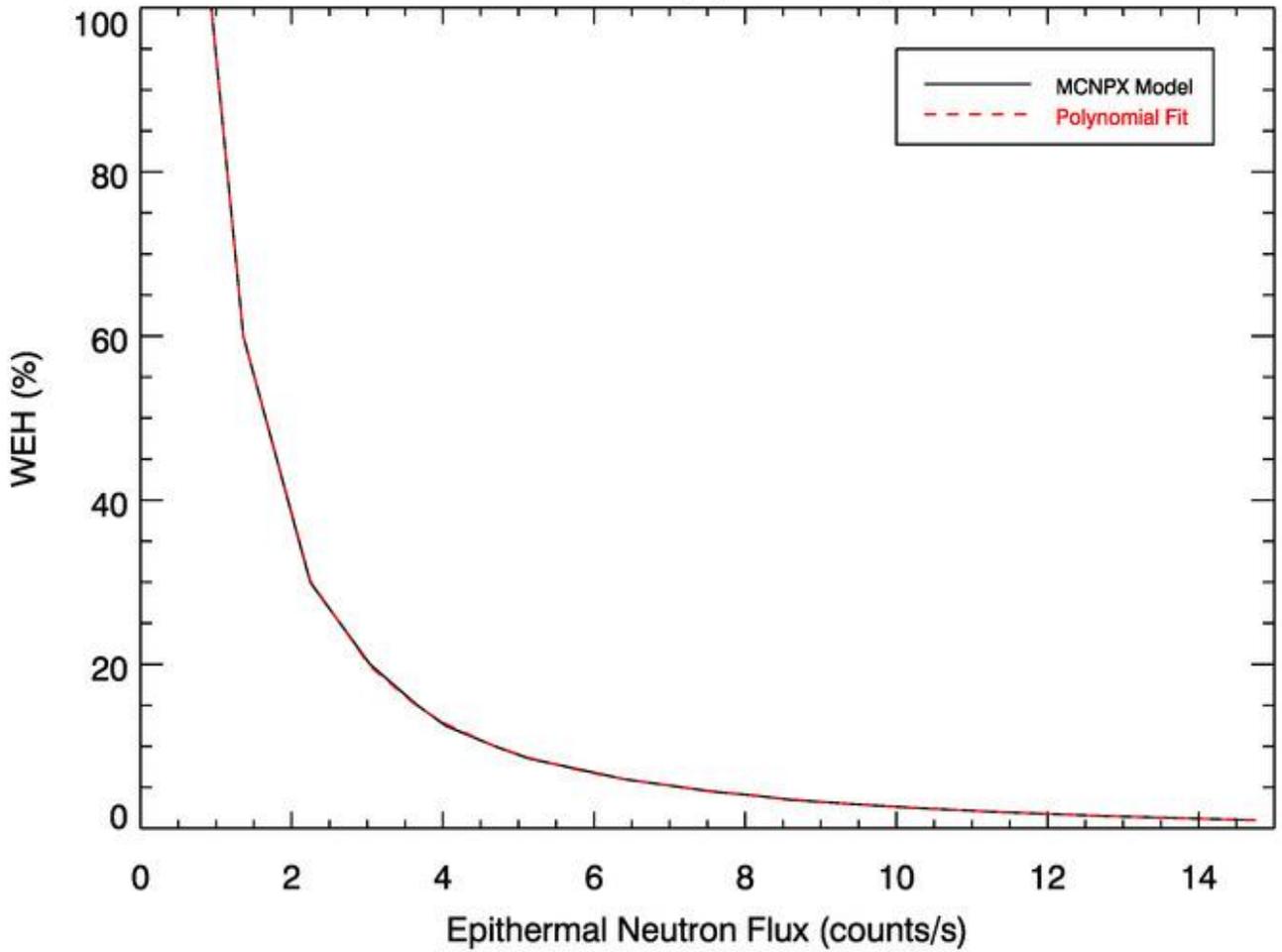

**Figure 4a:** MCNPX model zero-depth 1D WEH solutions (WEHD0) and polynomial fits (via Eq. 4 and Table 1) for epithermal neutron flux.

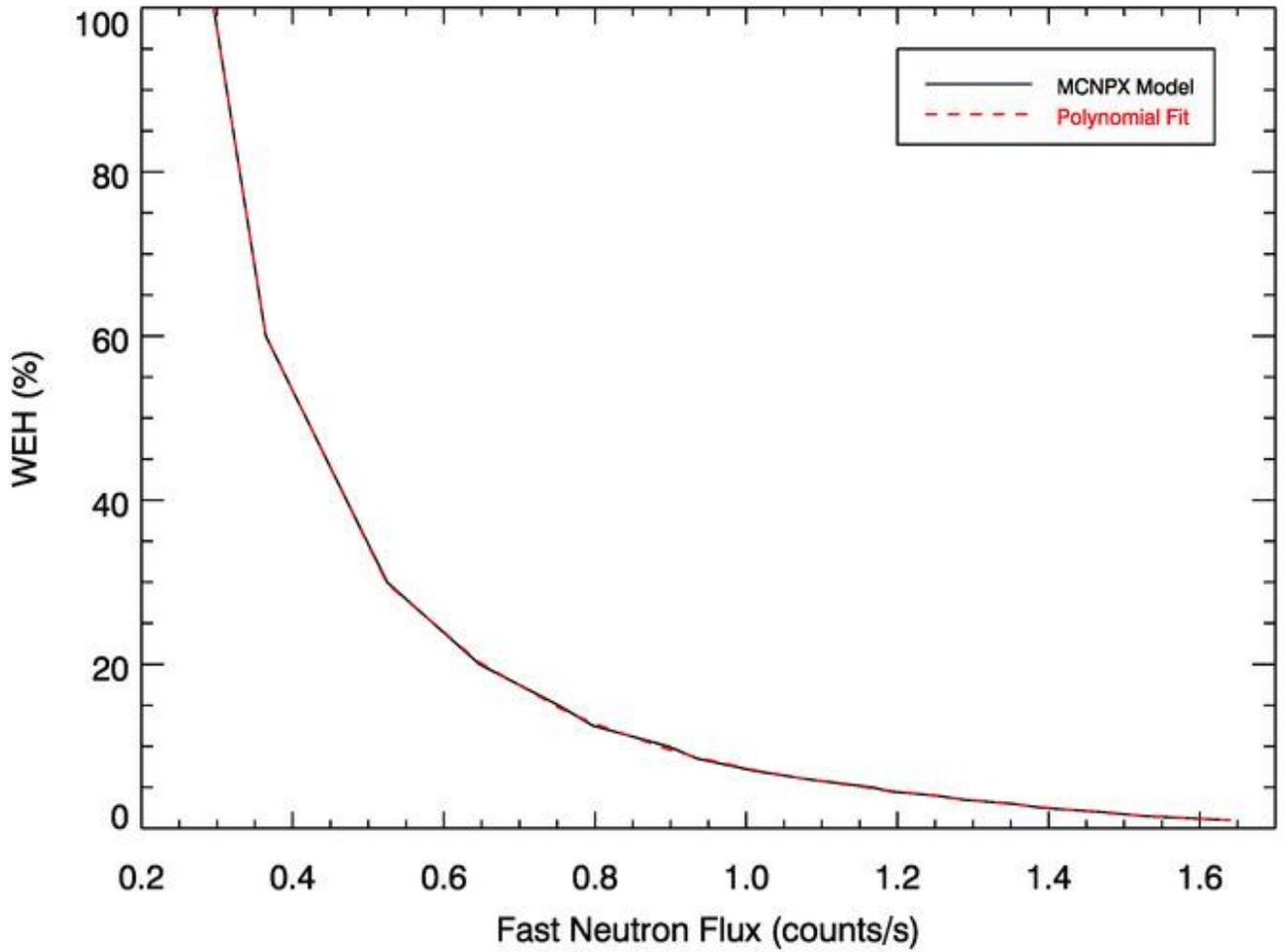

**Figure 4b:** MCNPX model zero-depth 1D WEH solutions (WEHD0) and polynomial fits (via Eq. 4 and Table 1) for fast neutron flux.

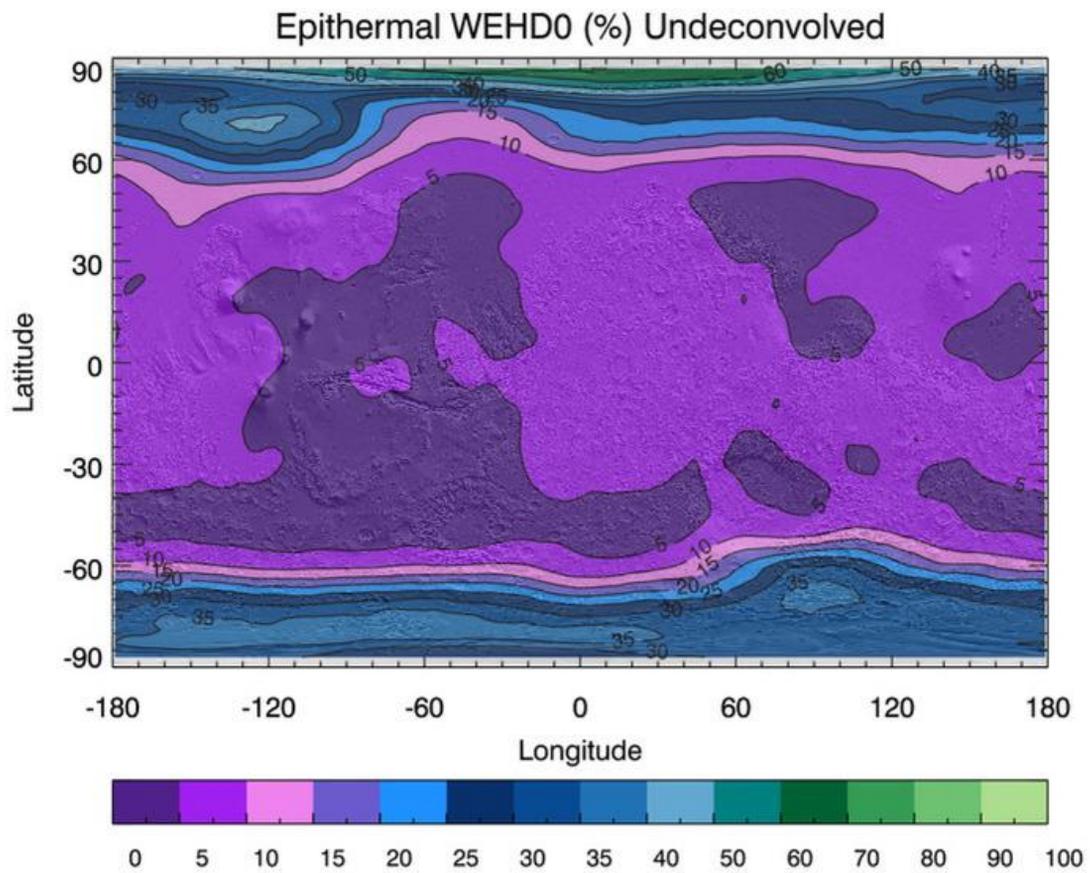

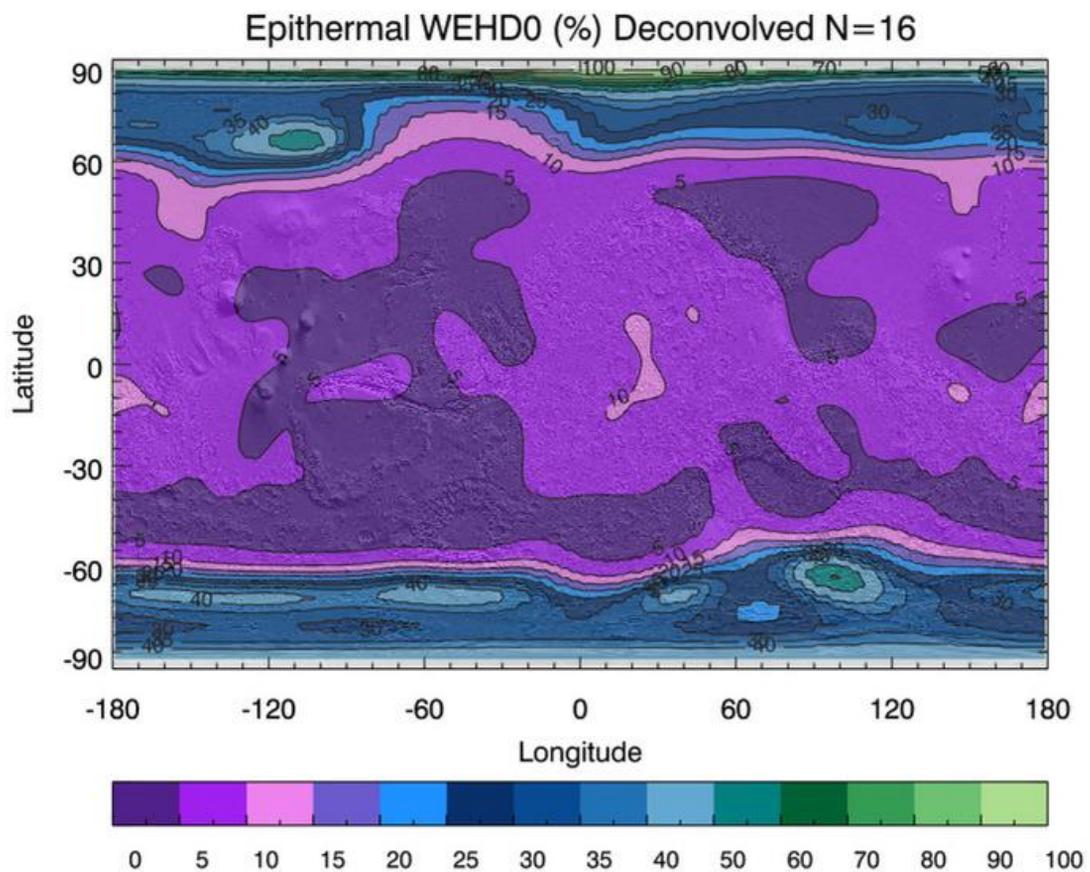

**Figure 5:** Global frost-free epithermal neutron 1D WEH zero-depth (WEHD0) maps, expressed as a wt. % relative to pure ice, for **(A)** undeconvolved and **(B)** deconvolved {N = 16} solutions.

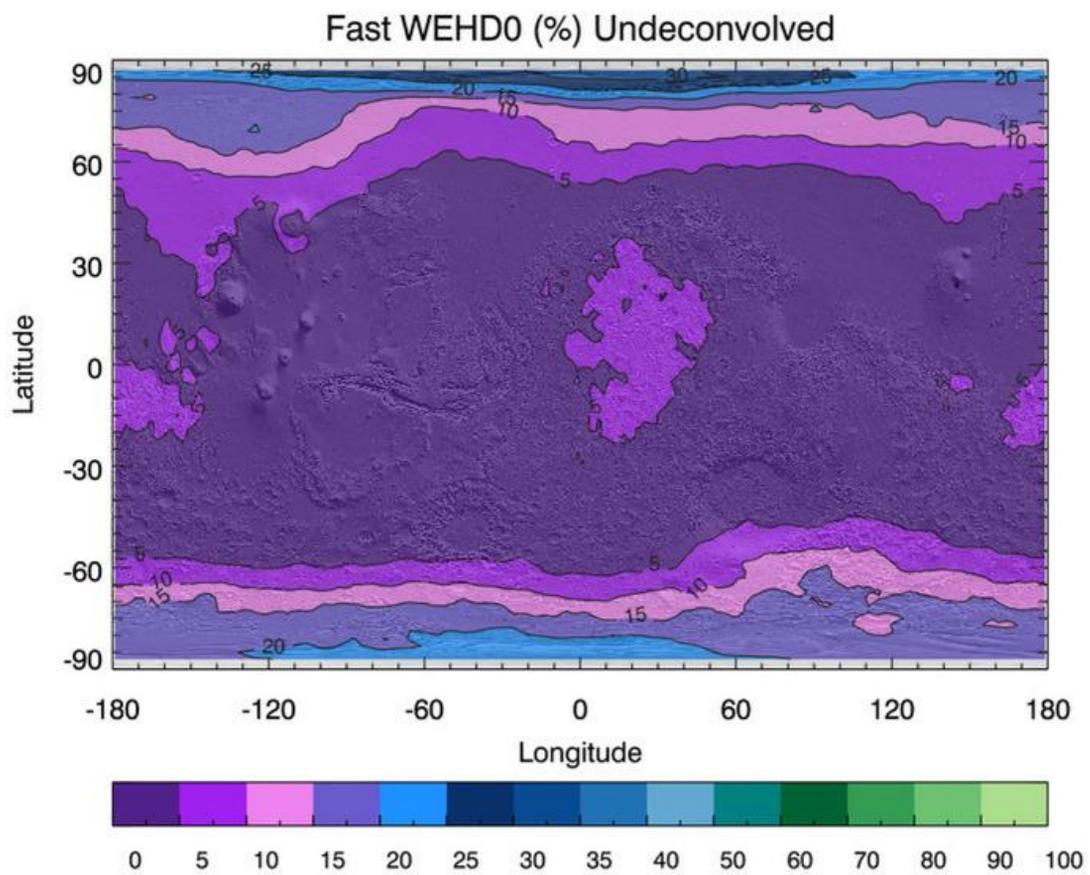

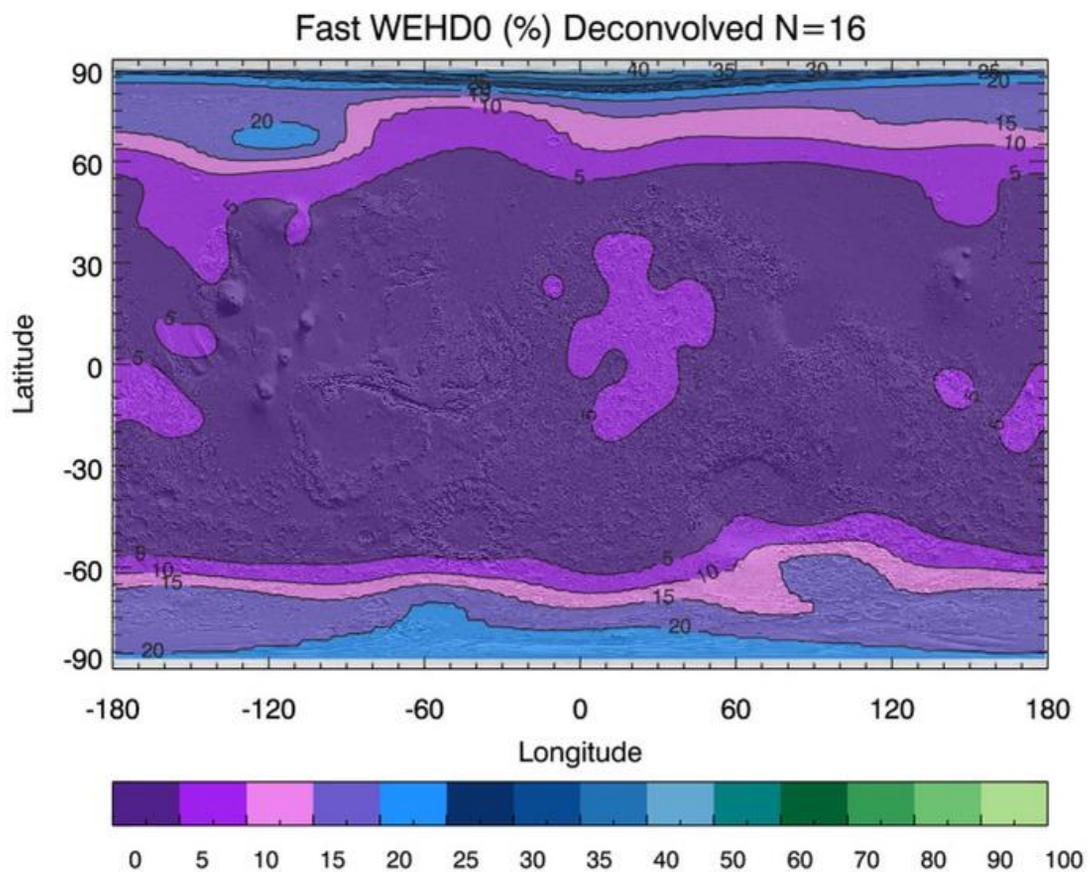

**Figure 5:** Global frost-free fast neutron 1D WEH zero-depth (WEHD0) maps, expressed as a wt. % relative to pure ice, for **(C)** undeconvolved and **(D)** deconvolved {N = 16} solutions.

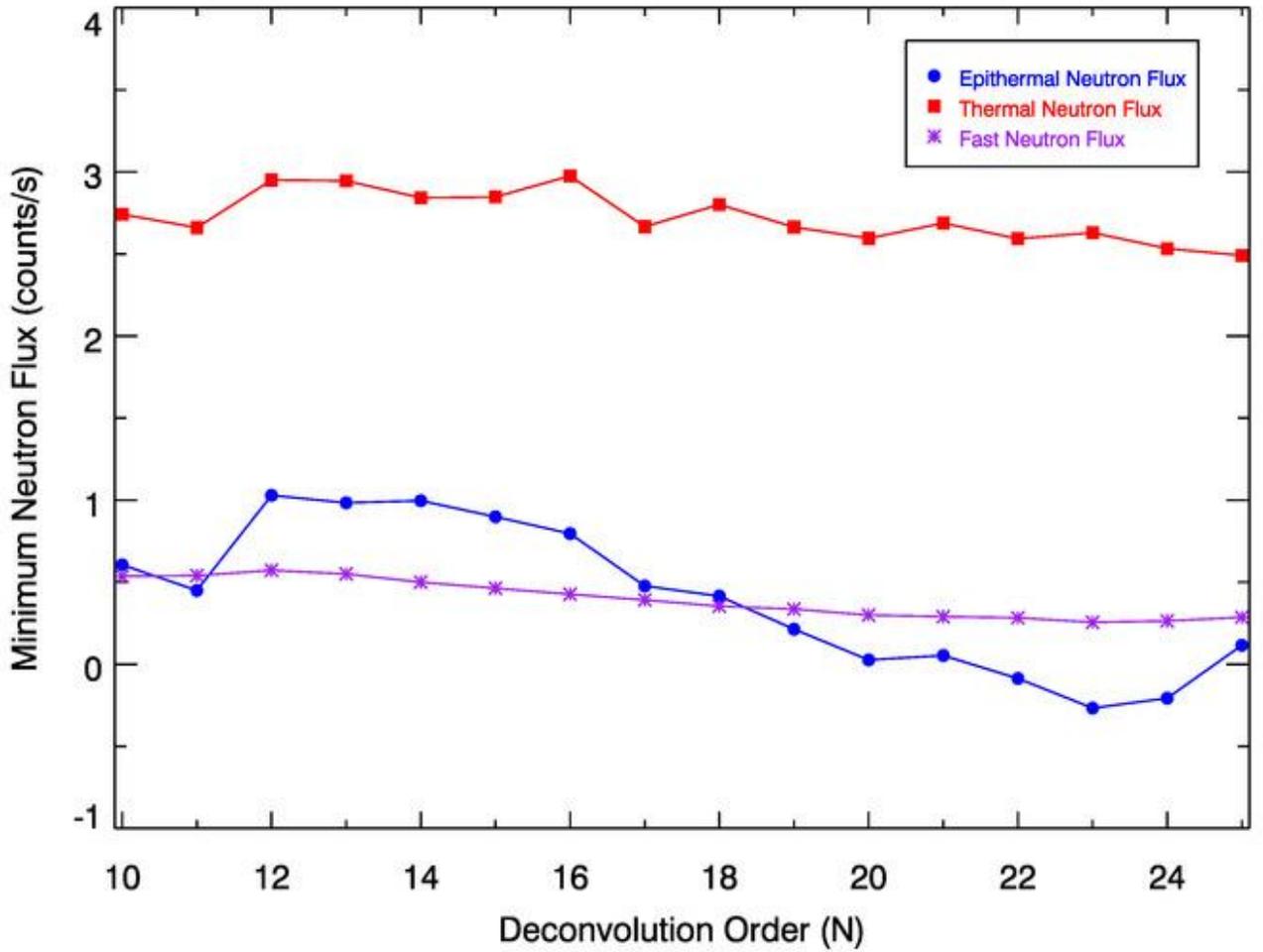

**Figure 6a:** Minimum neutron flux vs. deconvolution order (N) for epithermal (circles), thermal (squares), fast (asterisks) neutron fluxes.

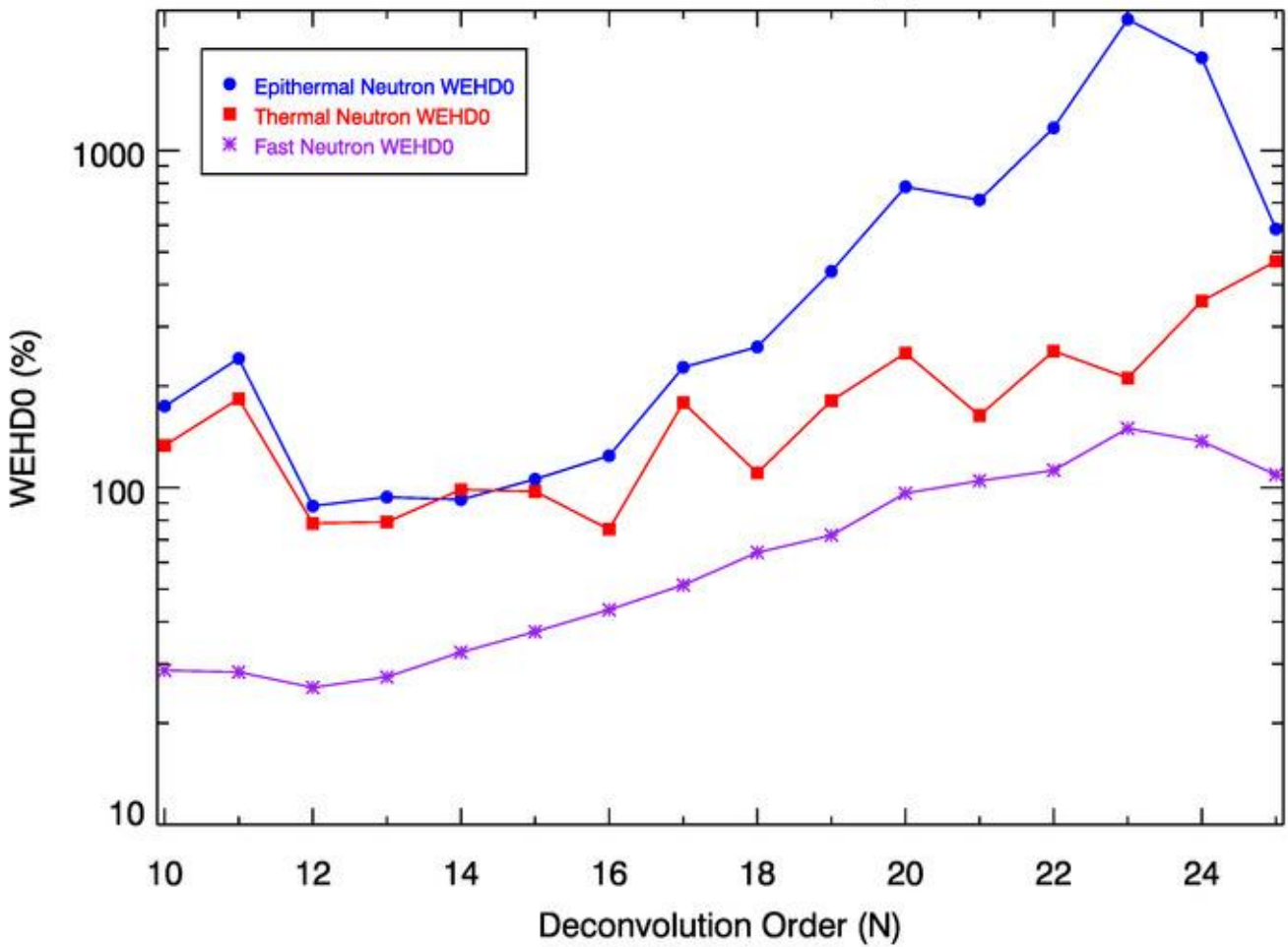

**Figure 6b:** Maximum WEHD0 vs. deconvolution order (N) derived from epithermal (circles), thermal (squares), and fast (asterisks) neutron fluxes.

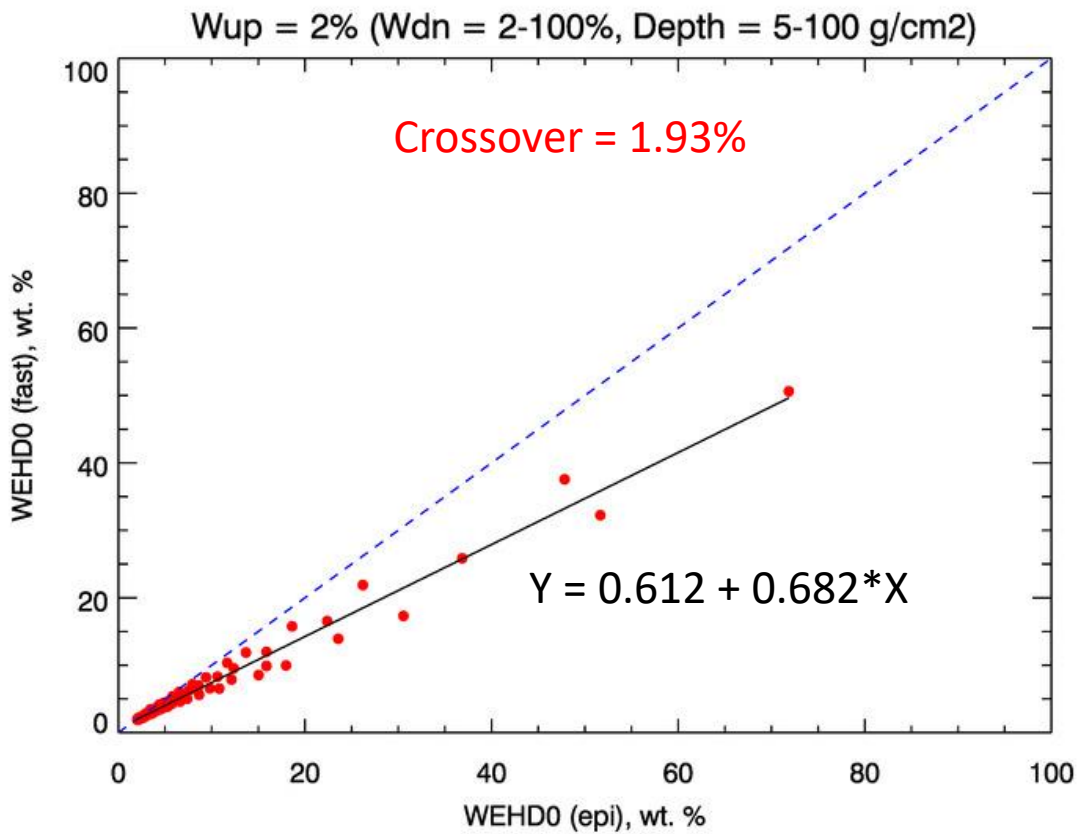

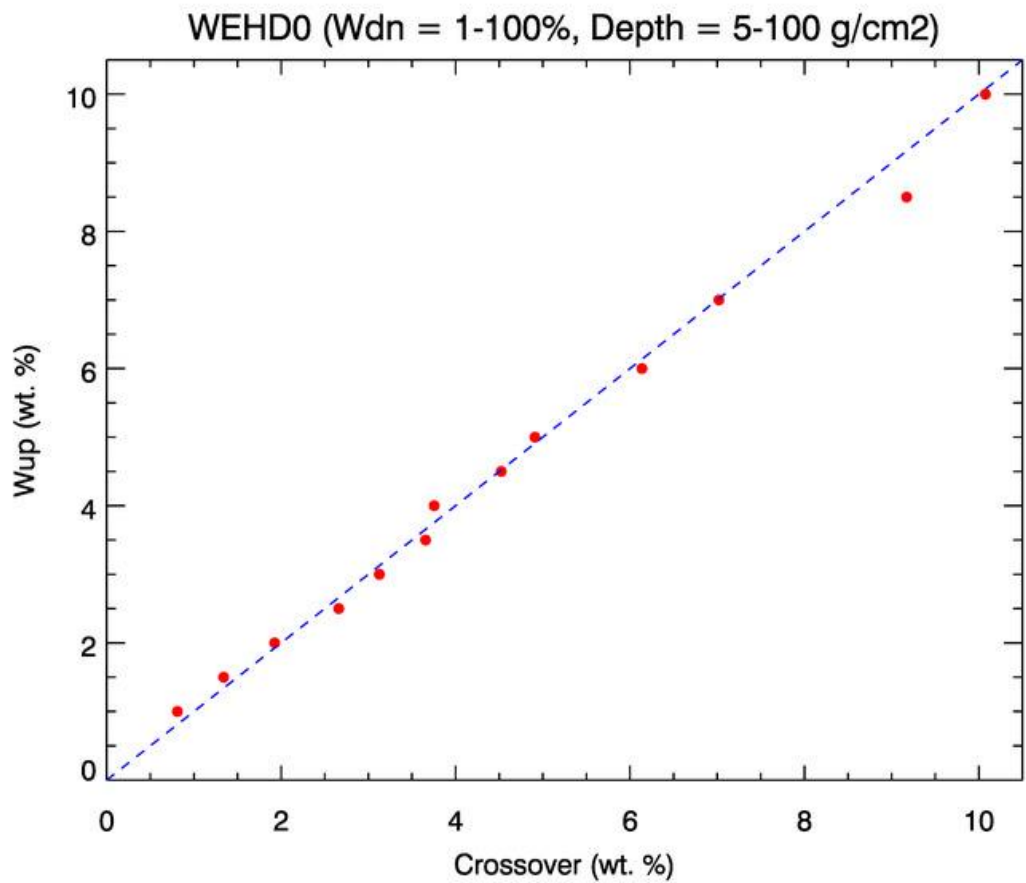

**Figure 7: (A)** Sample crossover calculation for MCNPX $W_{up}$ = 2% simulations. Best fit line (solid) crosses over with x=y unity line (dashed) at WEHD0 = 1.93%. **(B)** Crossover results for MCNPX $W_{up}$ = 1, 1.5, 2, 2.5, 3, 3.5, 4, 4.5, 5, 6, 7, 8.5, 10 % simulations.

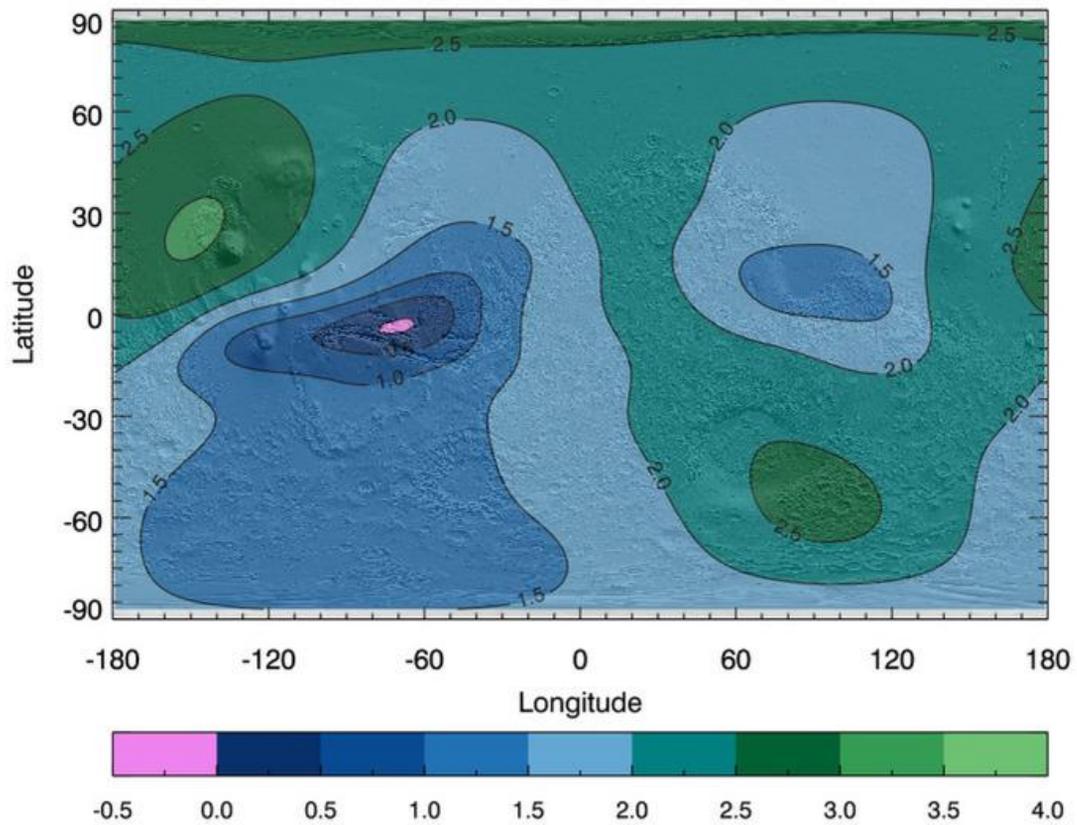

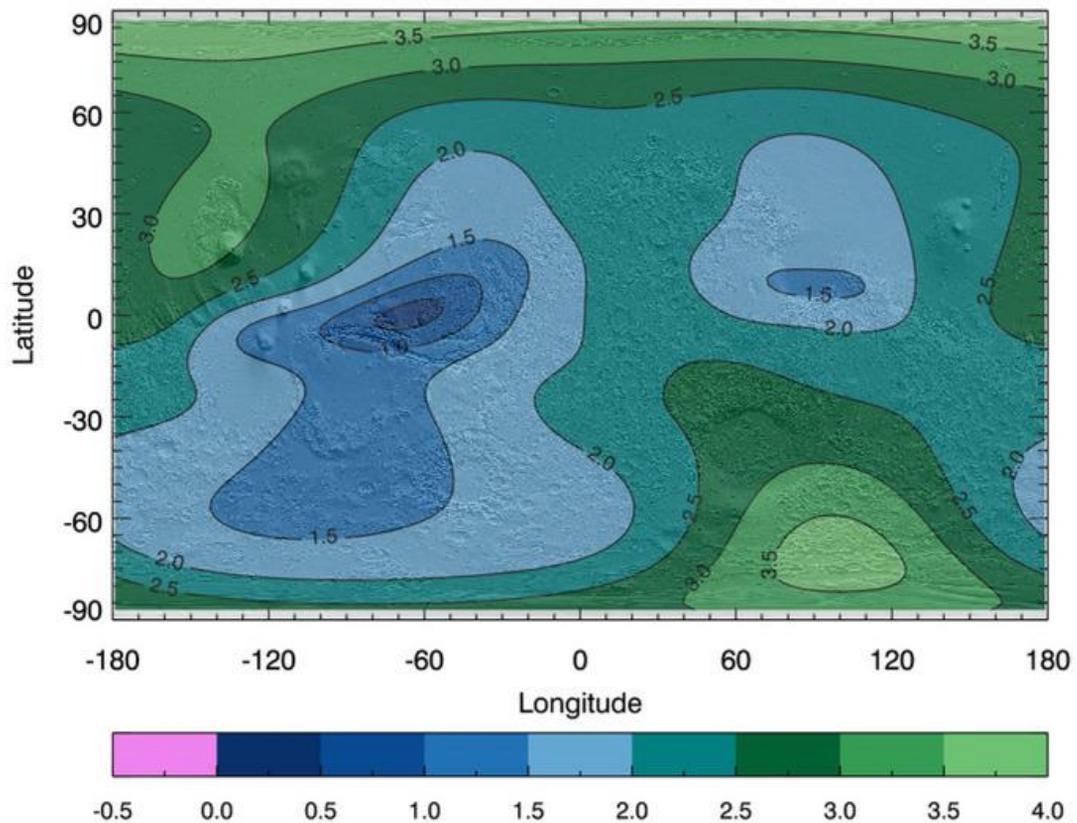

**Figure 8:** Global frost-free $W_{up}$ maps for **(A)** undeconvolved and **(B)** deconvolved {N = 16} solutions, calculated using Gaussian-weighted least squares fitting normalized to $R_o$ = 1300 km.

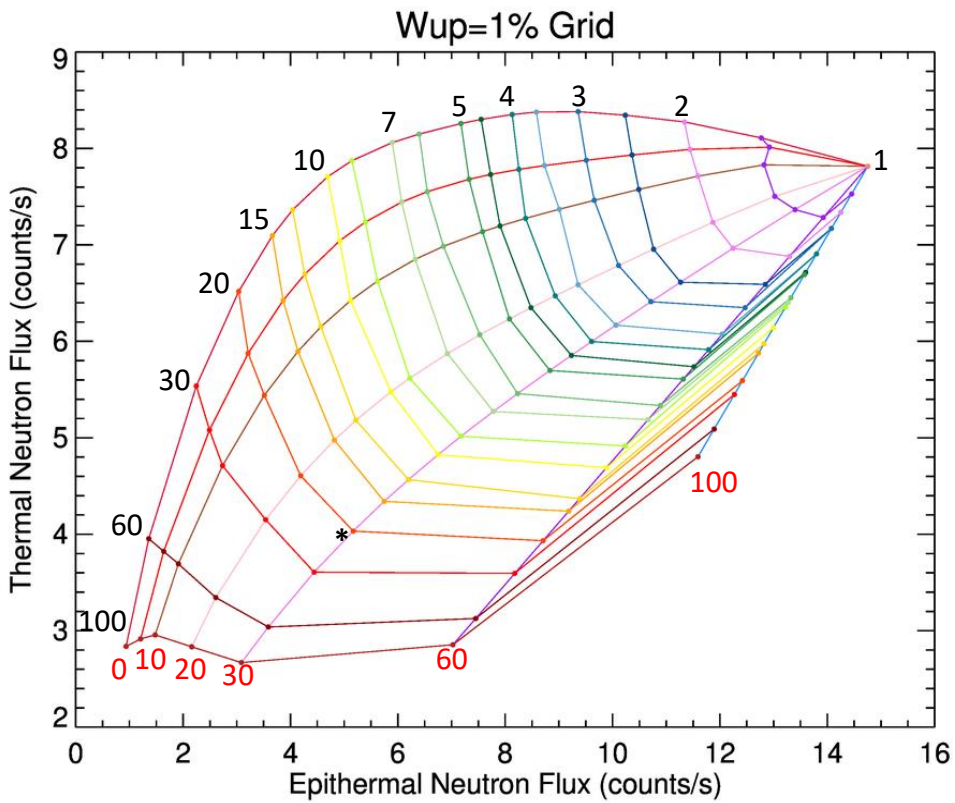

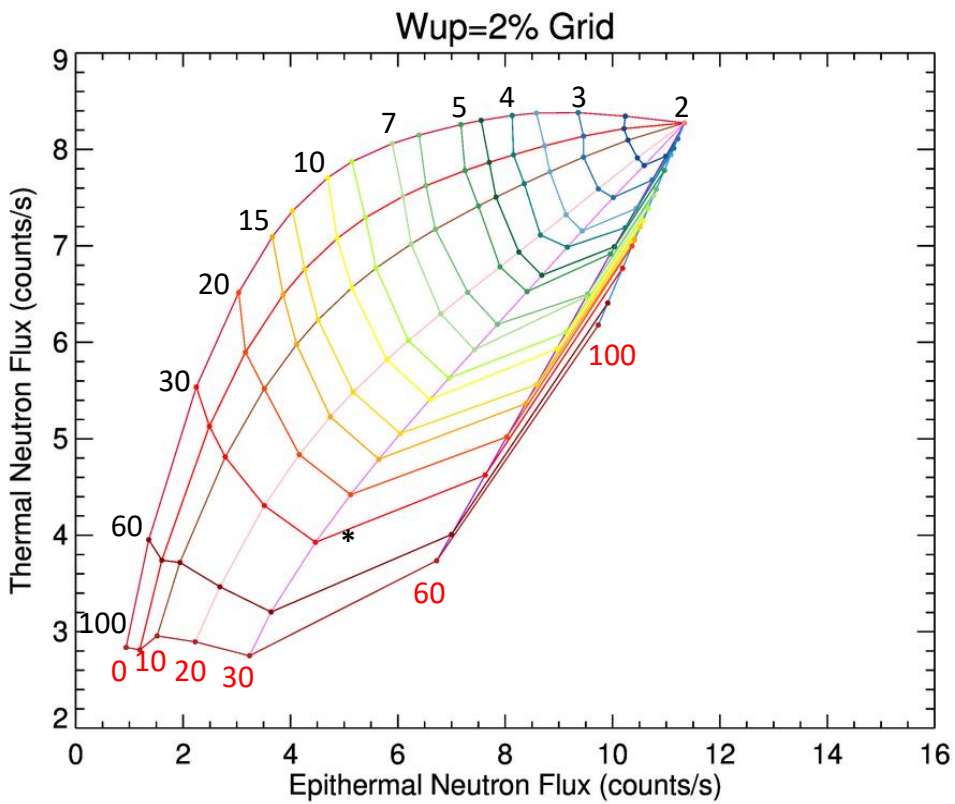

**Figure 9:** MCNPX-modeled thermal vs. epithermal grids for **(A)** $W_{up}$ = 1% and **(B)** $W_{up}$ = 2%. Contours indicate constant values of $W_{dn}$ (**black** labels [%]) and Depth (**red** labels [g/cm²]). For each $W_{dn}$ contour, thermal counting rates have been smoothed as a function of $C_E$ using a 6-degree polynomial fit.

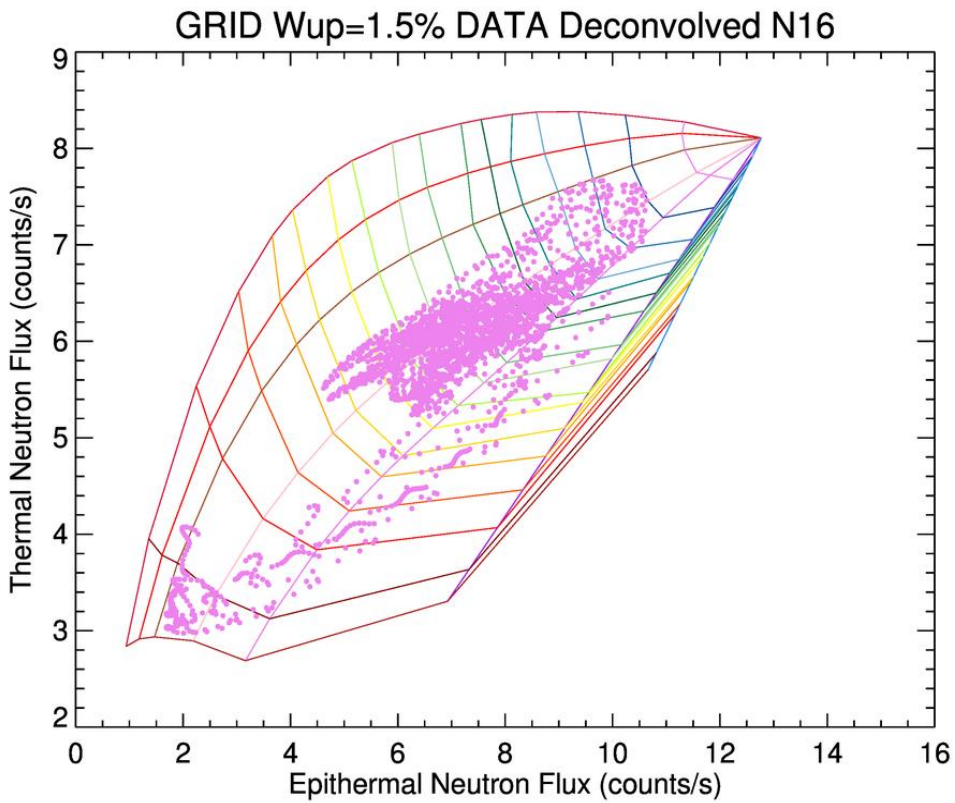

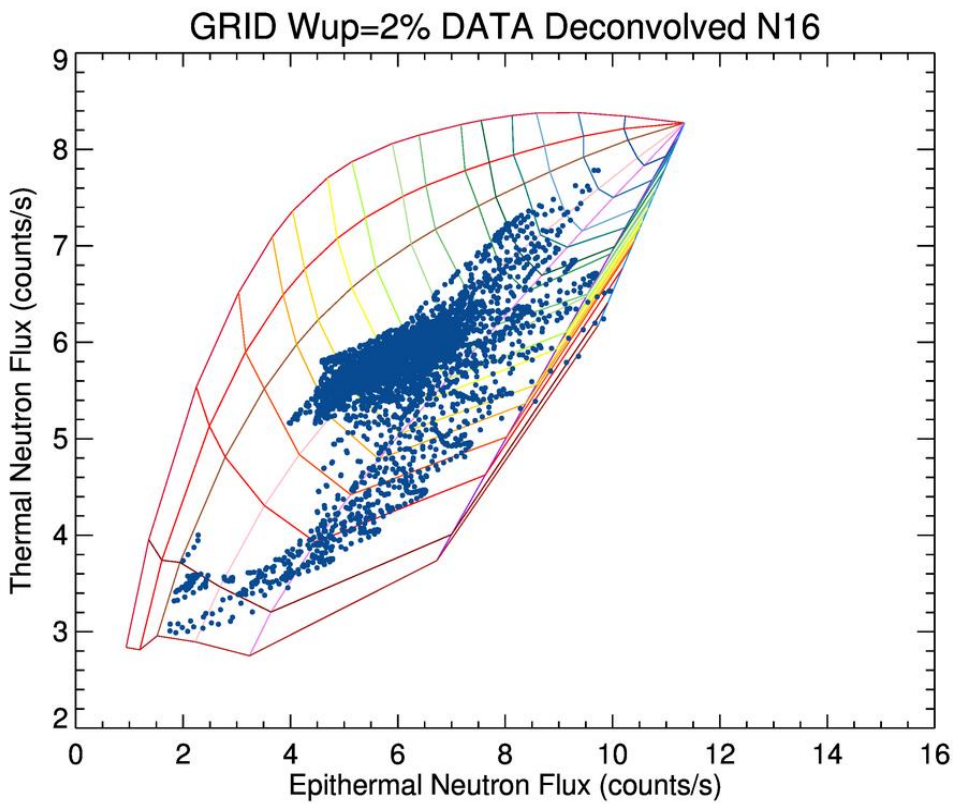

**Figure 10: (A)** Deconvolved {N = 16} MONS data spanning $1.5\% < W_{up} < 2\%$ on $W_{up} = 1.5\%$ model grid. **(B)** Deconvolved {N = 16} MONS data spanning $2\% < W_{up} < 2.5\%$ on $W_{up} = 2\%$ model grid. *Data poleward of 75°S have been excluded.*

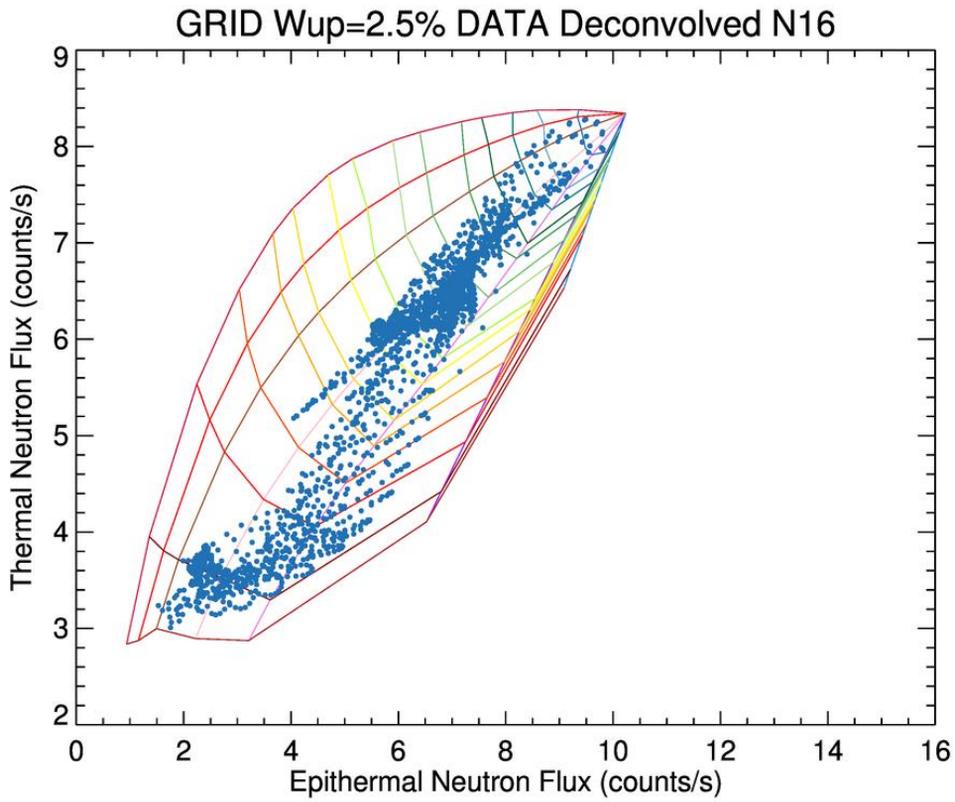

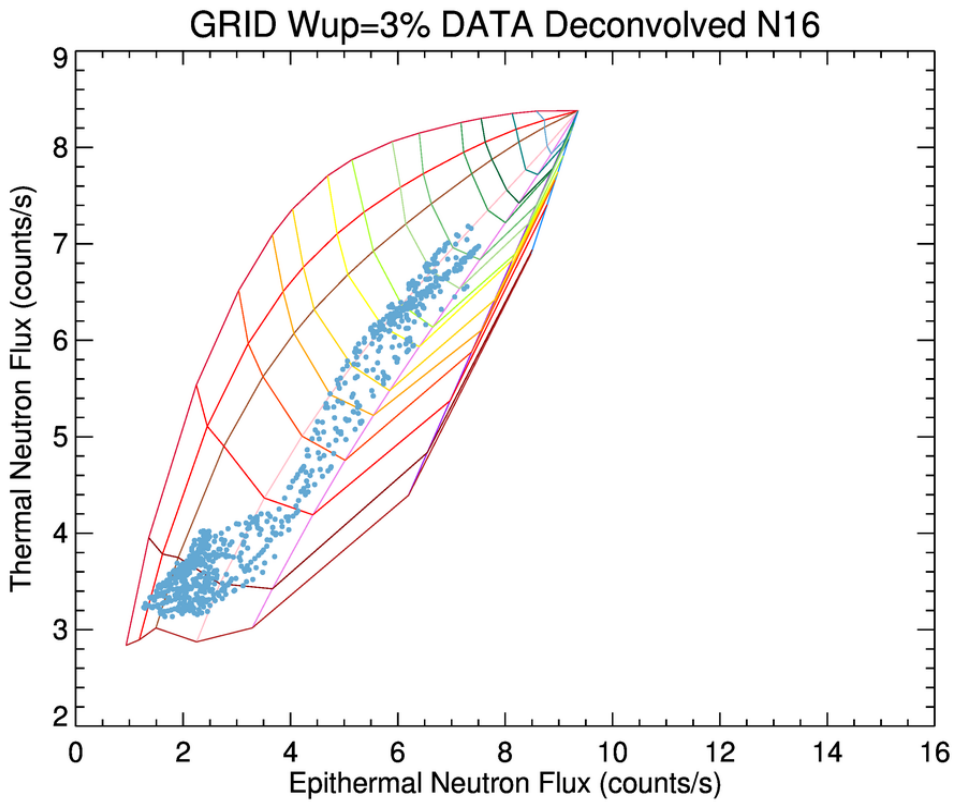

**Figure 10: (C)** Deconvolved {N = 16} MONS data spanning 2.5% < $W_{up}$ < 3% on $W_{up}$ = 2.5% model grid. **(D)** Deconvolved {N = 16} MONS data spanning 3% < $W_{up}$ < 3.5% on $W_{up}$ = 3% model grid. *Data poleward of 75°S have been excluded.*

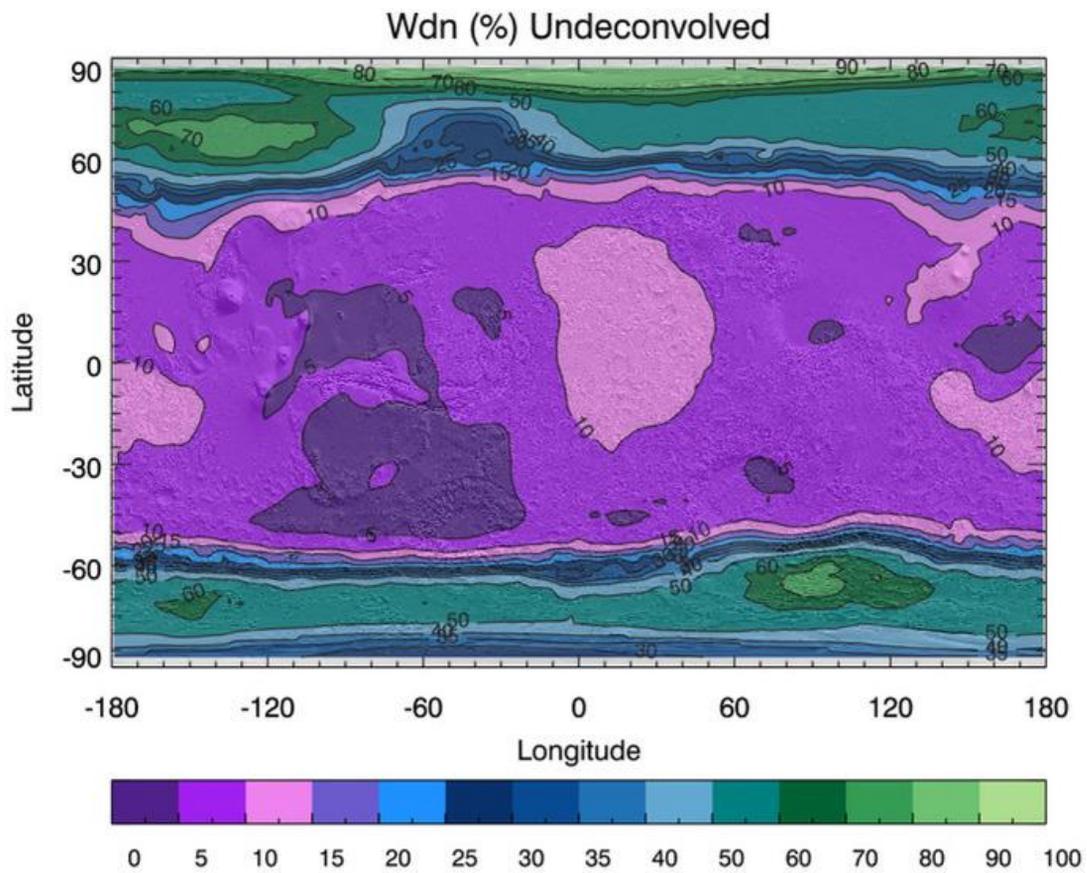
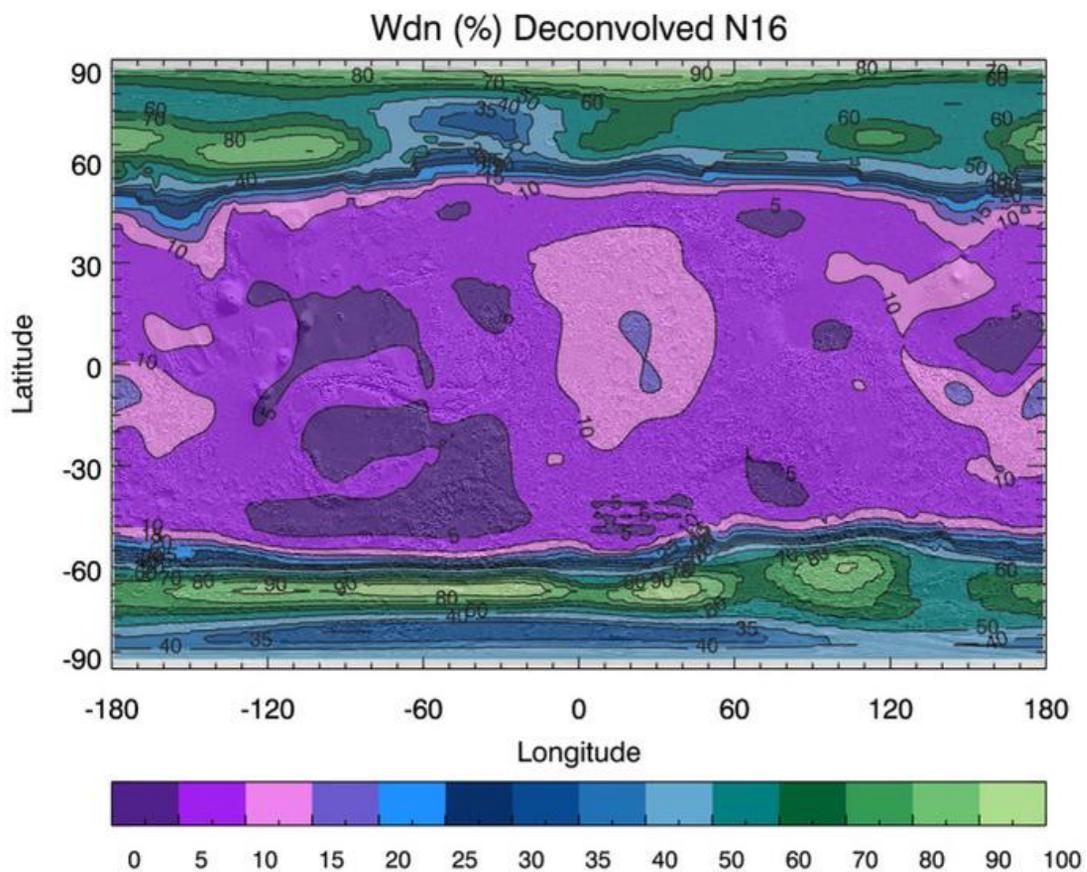

**Figure 11:** Global frost-free $W_{dn}$ maps for **(A)** undeconvolved and **(B)** deconvolved {N = 16} solutions.

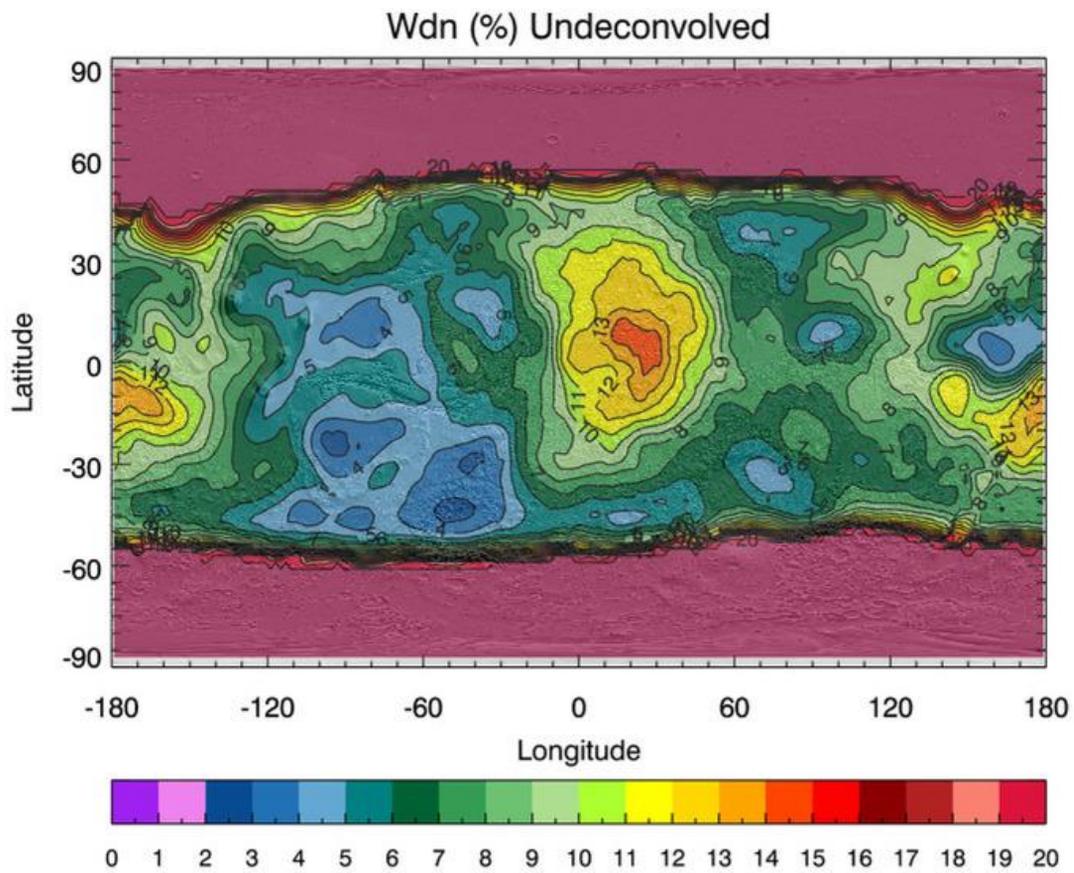
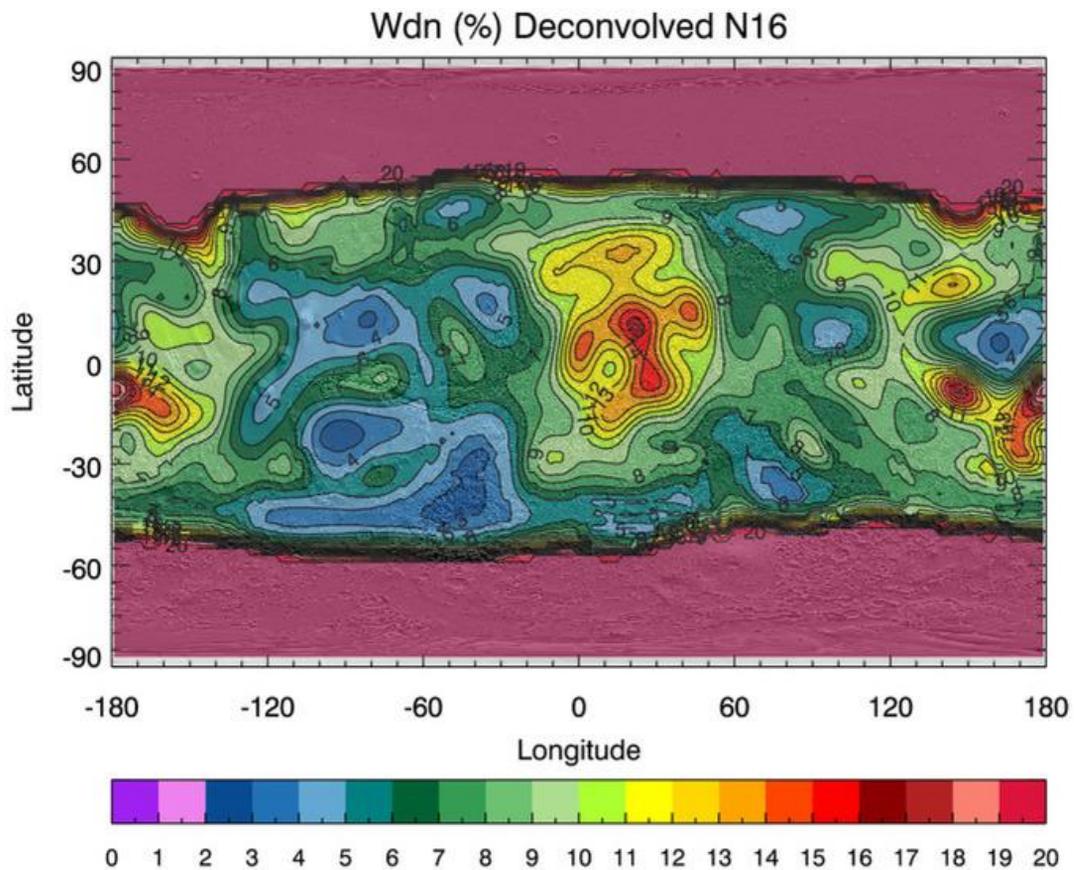

**Figure 11:** Global frost-free $W_{dn}$ maps for **(C)** undeconvolved and **(D)** deconvolved {N = 16} solutions, re-scaled to emphasize $W_{dn}$ variations at low latitudes.

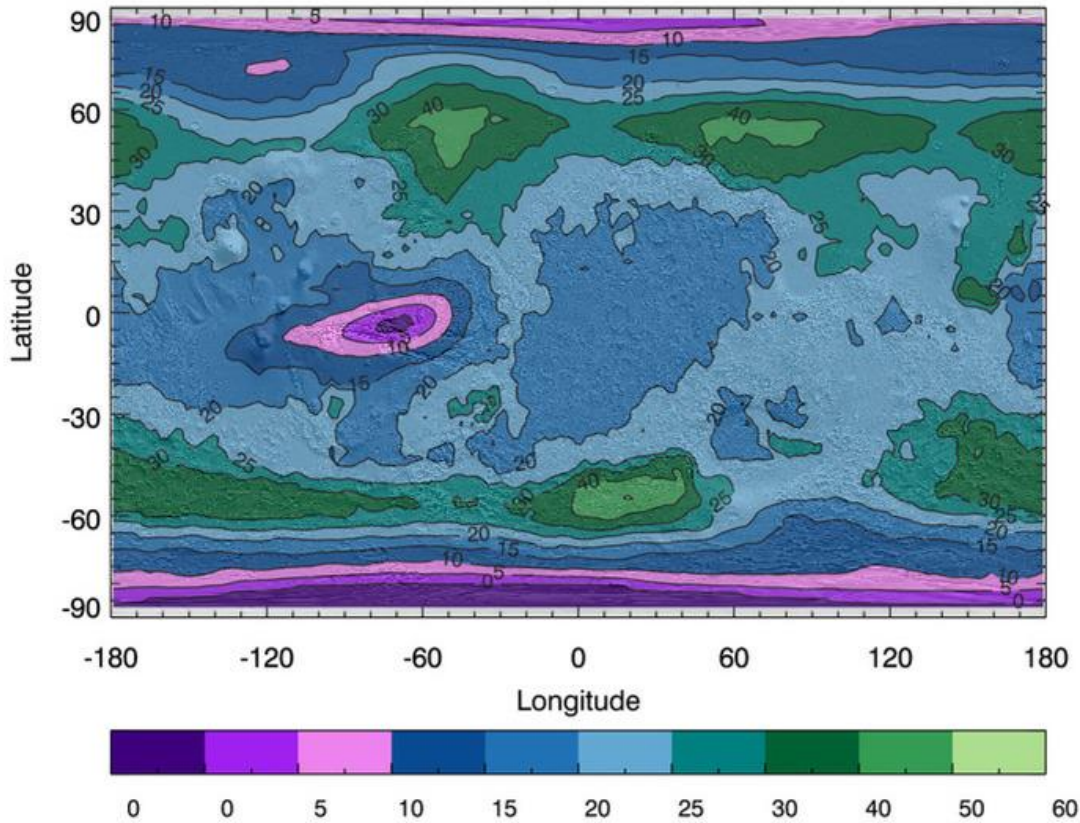
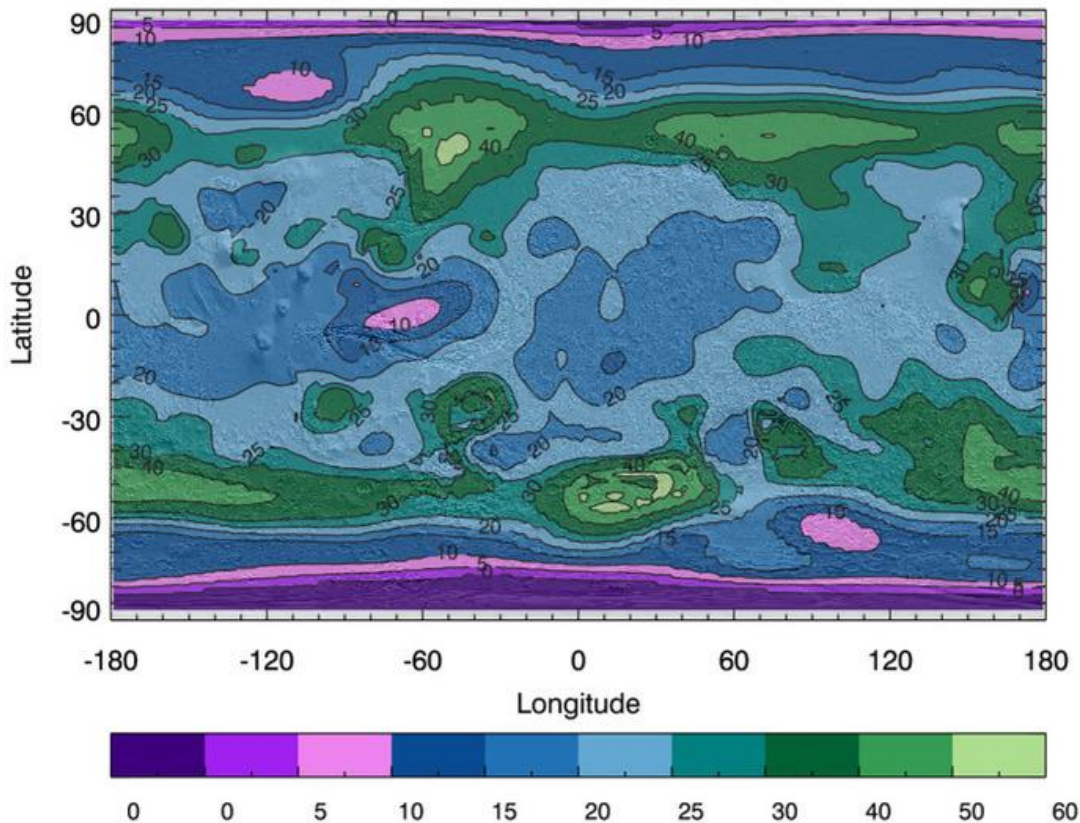

**Figure 12:** Global frost-free Depth maps for **(A)** undeconvolved and **(B)** deconvolved {N = 16} solutions. The darkest purple contour corresponds to $D = 0$, indicating inapplicability of standard two-layer model.

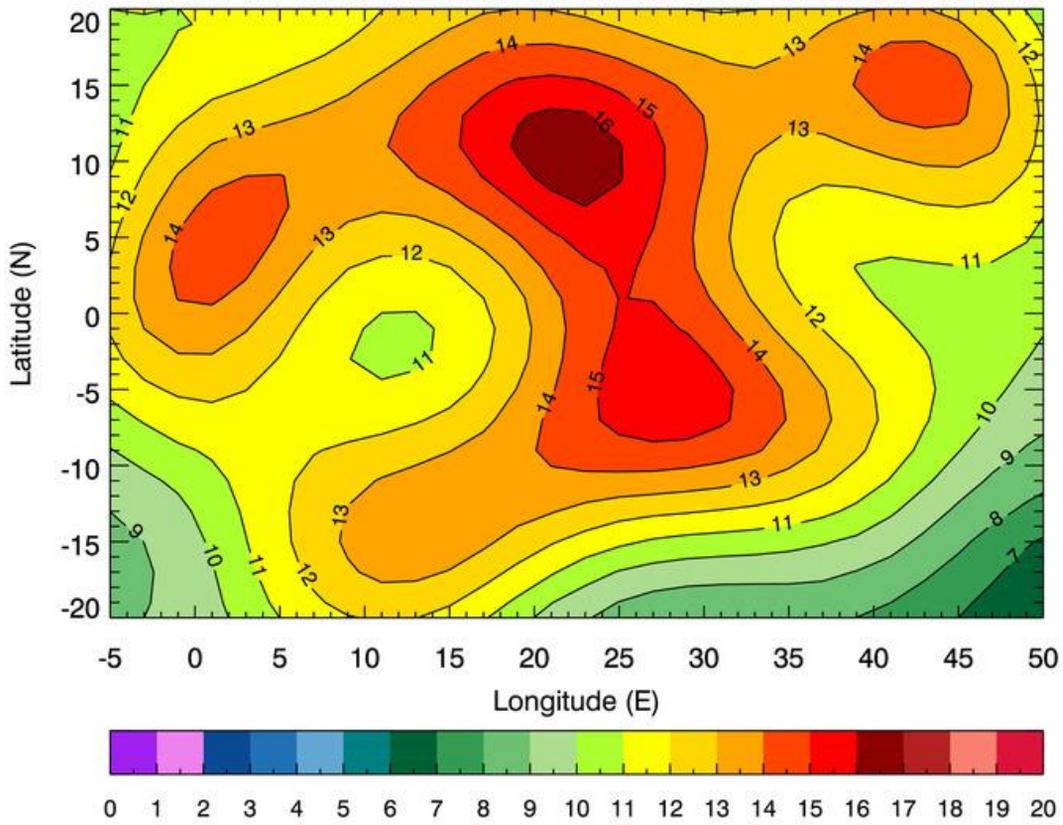
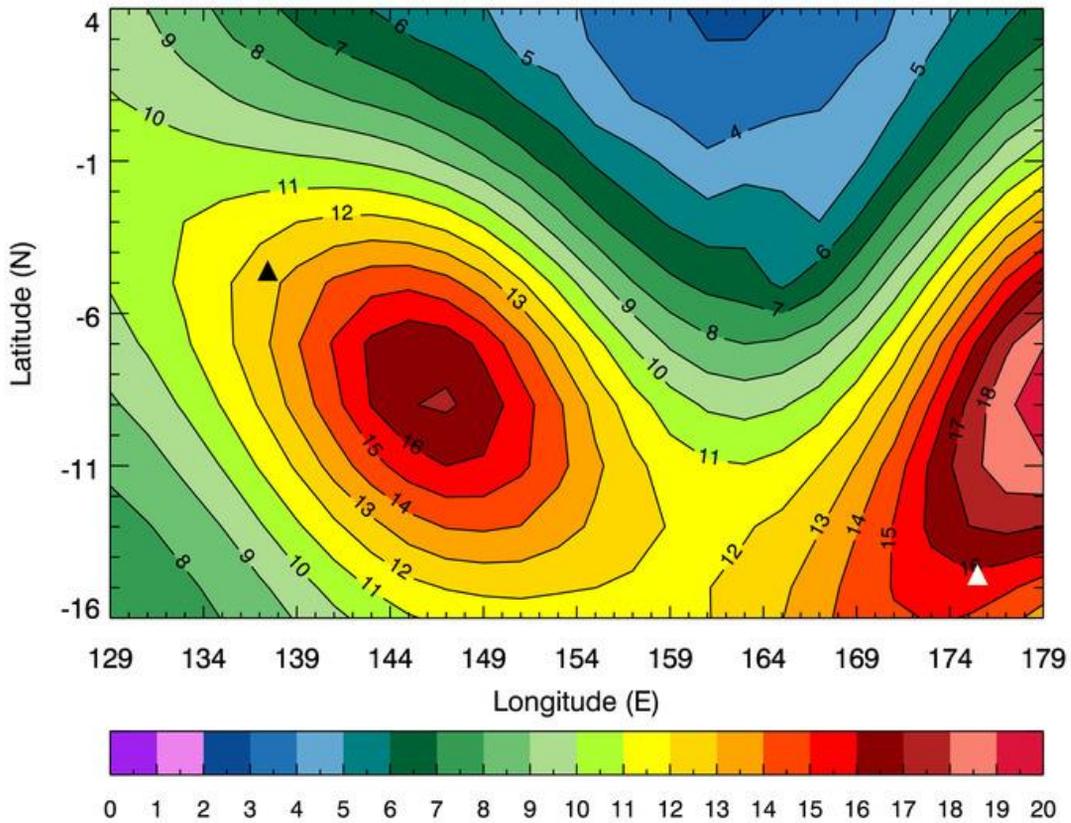

**Figure 13:** Regional maps of $W_{dn}$ for deconvolved {N=16} solution in **(A)** southern Arabia Terra and **(B)** vicinity of Curiosity (black triangle) and Spirit (white triangle) landing sites.

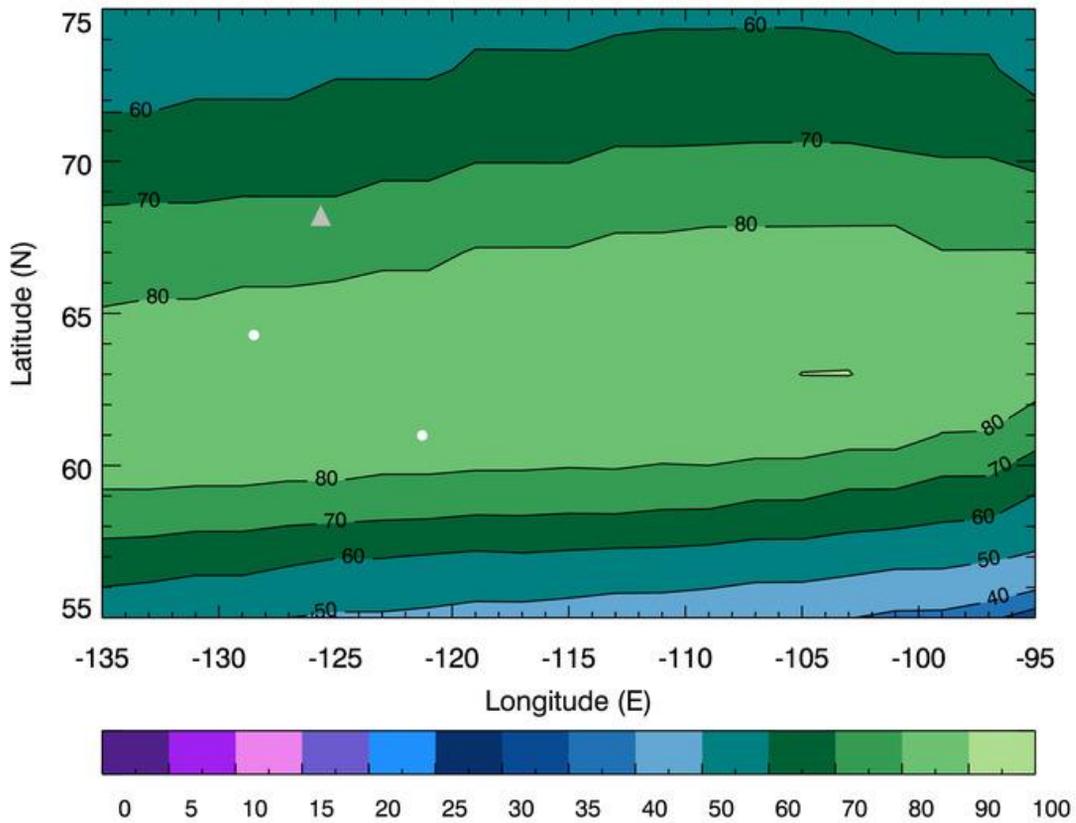
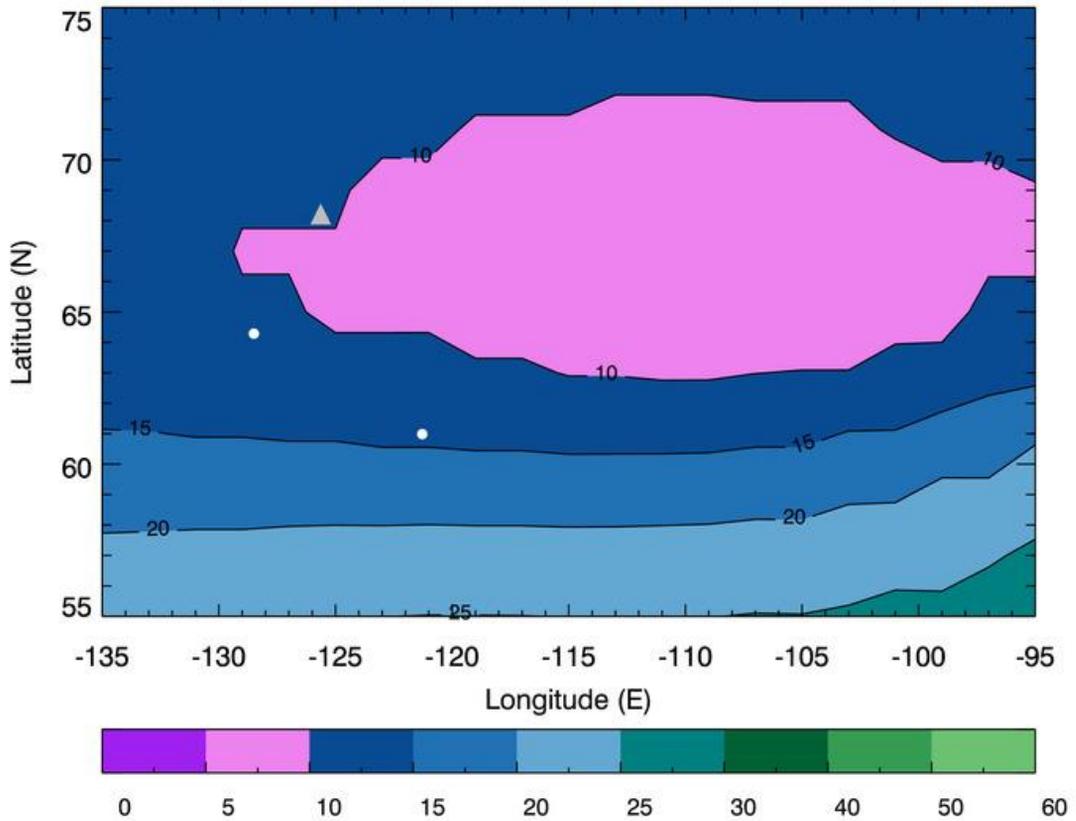

**Figure 14:** Regional maps of WEH for deconvolved {N = 16} solution: (A) $W_{dn}$ and (B) Depth. Grey triangle indicates Phoenix landing site ($W_{dn}$ = 72%, D = 10 g/cm$^2$); white circles denote recent ice-exposing craters.

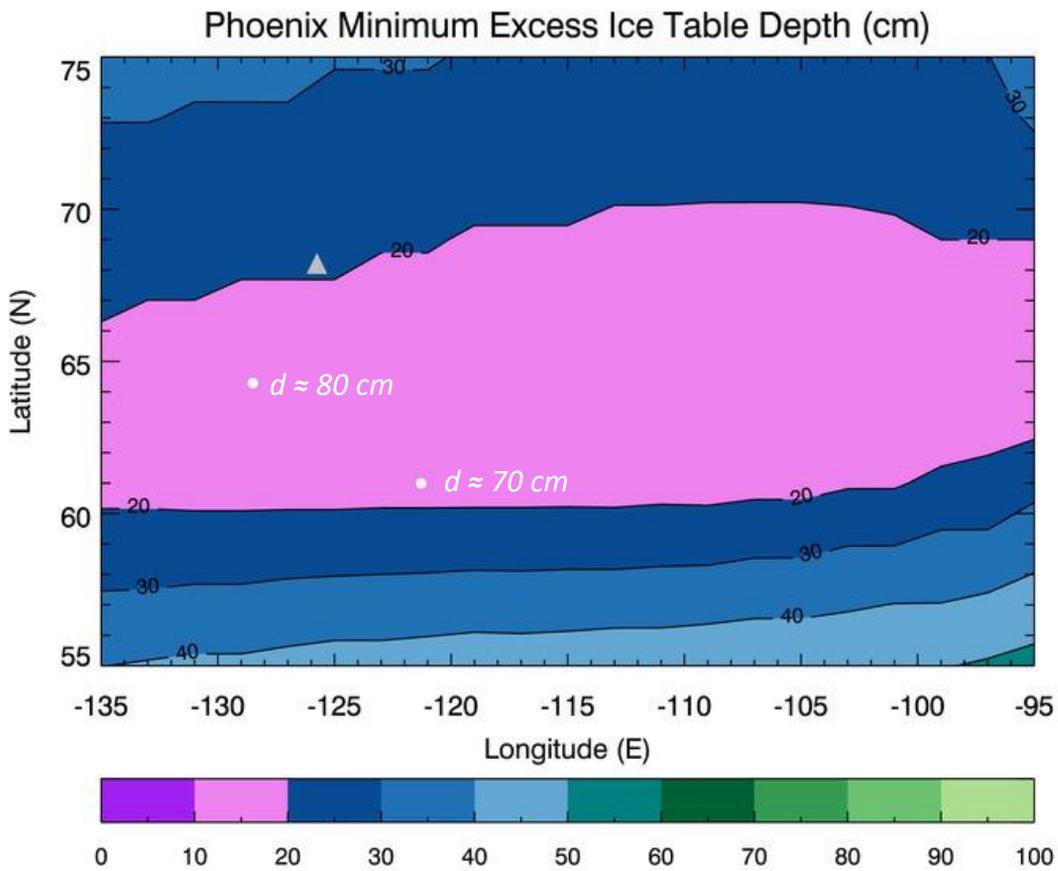

**Figure 14:** Regional map of WEH for deconvolved {N = 16} solution: (C) Minimum Excess Ice Table Depth. Grey triangle indicates Phoenix landing site; white circles denote recent ice-exposing craters.

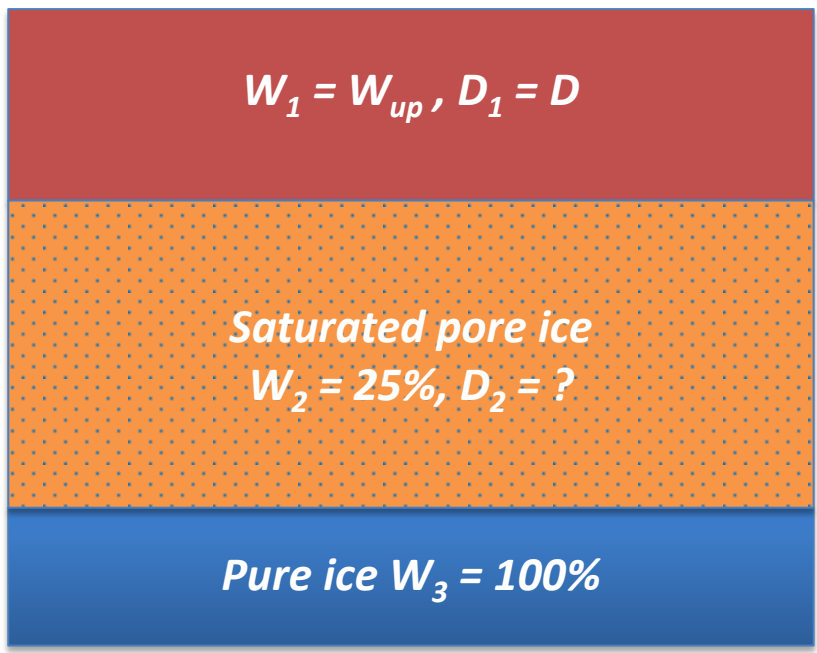

**Figure 15:** Schematic of five parameter, three-layer subsurface model.

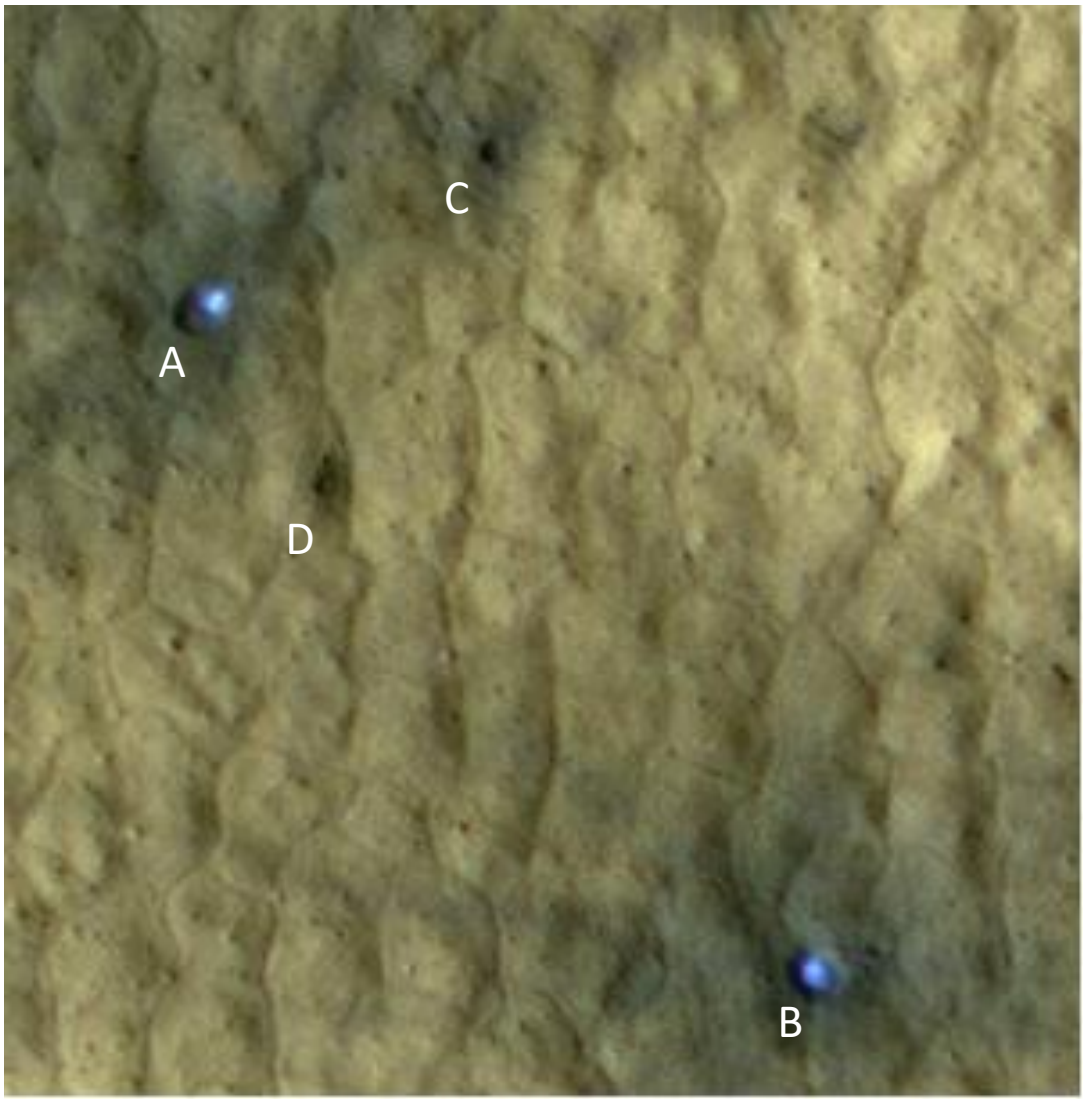

**Figure 16:** Recent ice-exposing impact crater cluster at 46.3°N, 176.9°E, corresponding to Site 1 of *Byrne et al.* (2009). This portion of HiRISE image PSP_009978_2265 is 35 m across. Note the two larger $D$ = 4 m / $d$ = 80 cm impacts (denoted by A & B) in the cluster appear to penetrate to the excess ice table, whereas the smaller $D$ = 2.5 m / d = 50 cm (denoted by C & D) impacts do not. Image Credit: NASA/JPL-Caltech/University of Arizona.

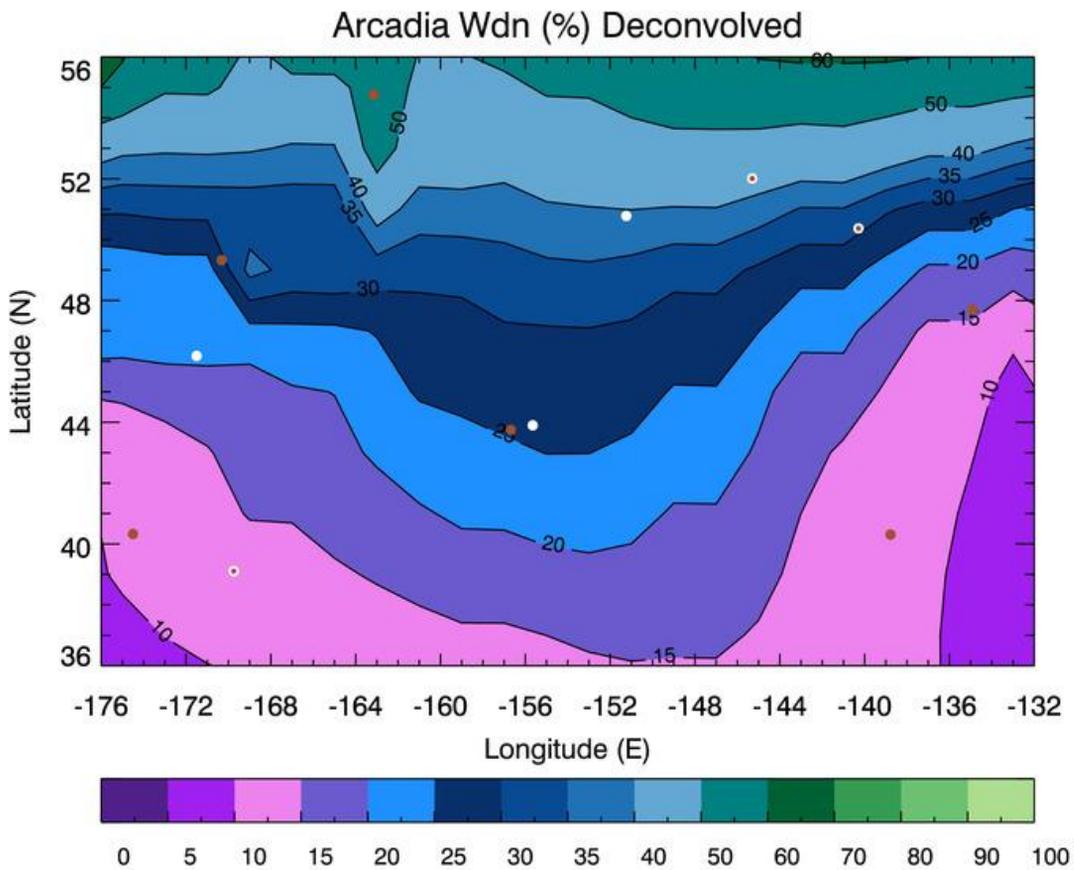
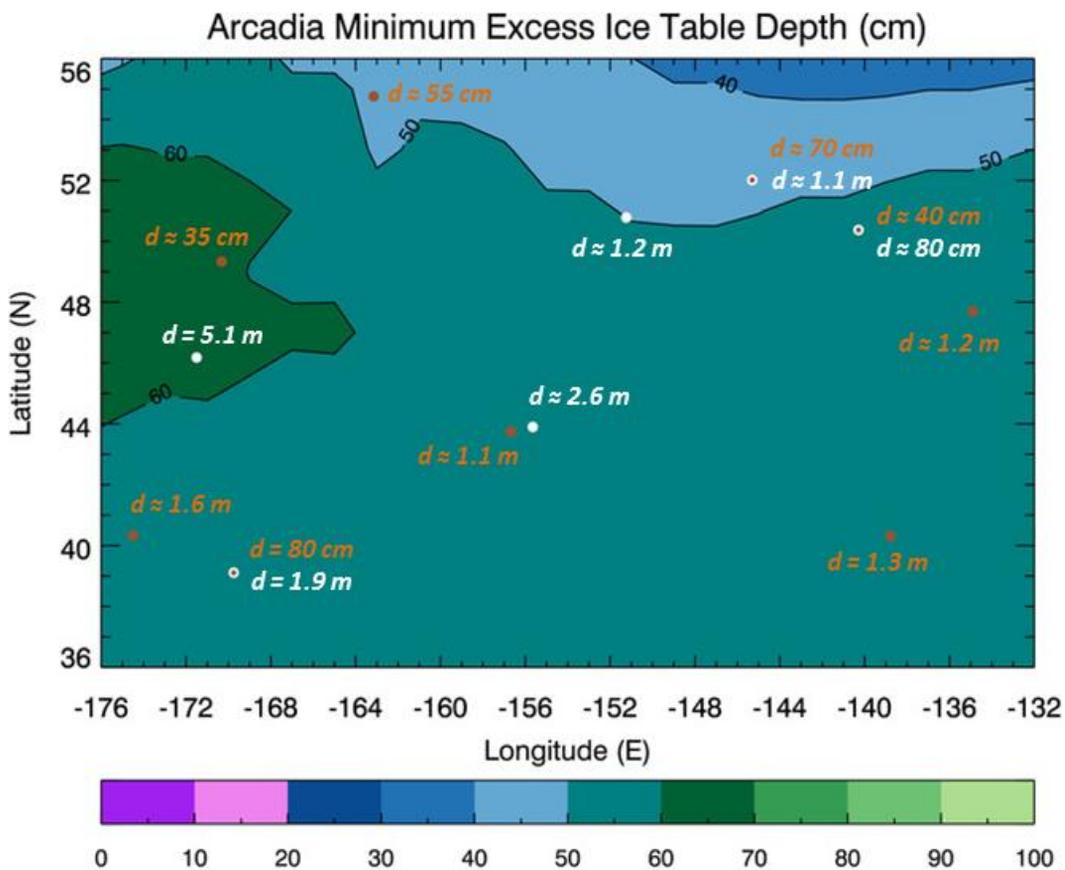

**Figure 17:** Regional maps of WEH for preferred {N = 16} deconvolved solution: (A) $W_{dn}$ and (B) Minimum Excess Ice Table Depth. White/brown circles denote recent ice-exposing/ice-free craters.

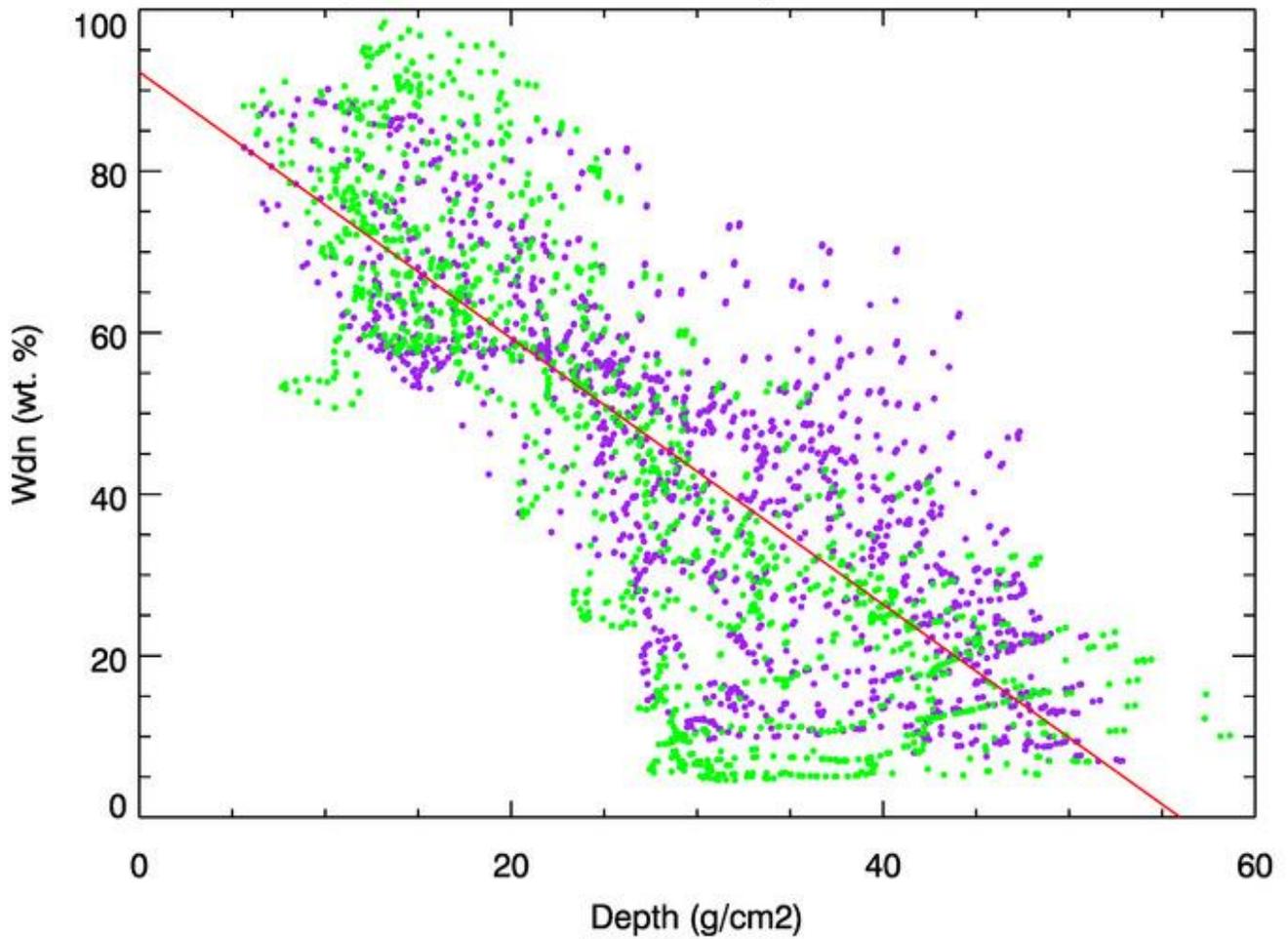

**Figure 18:** Anticorrelation (R = -0.80) of $W_{dn}$ and $D$ for deconvolved {N=16} solution at high latitudes (purple = 50°N-75°N, green = 50°S-75°S). Best fit line (red) given by $W_{dn} = 92.3 - 1.65*D$.

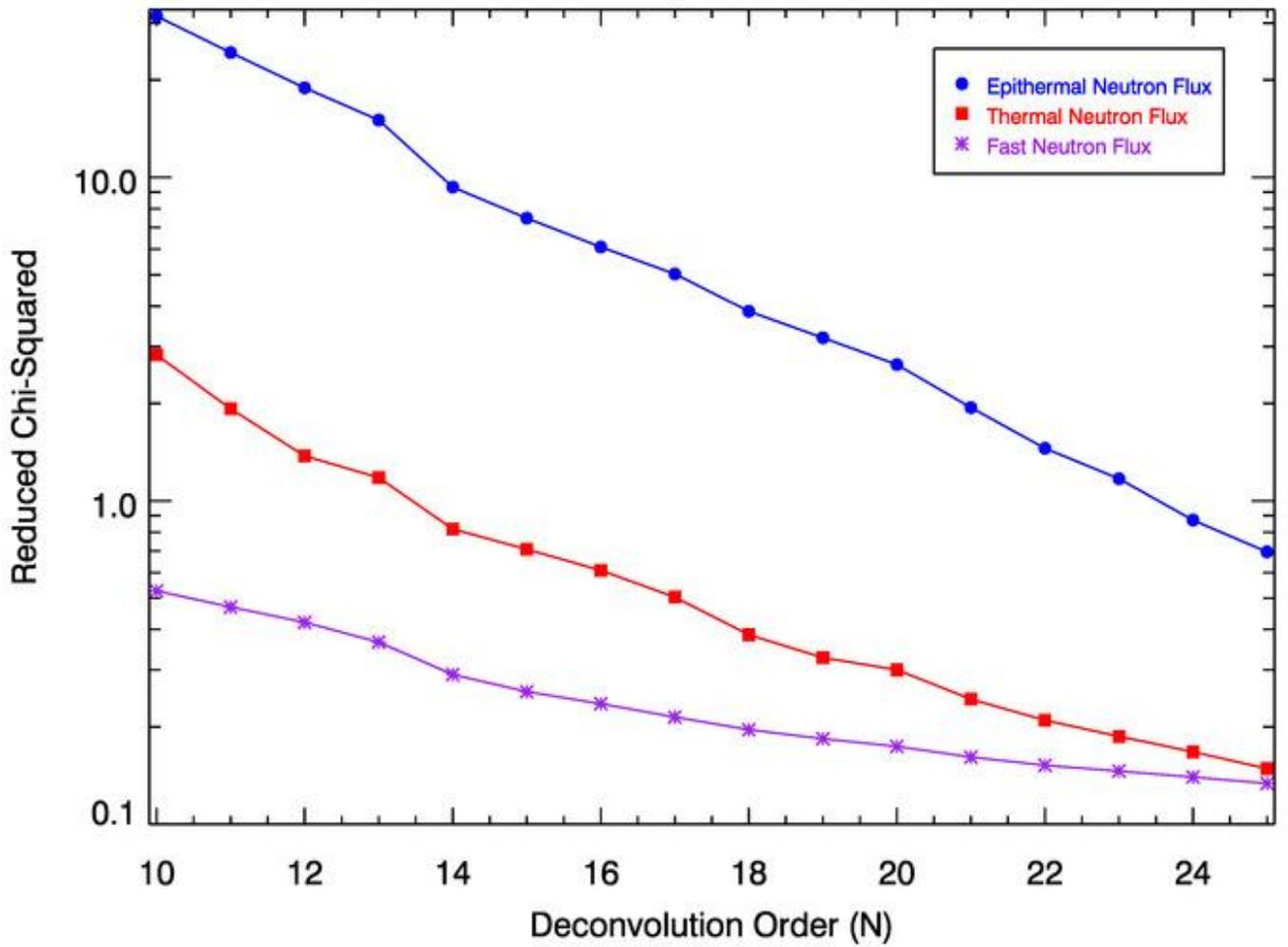

**Figure A1:** Reduced chi-squared vs. deconvolution order (N) for epithermal (circles), thermal (squares), and fast (asterisks) neutron fluxes.

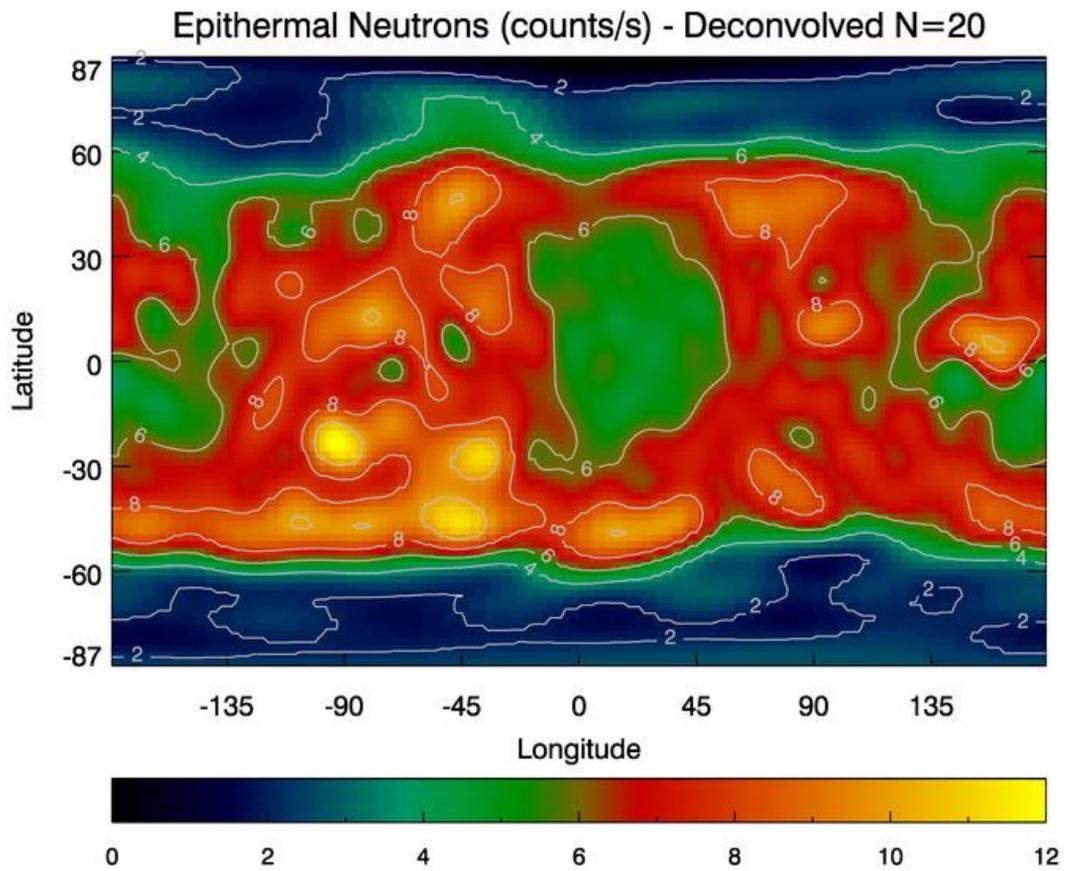
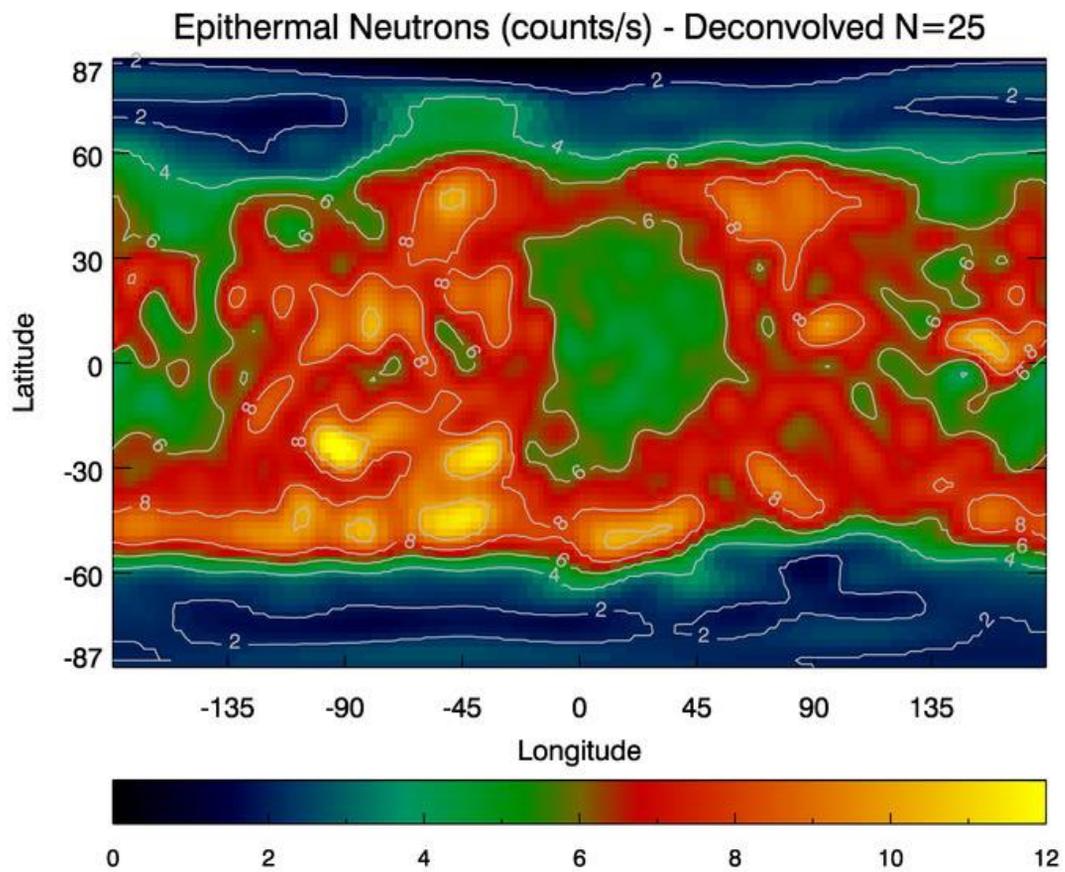

**Figure S1:** MONS frost-free map of epithermal neutron fluxes (counts/s) for deconvolved **(A)** {N = 20} and **(B)** {N = 25} solutions.

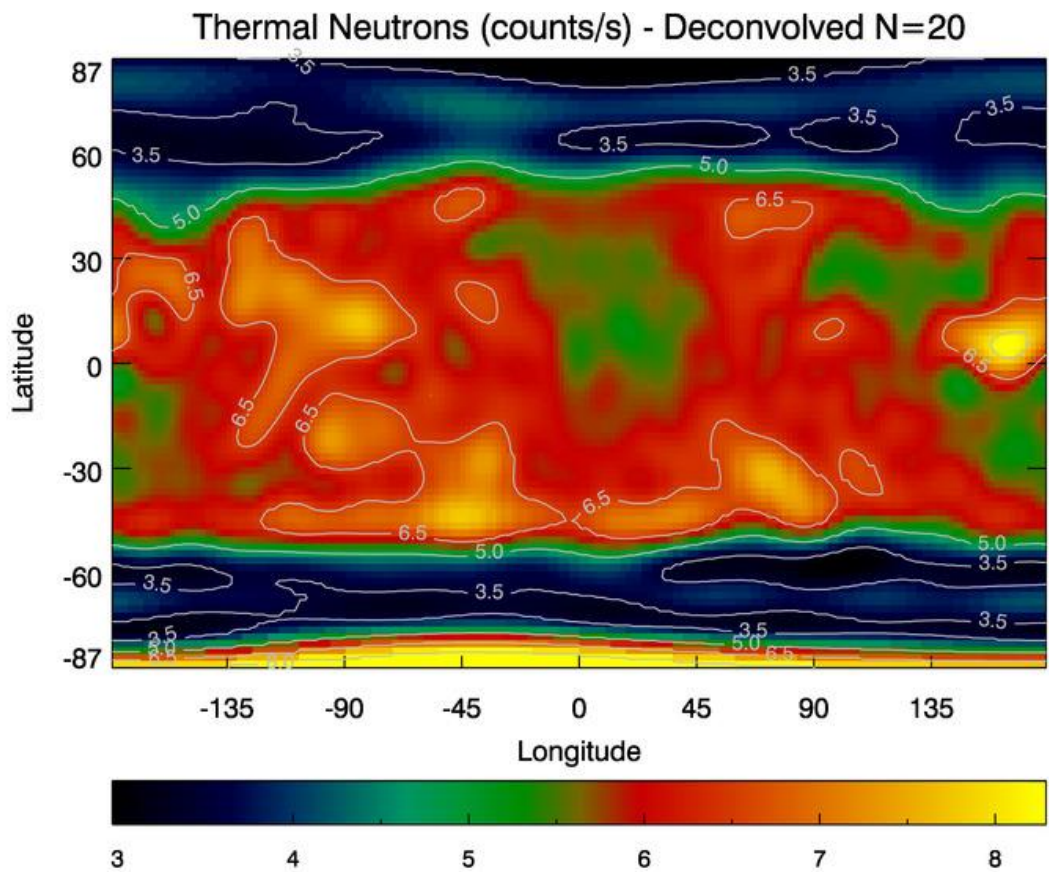

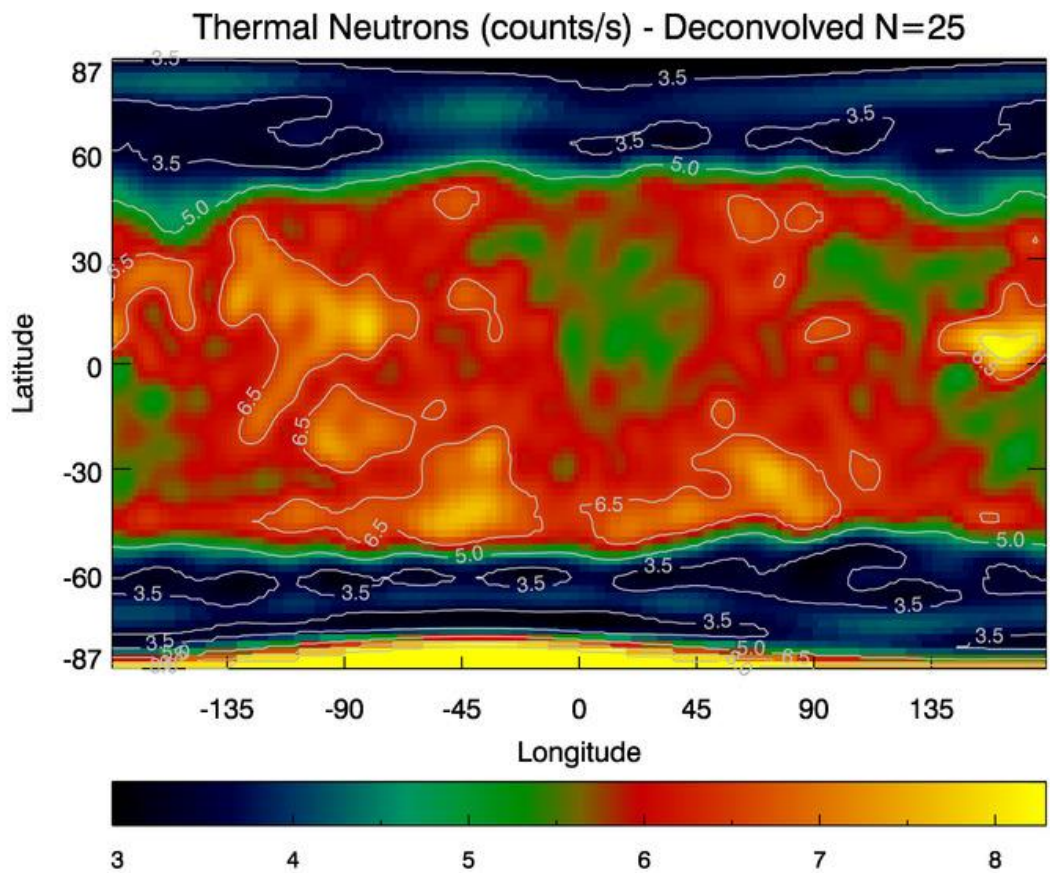

**Figure S2:** MONS frost-free map of thermal neutron fluxes (counts/s) for deconvolved **(A)** {N = 20} and **(B)** {N = 25} solutions.

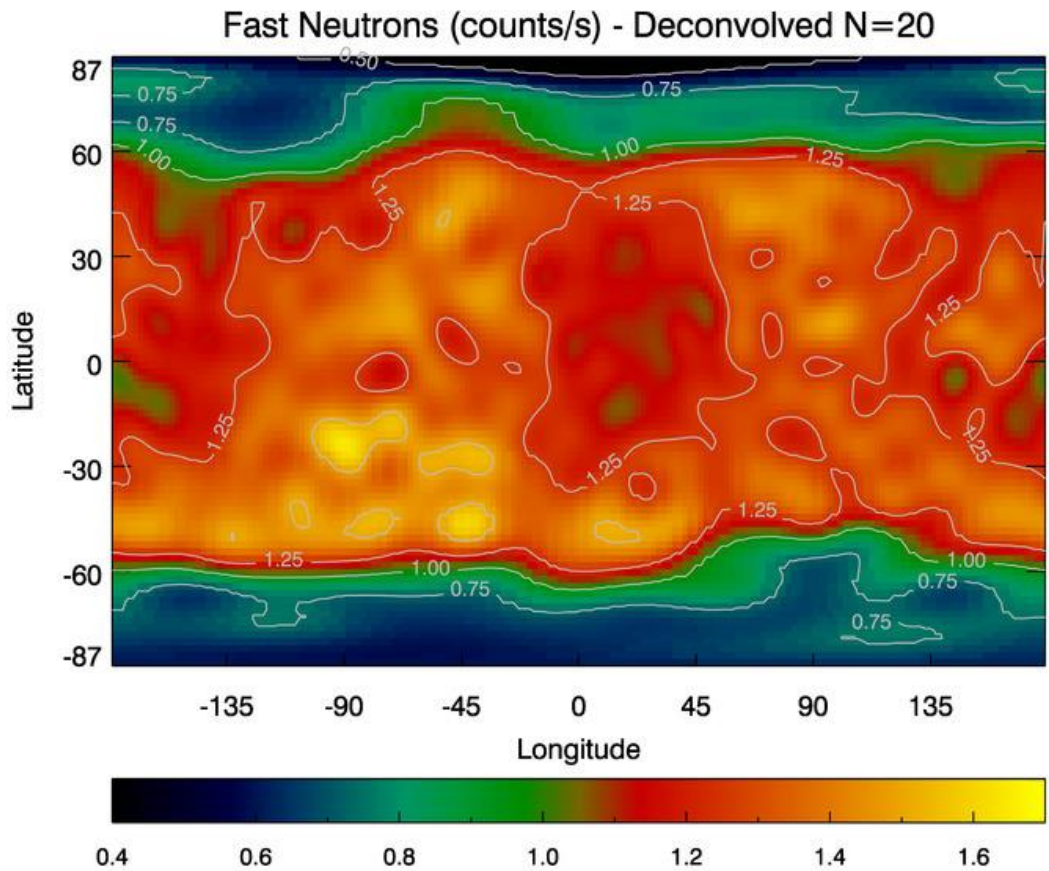
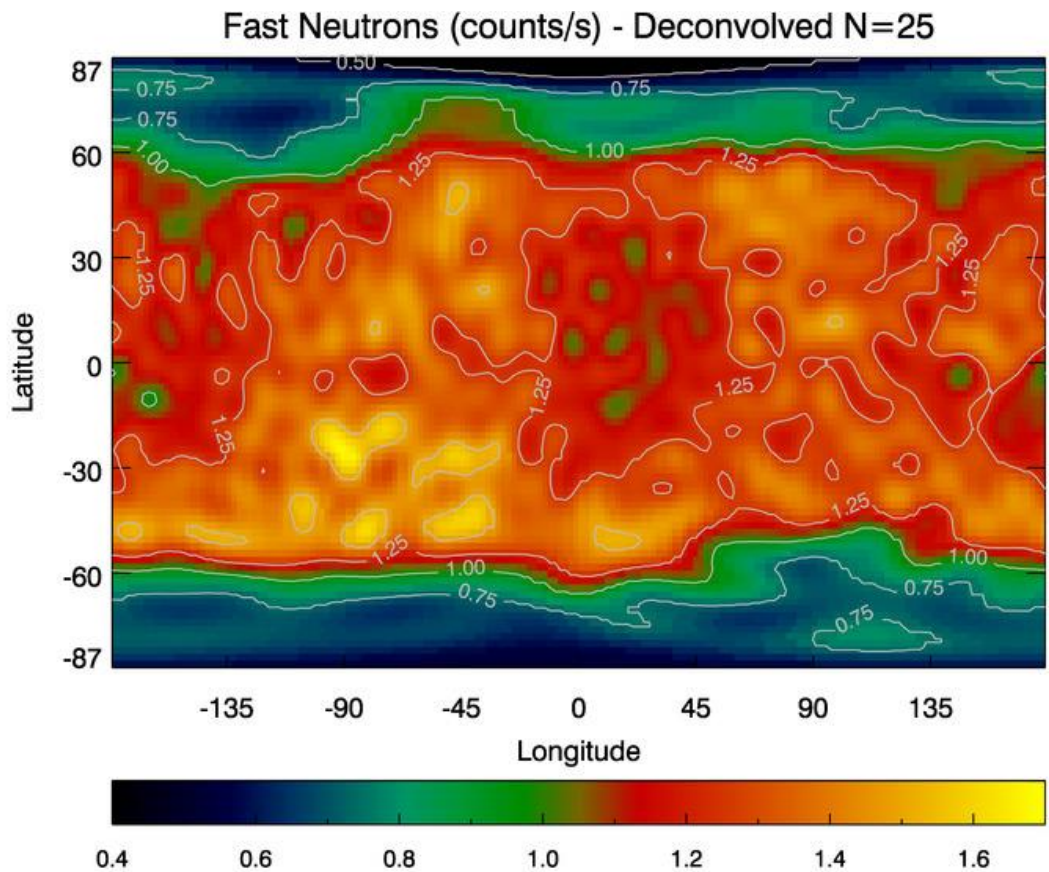

**Figure S3:** MONS frost-free map of fast neutron fluxes (counts/s) for deconvolved **(A)** {N = 20} and **(B)** {N = 25} solutions.

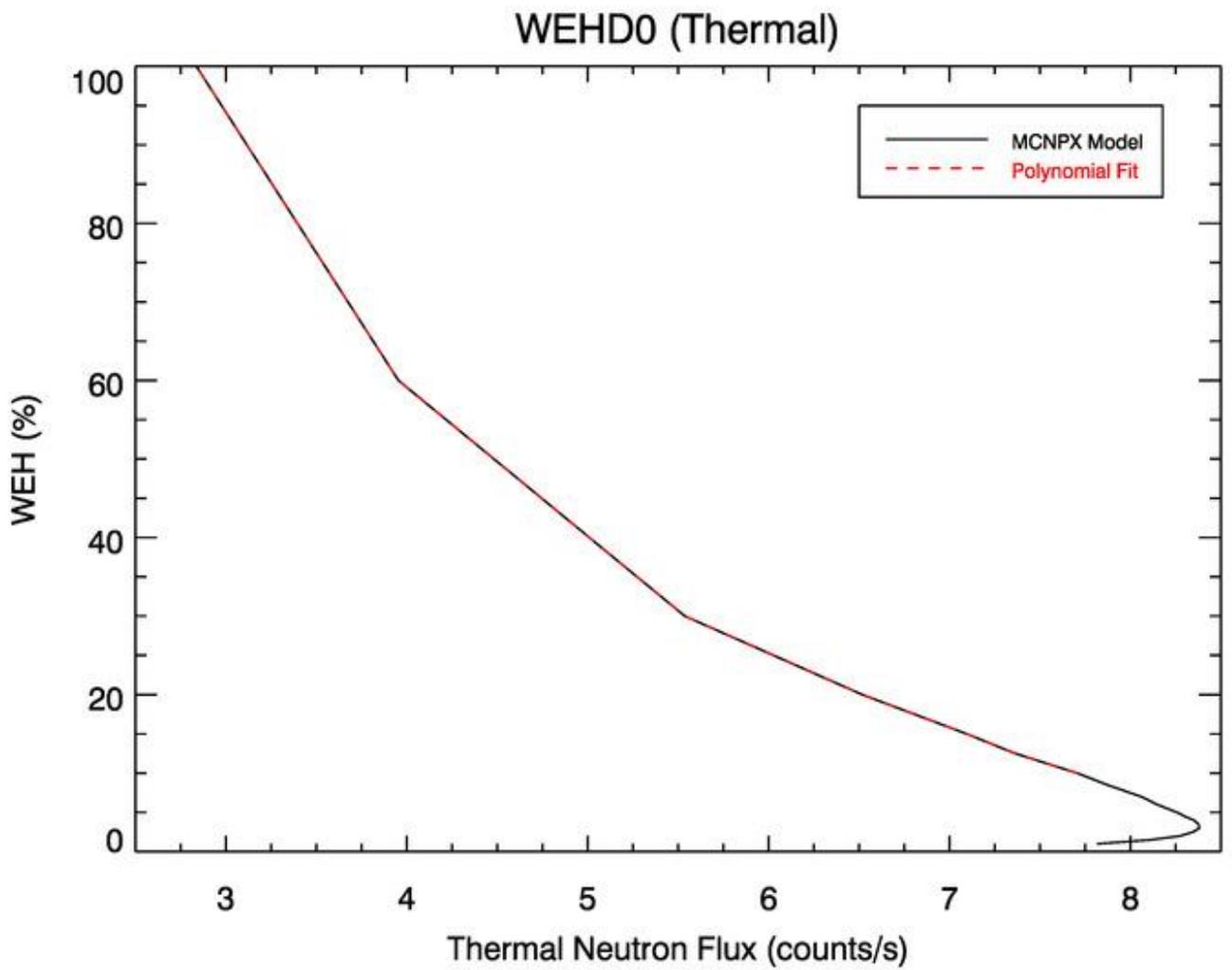

**Figure S4:** MCNPX model zero-depth 1D WEH solutions (WEHD0) and polynomial fits (via Eq. 4 and Table 1) for thermal neutron flux. Note that fit only applies to monotonically increasing portion of curve corresponding to WEHD0 > 10% ($C_T$ < 7.7 counts/s).

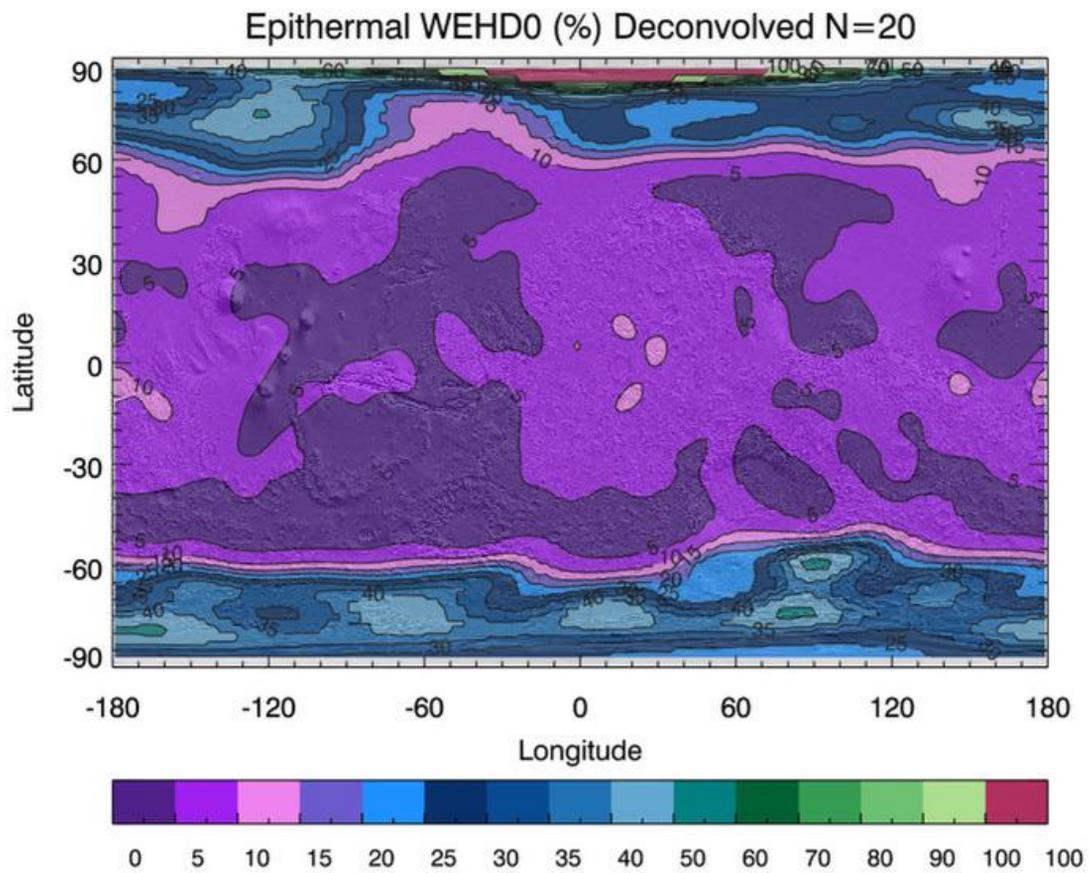

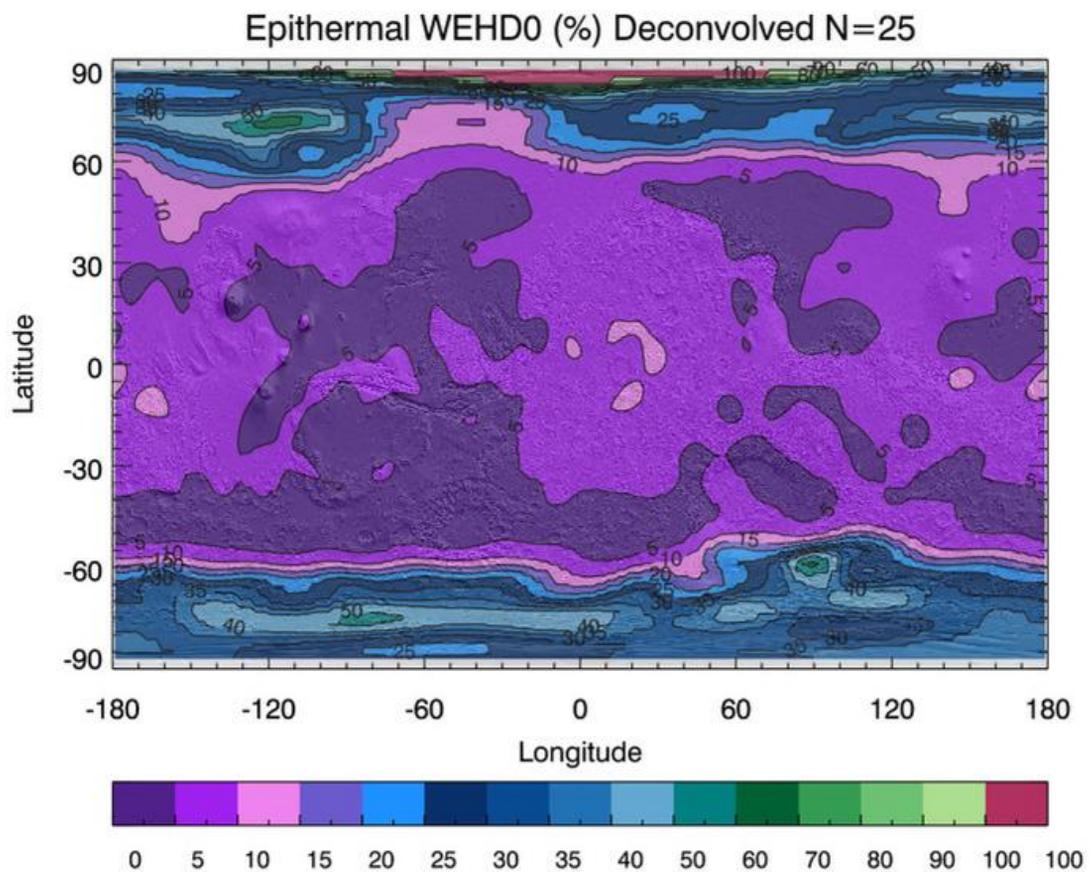

**Figure S5:** Global frost-free epithermal neutron 1D WEH zero-depth (WEHD0) maps for deconvolved **(A)** {N = 20} and **(B)** {N = 25} solutions. Maroon contours indicate unphysical WEHD0 > 100% results.

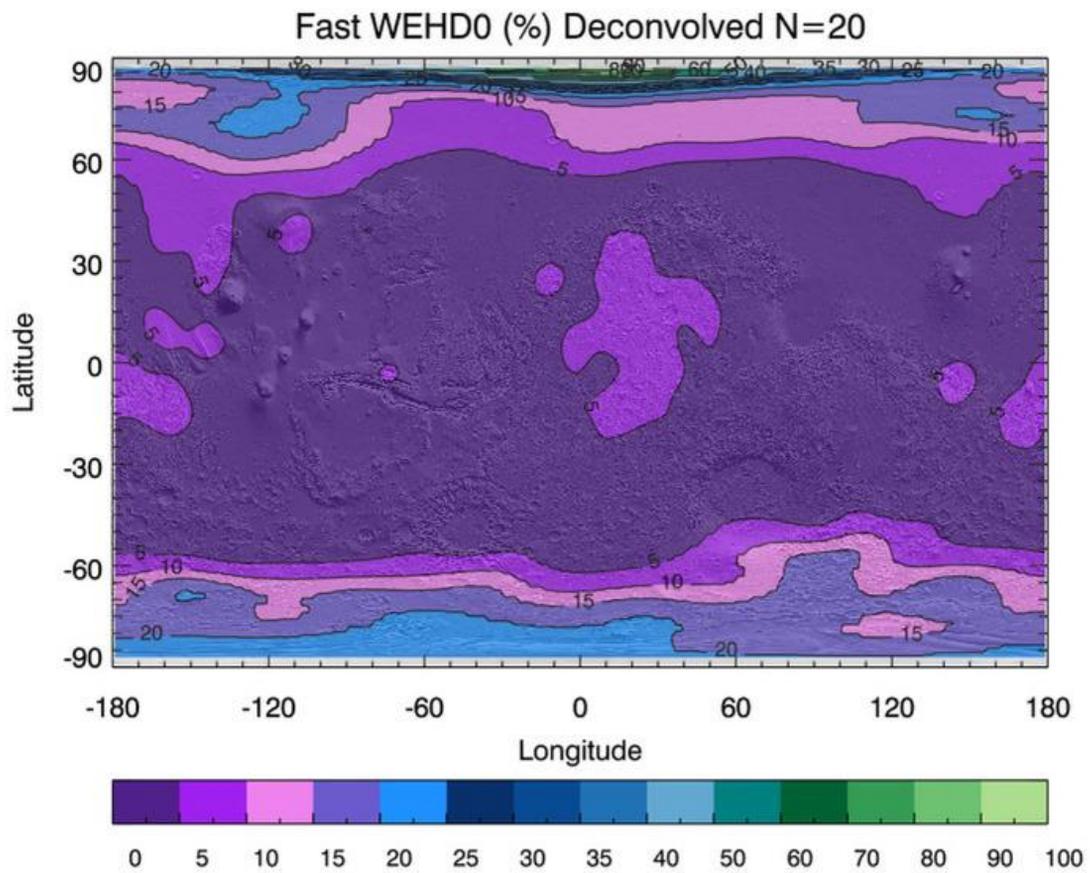

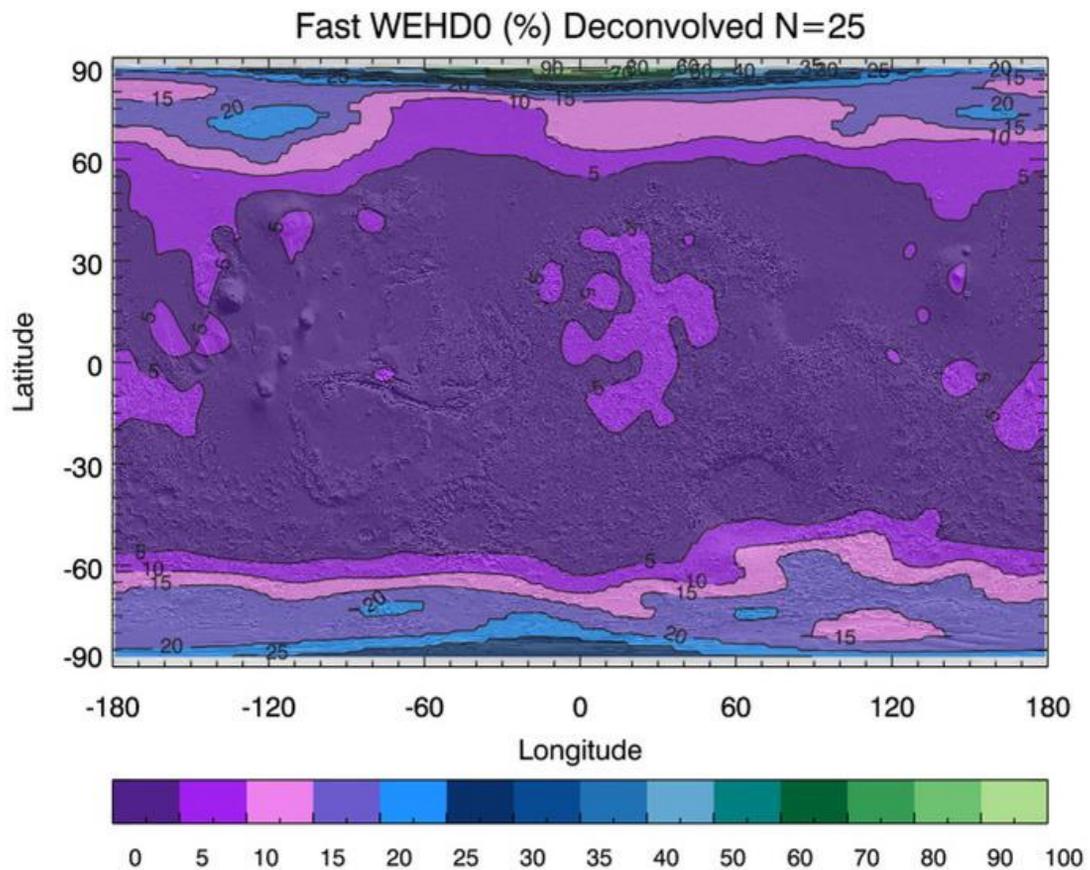

**Figure S5:** Global frost-free fast neutron 1D WEH zero-depth (WEHD0) maps, , expressed as a wt. % relative to pure ice, for deconvolved **(C)** {N = 20} and **(D)** {N = 25} solutions.

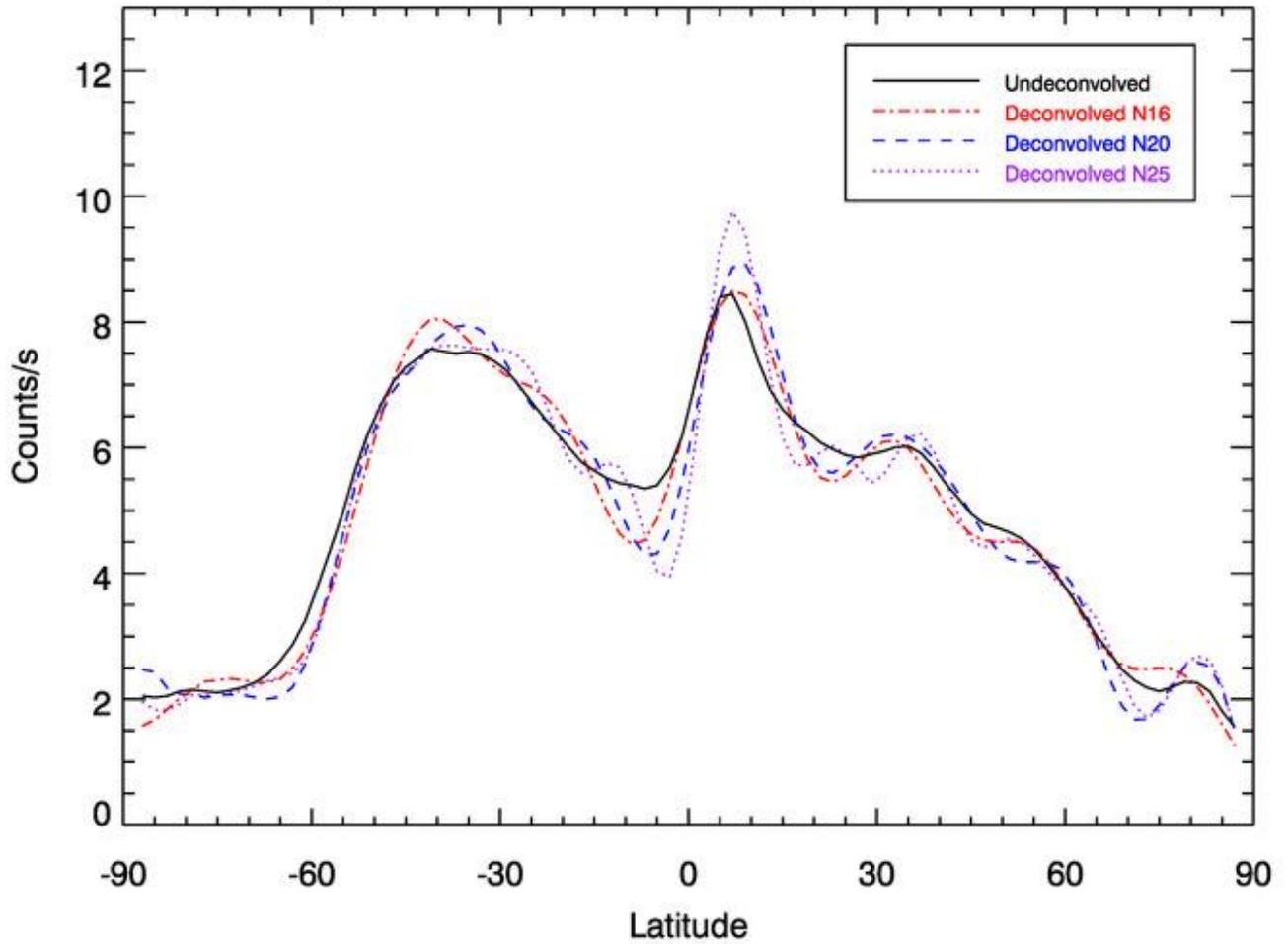

**Figure S6:** Longitudinal profiles of frost-free epithermal neutron flux (counts/s) for undeconvolved and deconvolved {N=16}, {N=20}, {N=25} solutions at 147°E.

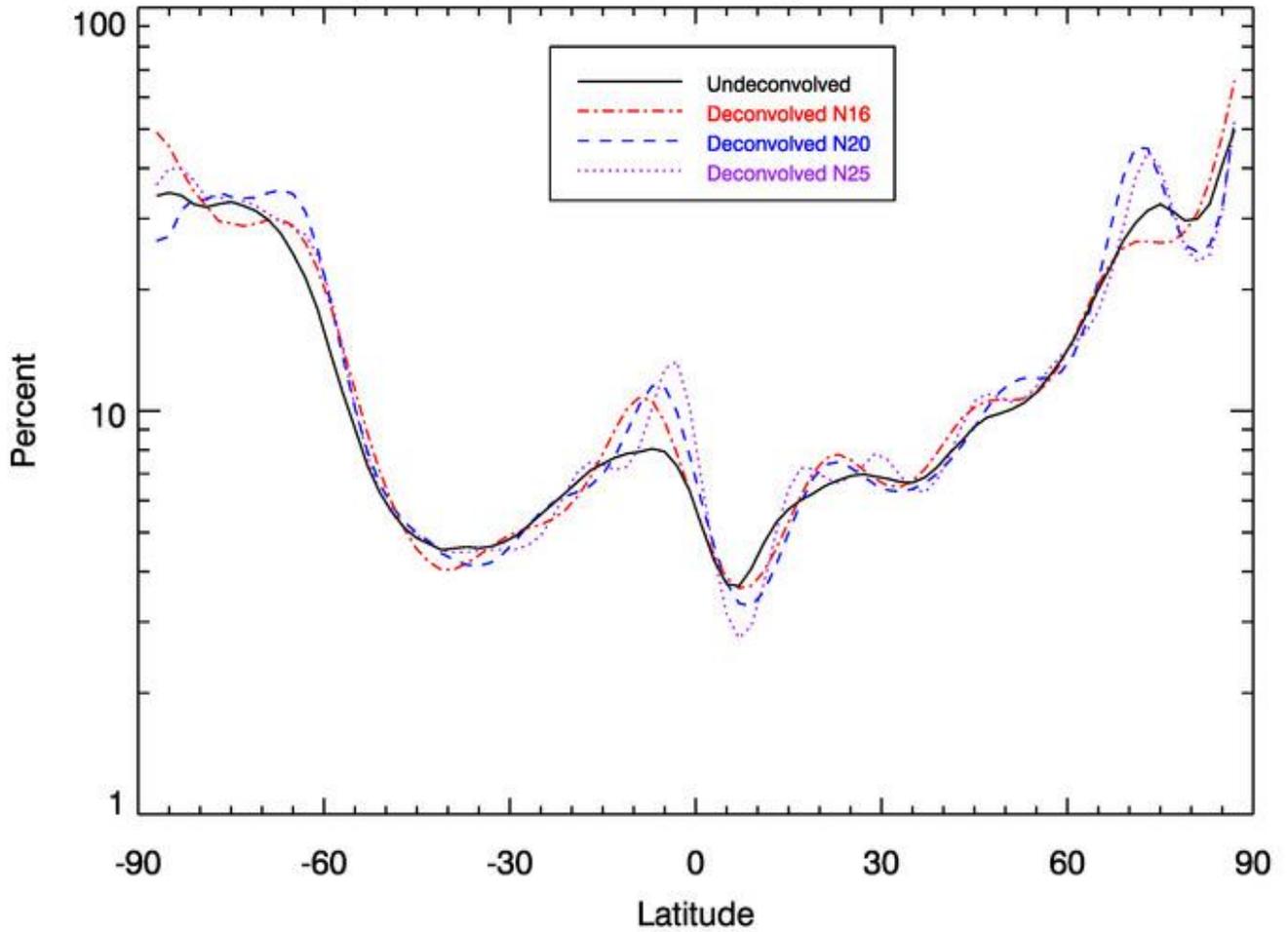

**Figure S7:** Longitudinal profiles of frost-free epithermal neutron WEHD0 (%) for undeconvolved and deconvolved {N=16}, {N=20}, {N=25} solutions at 147°E.

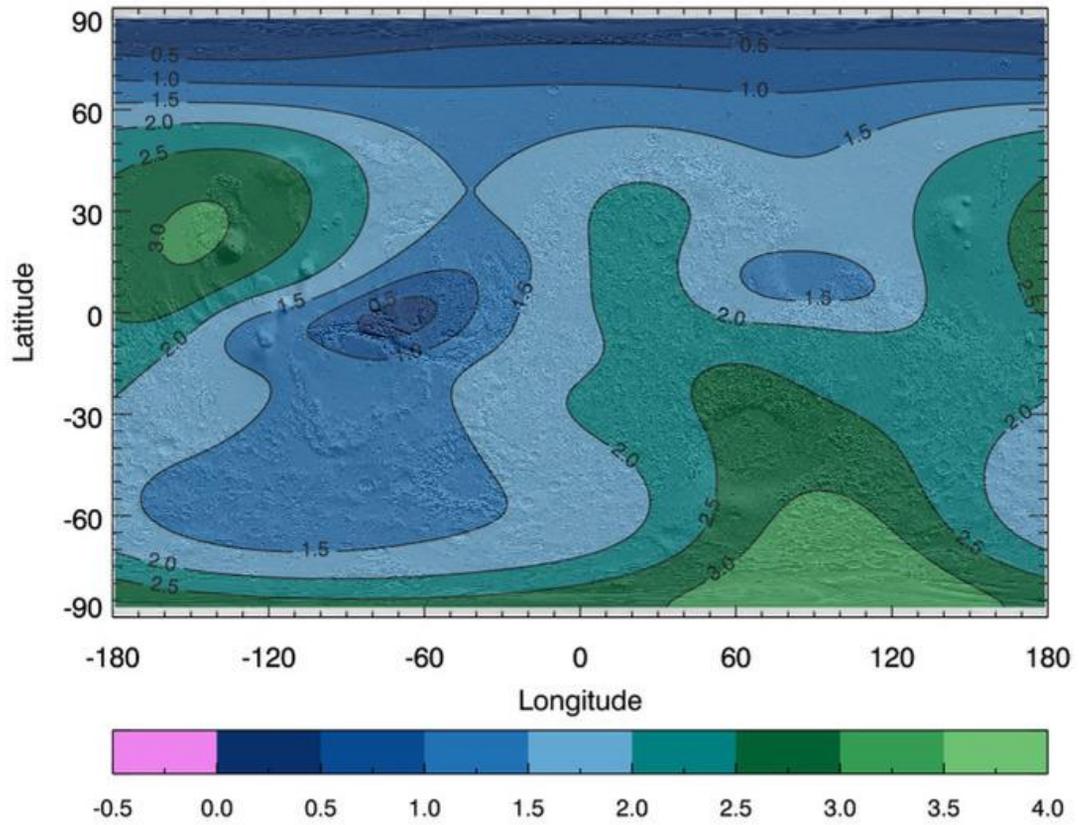
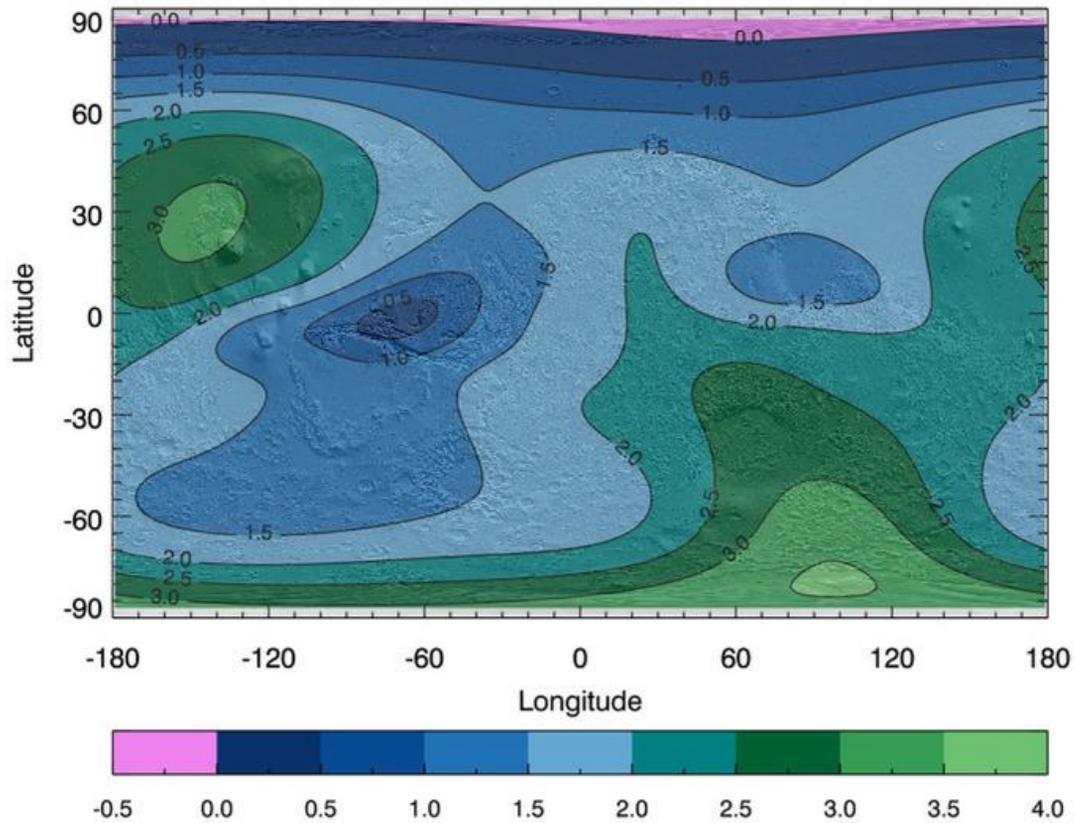

**Figure S8:** Global frost-free $W_{up}$ maps for deconvolved **(A)** {N = 20} and **(B)** {N = 25} solutions, calculated using Gaussian-weighted least squares fitting normalized to $R_o$ = 1300 km.

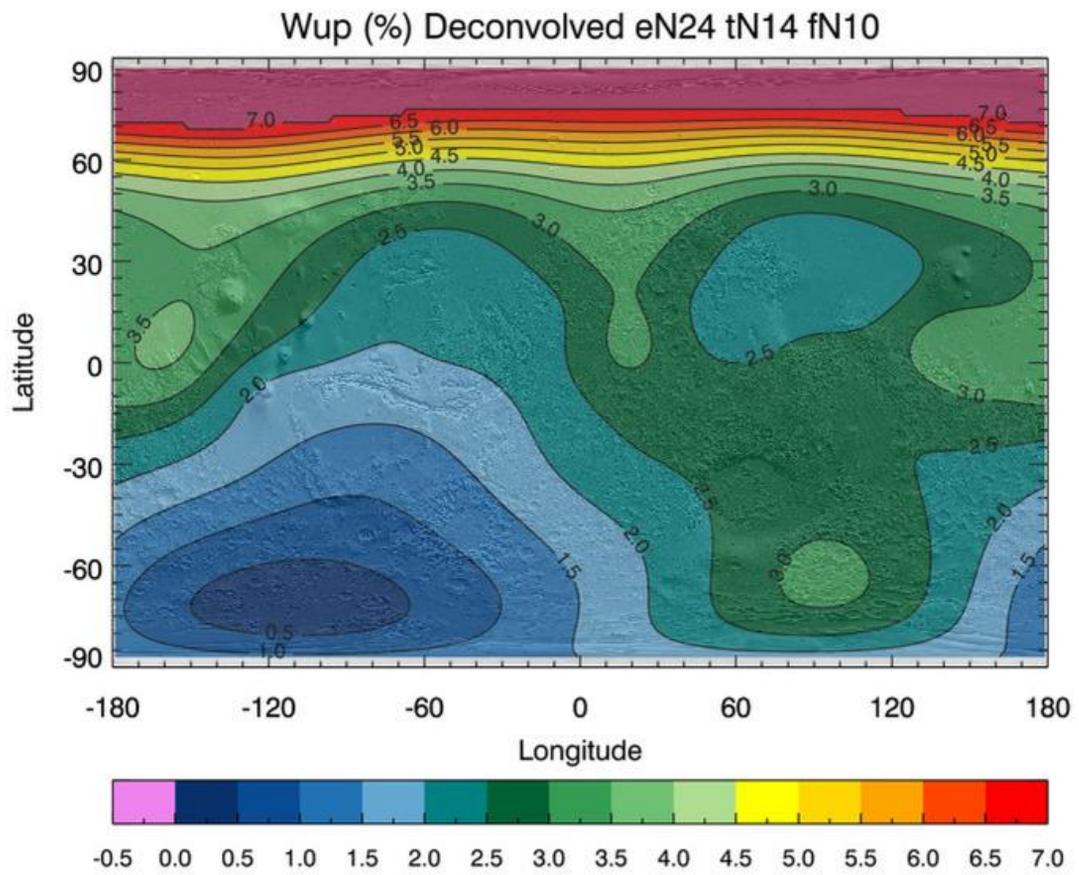

**Figure S9:** Global frost-free $W_{up}$ map for deconvolved epithermal {N = 24}, thermal {N = 14}, and fast {N=10} solution, calculated using Gaussian-weighted least squares fitting normalized to $R_o$ = 1300 km.

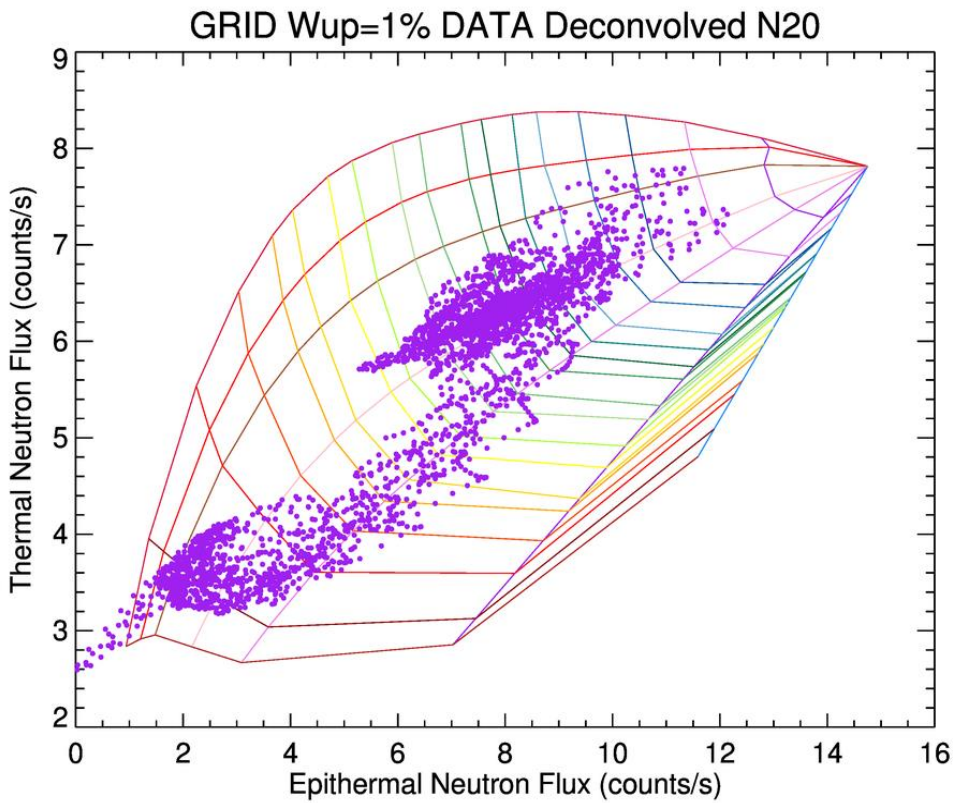

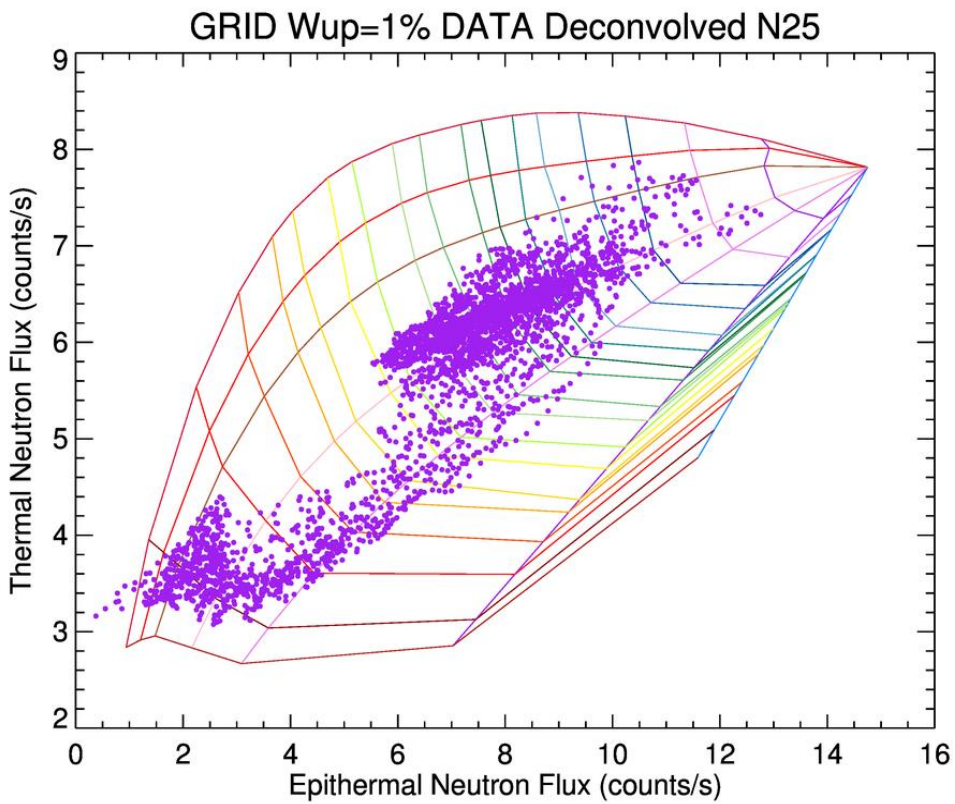

**Figure S10: (A)** Deconvolved {N = 20} MONS data for $W_{up} < 1.5\%$ on $W_{up} = 1\%$ model grid. **(B)** Deconvolved {N = 25} MONS data for $W_{up} < 1.5\%$ on $W_{up} = 1\%$ model grid. *Data poleward of 75°S have been excluded.*

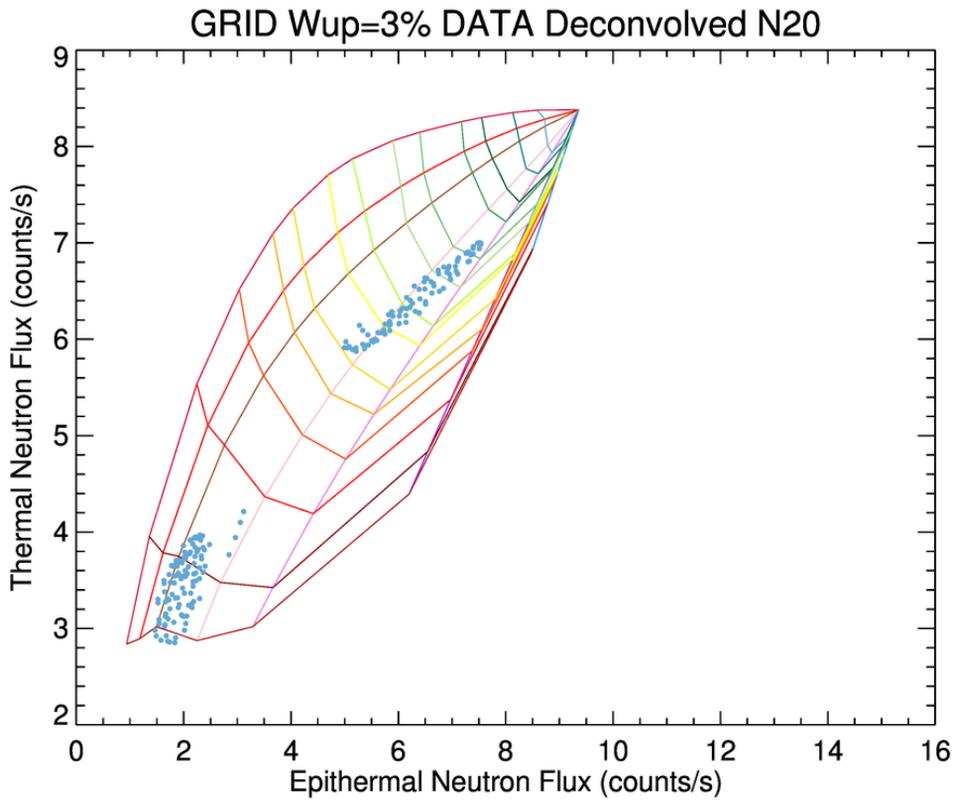

**Figure S10: (C)** Deconvolved {N = 20} MONS data for 3 < $W_{up}$ < 3.5% on $W_{up}$ = 3% model grid. *Data poleward of 75°S have been excluded.*

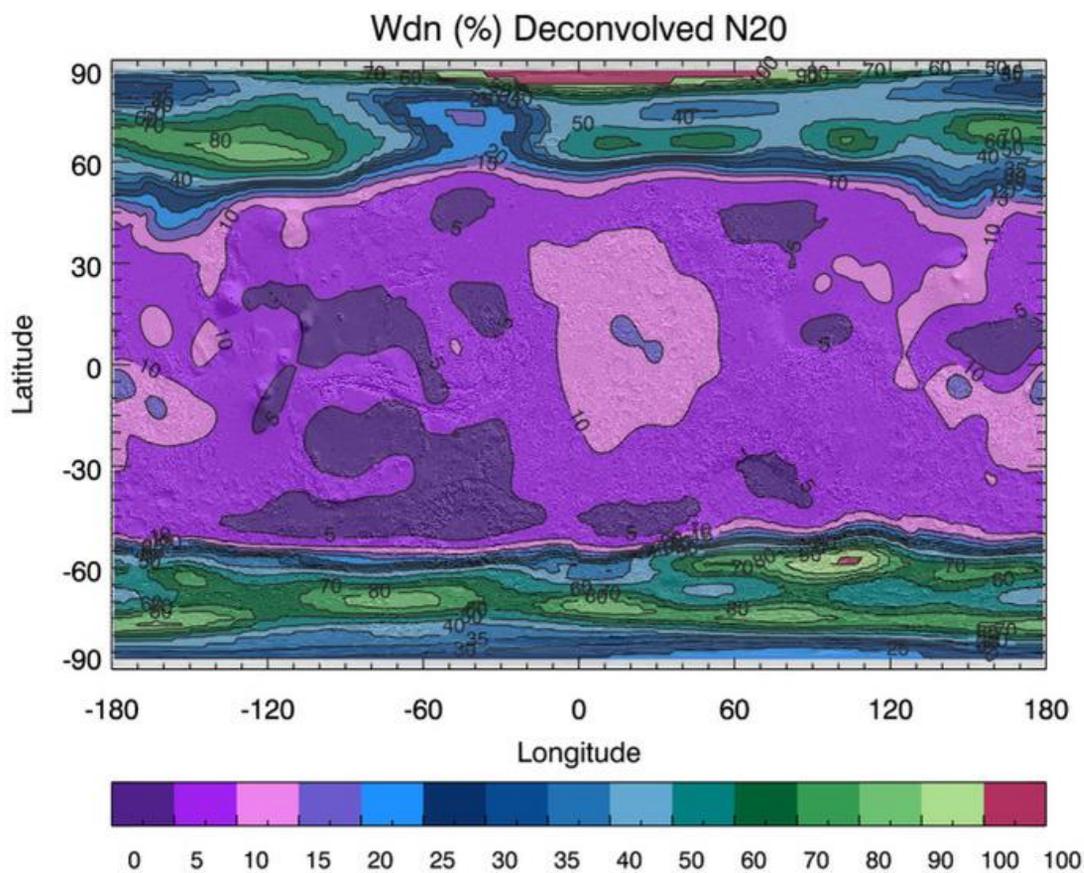
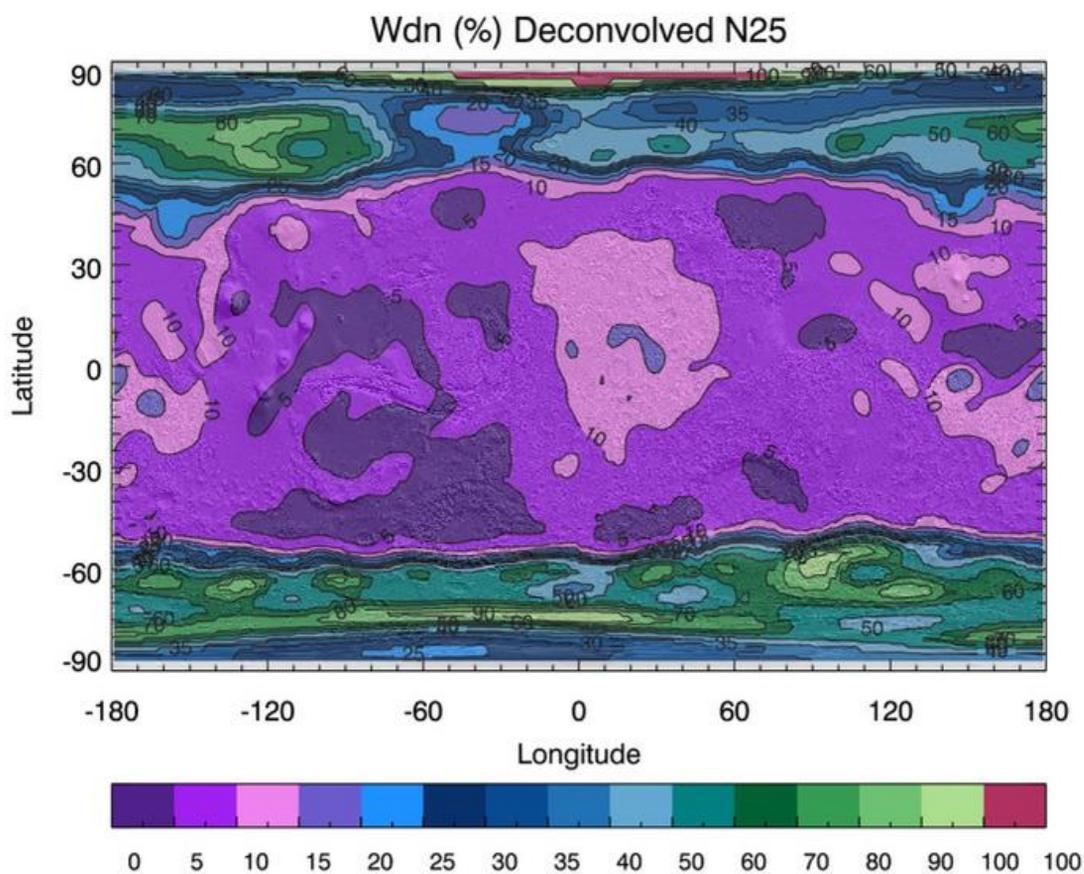

**Figure S11:** Global frost-free $W_{dn}$ maps for deconvolved **(A)** {N = 20} and **(B)** {N = 25} solutions. Maroon contours indicate unphysical WEHD0 > 100% results.

C

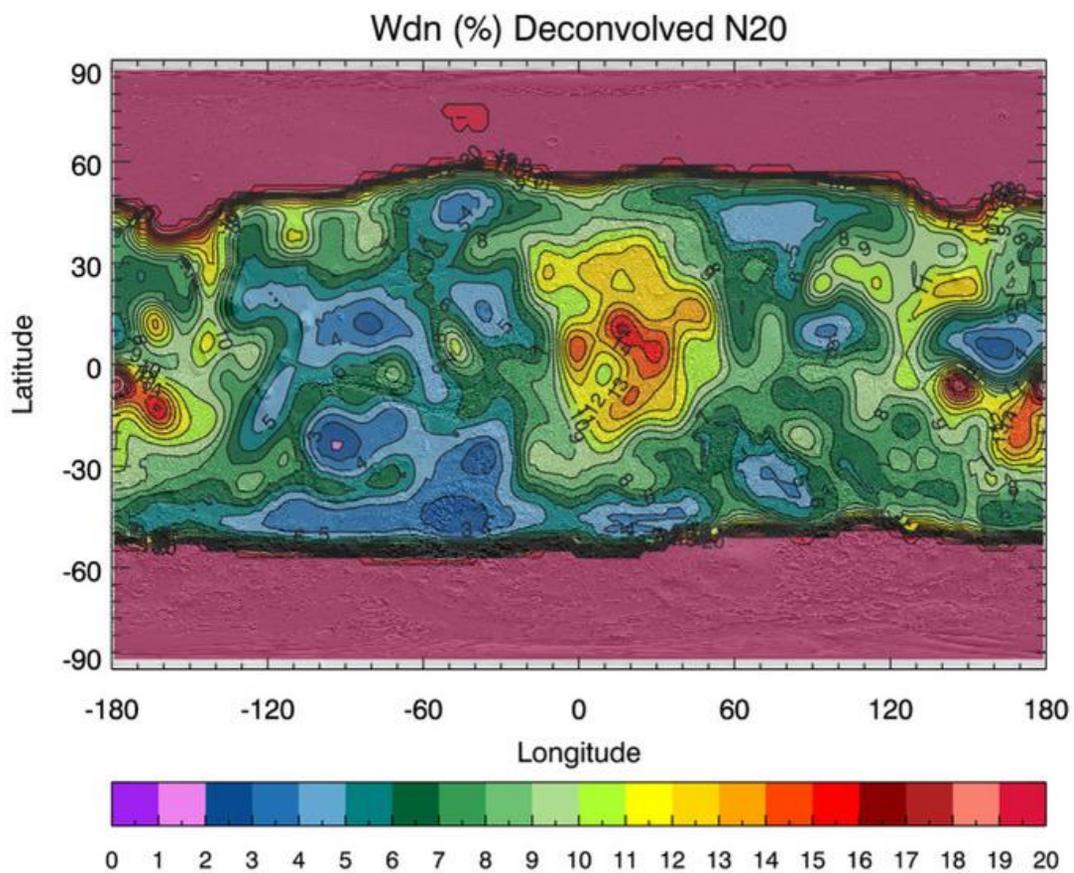

D

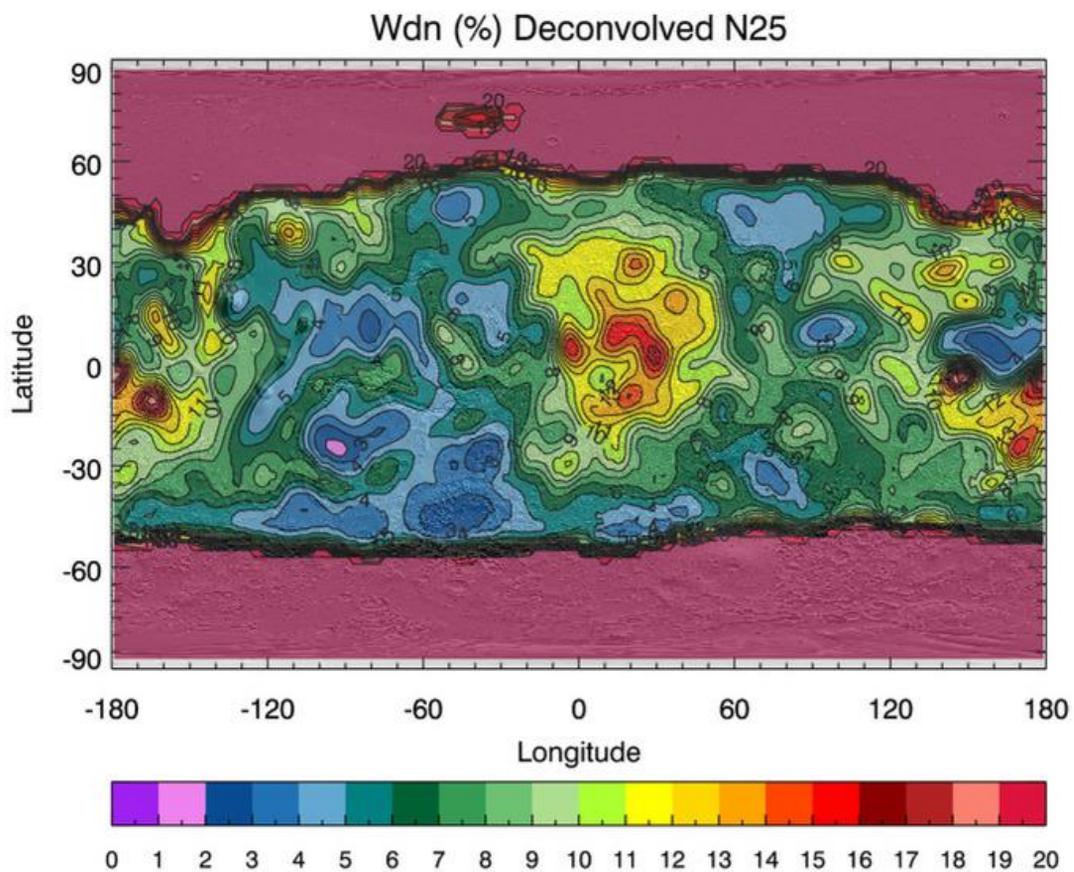

**Figure S11:** Global frost-free $W_{dn}$ maps for deconvolved **(A)** {N = 20} and **(B)** {N = 25} solutions, re-scaled to emphasize $W_{dn}$ variations at low latitudes.

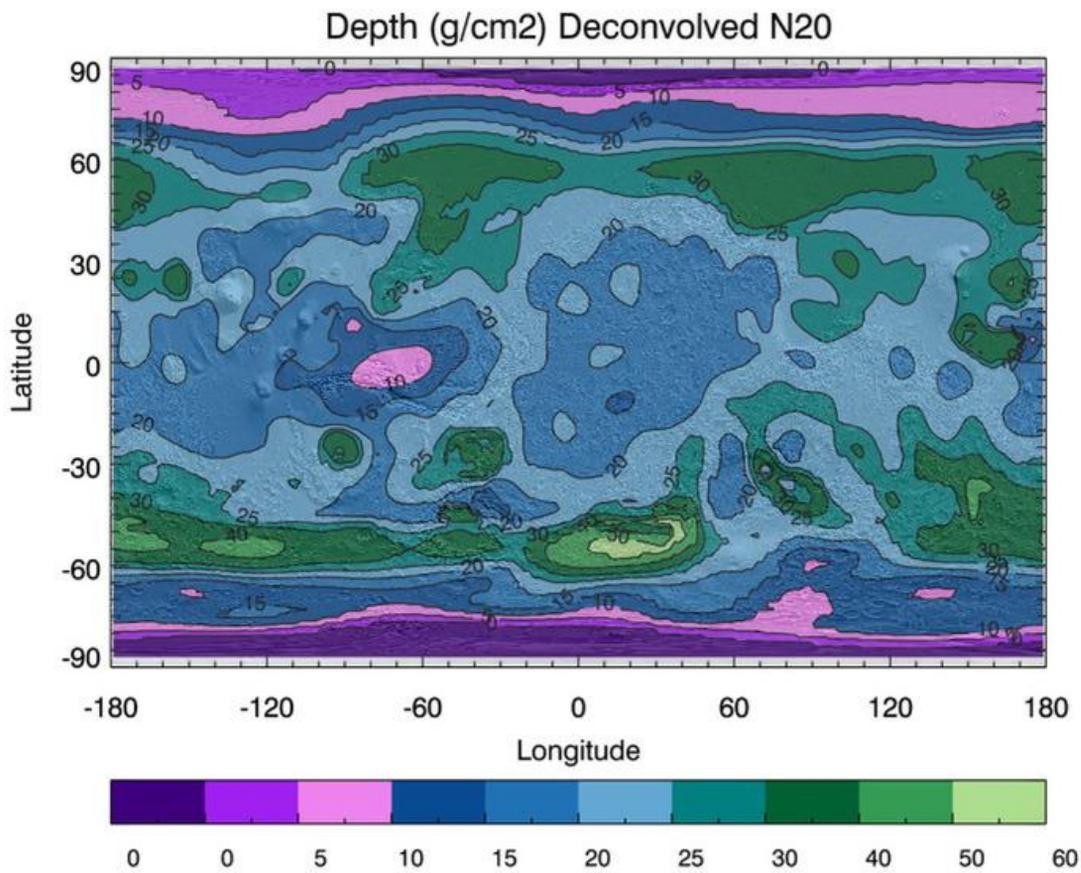
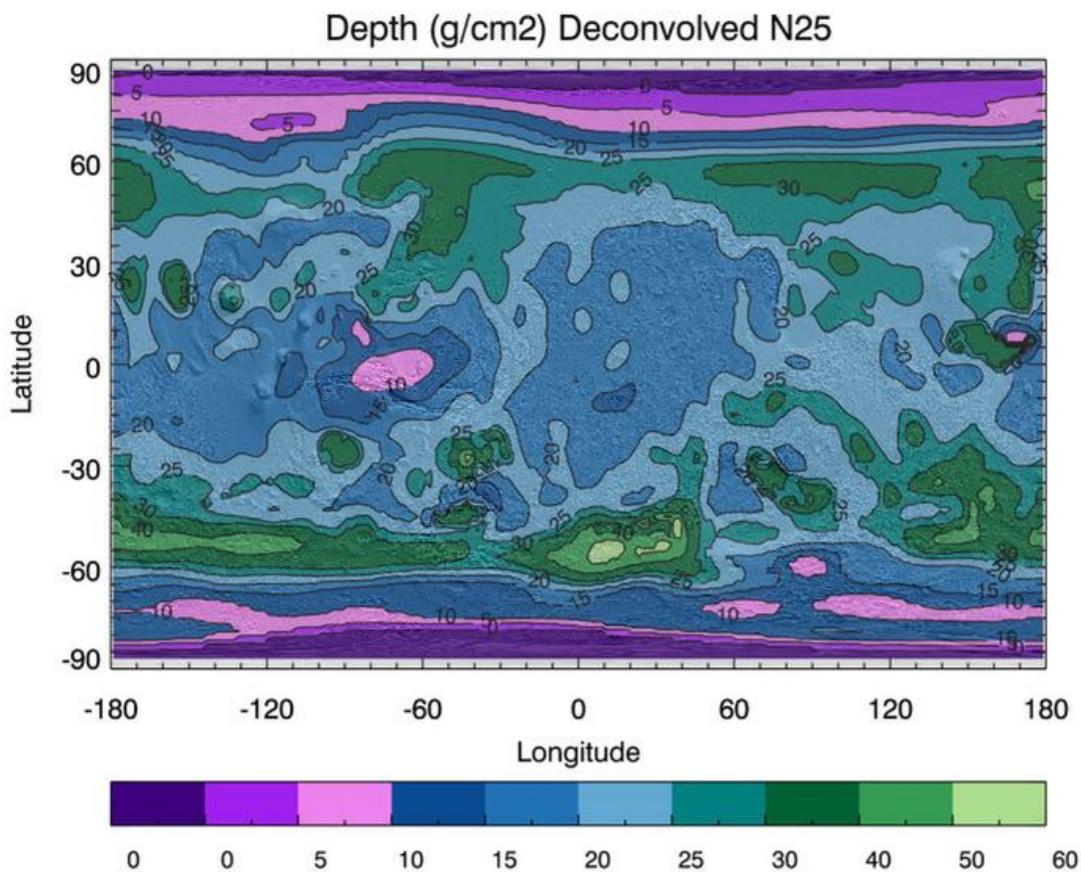

**Figure S12:** Global frost-free Depth maps for deconvolved **(A)** {N = 20} and **(B)** {N = 25} solutions. The darkest purple contour corresponds to *D* = 0, indicating inapplicability of standard two-layer model.

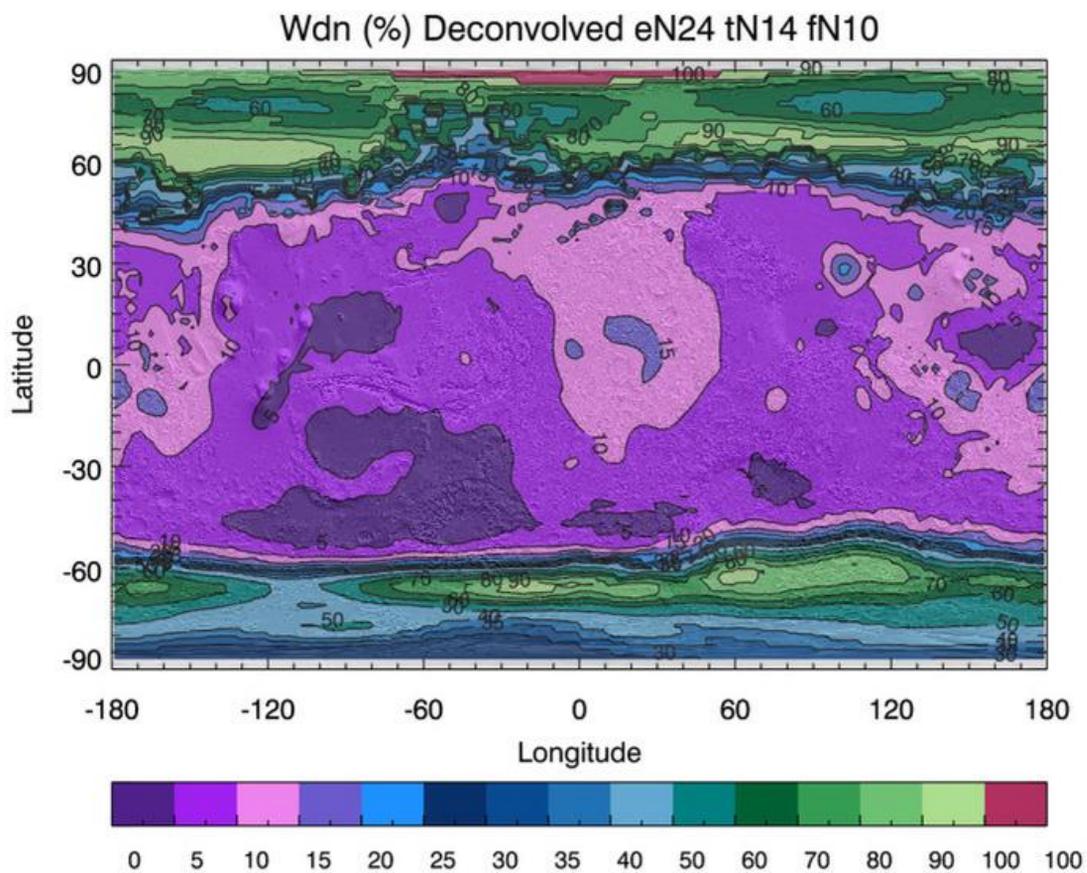

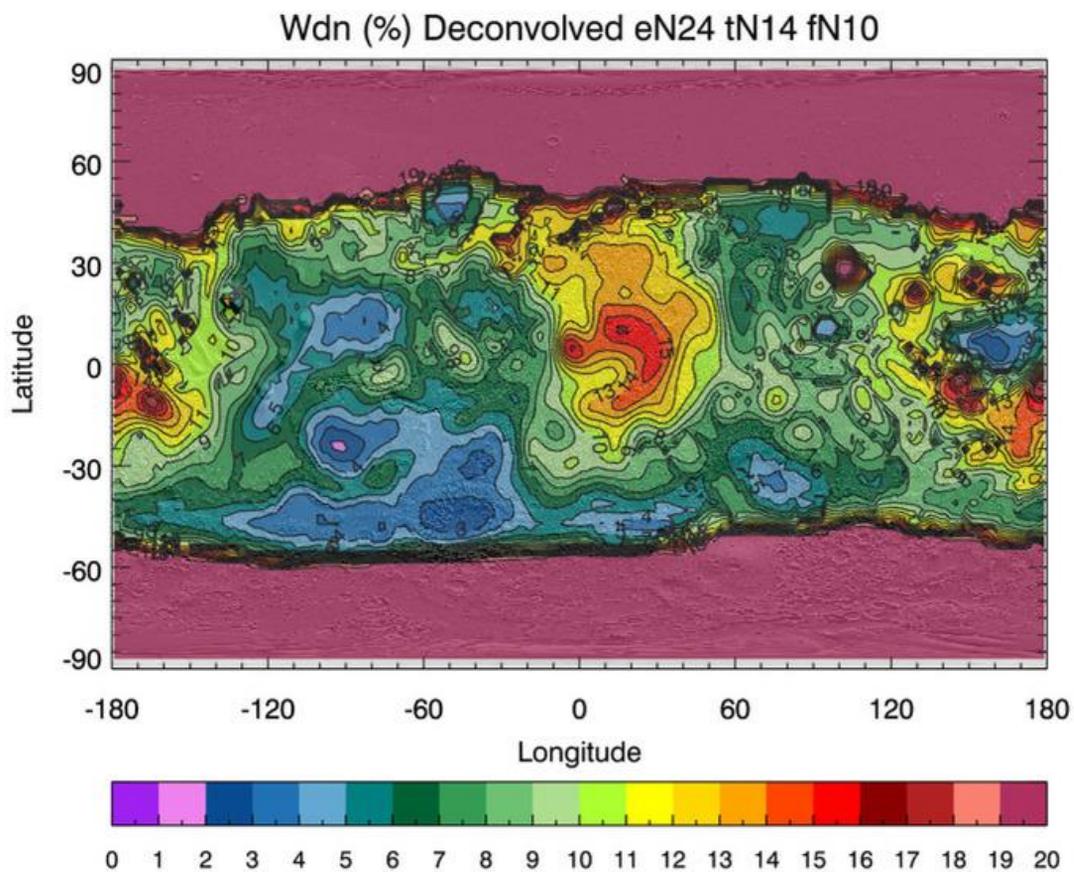

**Figure S13:** Global frost-free $W_{dn}$ map for deconvolved epithermal {N = 24}, thermal {N = 14}, and fast {N=10} solution, scaled from **(A)** 0 – 100 wt. % and **(B)** 0 – 20 wt. %.

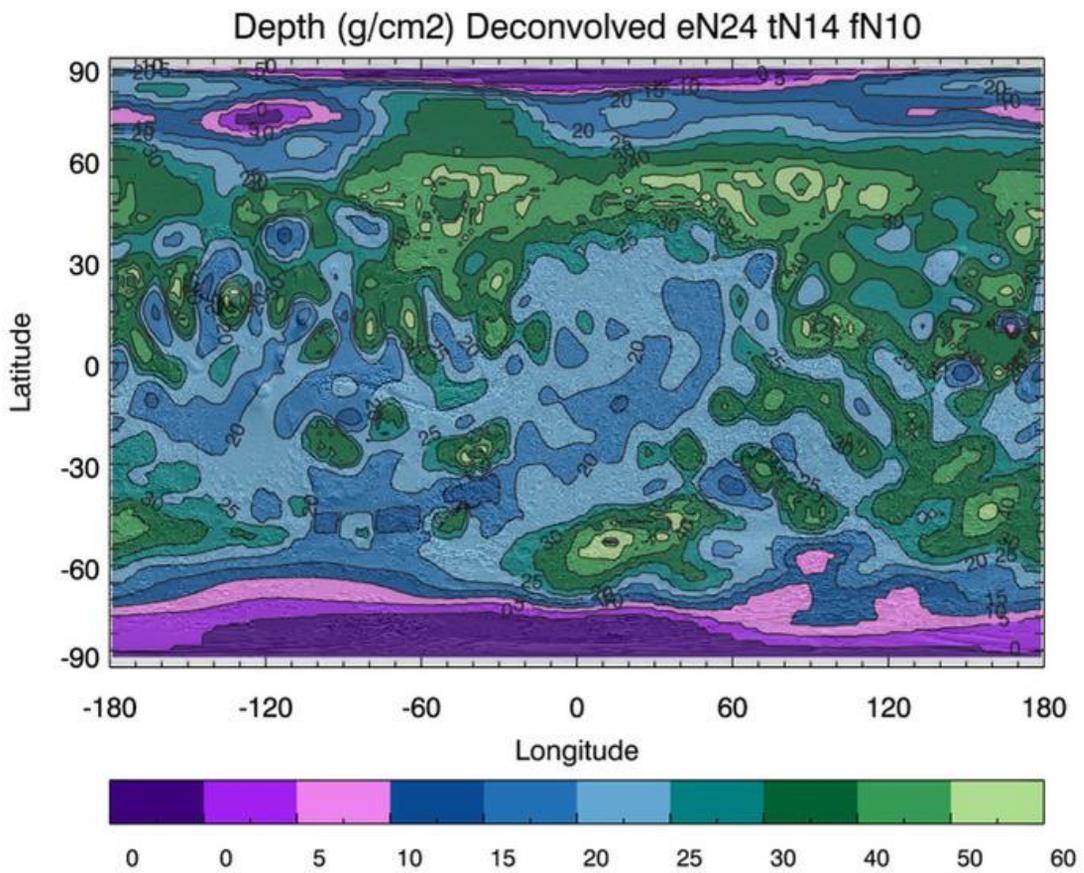

**Figure S14:** Global frost-free Depth maps for deconvolved epithermal {N = 24}, thermal {N = 14}, and fast {N=10} solution. The darkest purple contour corresponds to $D = 0$, indicating inapplicability of standard two-layer model.